\documentclass[a4paper,twocolumn,11pt,accepted=2022-02-25]{quantumarticle}
\pdfoutput=1
\usepackage[utf8]{inputenc}
\usepackage[english]{babel}
\usepackage[T1]{fontenc}
\usepackage{amsmath}
\usepackage{hyperref}

\usepackage{tikz}
\usepackage{lipsum}

\usepackage{graphicx}
\usepackage{color}
\usepackage{dcolumn}
\usepackage{bm}
\usepackage[mathlines]{lineno}

\usepackage[numbers,sort&compress]{natbib}
\usepackage{amssymb}

\newtheorem{definition}{Definition}
\newtheorem{corollary}{Corollary}
\newtheorem{lemma}{Lemma}
\newtheorem{theorem}{Theorem}

\newcommand{\Imag}{\mathrm{Im}}
\newcommand{\Real}{\mathrm{Re}}
\newcommand{\B}{\mathcal{B}}
\newcommand{\E}{\mathcal{E}}
\newcommand{\hil}{\mathcal{H}}
\newcommand{\ketbra}[1][]{| #1 \rangle \langle #1 |}
\newcommand{\ket}[1][]{| #1 \rangle}
\newcommand{\bra}[1][]{\langle #1 |}
\newcommand{\vect}[1][]{\mathbf{ #1 }}
\newcommand{\vecn}{{\mathbf{n}}}

\newcommand{\trace}{\mathrm{Tr}}
\newcommand{\id}{\mathbb{I}}
\newcommand{\dif}{\mathrm{d}}

\newcommand{\state}{\rho_{\bm x}}
\newcommand{\param}{{\bm x}}
\newcommand{\region}{X}
\newcommand{\transpose}{\mathrm{T}}
\newcommand{\model}{\{ \rho_{\bm x} | {\bm x} \in X \}}
\newcommand{\estimator}{\check{\bm x}}
\newcommand{\msem}{{\bm \Sigma}_{\bm x}}

\newcommand{\pspace}{{\bm x} \in X}
\newcommand{\eqspure}{| \Psi_{\bm x}^\mathrm{EQS} \rangle}
\newcommand{\eqsmix}{\rho_{\bm x}^\mathrm{EQS}}
\newcommand{\sld}[1][]{L_{{\bm x}, #1 }}
\newcommand{\antiu}{\Theta}
\newcommand{\conj}{\theta}
\newcommand{\fisher}{\mathcal{F}}
\newcommand{\swap}{S}
\newcommand{\nq}{N_Q}
\newcommand{\nc}{N_C}

\newcommand{\qed}{\hfill\blacksquare}

\begin{document}
\title{Imaginarity-free quantum multiparameter estimation}
\author{Jisho Miyazaki}
\email{miyzakijisho@gmail.com}
\affiliation{Shinpouin, Mikataharacho, Kita-ku, Hamamatsu, Shizuoka, 433-8105, Japan}
\author{Keiji Matsumoto}
\email{keiji@nii.ac.jp}
\affiliation{Quantum Computation Group, National Institute of Informatics,
2-1-2 Hitotsubashi, Chiyoda-ku, Tokyo 101-8430}

\begin{abstract}
Multiparameter quantum estimation is made difficult by the following three obstacles.
First, incompatibility among different physical quantities poses a limit on the attainable precision.
Second, the ultimate precision is not saturated until you discover the optimal measurement.
Third, the optimal measurement may generally depend on the target values of parameters, and thus may be impossible to perform for unknown target states.

We present a method to circumvent these three obstacles.
A class of quantum statistical models, which utilizes antiunitary symmetries or, equivalently, real density matrices, offers compatible multiparameter estimations.
The symmetries accompany the target-independent optimal measurements for pure-state models.
Based on this finding, we propose methods to implement antiunitary symmetries for quantum metrology schemes.
We further introduce a function which measures antiunitary asymmetry of quantum statistical models as a potential tool to characterize quantumness of phase transitions.

\end{abstract}

\maketitle

\section{Introduction}
Incompatibility residing in simultaneous measurements of different quantities has been a widely acknowledged character of quantum mechanics.
Besides the philosophical implications, the measurement incompatibility imposes practical limitations on quantum information processing.

The measurement incompatibility is particularly troubling for quantum metrological tasks which aim at estimating parameters with ever increasing precision \cite{Giovanettietal2006,Giovanettietal2011}.
In its early stages, quantum metrology focused on single-parameter estimations.
Quantum correlations unlock a region of precision which could never be reached by classical mechanics \cite{Demkowiczetal2015,Degenetal2017,Pezzeetal2018,Braunetal2018,Pirandolaetal2018,BercheraDegiovanni2019}.
The ultimately attainable precision limits are given by quantum Cram\'{e}r-Rao bounds (QCRBs). \cite{Helstrom1967,Helstrom1968,Holevo2011,Hayashi2017quantum}.

Quantum correlations remain advantageous for various quantum multiparameter metrological schemes \cite{Szczykulskaetal2016,Albarellietal2019}.
Simultaneous estimations outperform combinations of individual estimations for sensing multi-dimensional fields \cite{BaumgratzDatta2016} and imaging \cite{Humphreys2013,Gagatsosetal2016}.
This motivates experimental realizations of multi-parameter metrology \cite{Vidrighin2014joint,Rocciaetal2018,Parniaketal2018,Polinoetal2019}.

Despite the rapid growth of multiparameter estimation, the status of multiparameter QCRBs \cite{Helstrom1976,Holevo1982} has always been under doubt.
There are systems which do not allow compatible estimation, rendering QCRBs unattainable \cite{Crowleyetal2014}.
Therefore, conditions to realize compatible multiparameter estimation have been under intense scrutiny \cite{Matsumoto2002,Vanephetal2013,Vidrighin2014joint,Crowleyetal2014,Suzuki2016,Ragyetal2016,BaumgratzDatta2016,Pezzeetal2017optimal,Yang2019partialcommutativity,Napolietal2019,Kukita2020,BelliardoGiovannetti2021}.
Attainable precision limits for incompatible estimations are also avidly studied \cite{Holevo1976,Holevo2011,GutaKahn2006,HayashiMatsumoto2008,KahnGuta2009,Yamagata2013,Yangetal2019attaining,Suzuki2016,Matsumoto2002,Bradshawetal2017,Bradshawetal2018,Gorecki2020optimalprobeserror,Albarellietal2019,Carollo2019,Yamagata2021,Sidhuetal2021,Goldbergetal2021}.
Whether or not the QCRB is saturated, identifying attainable precision limit is the first step of quantum multiparameter metrology.

However, even if one could identify the best precision bound, optimal measurements to attain it can be difficult to perform, and further investigations have been carried out in order to study the implementation of the optimal measurements for given quantum statistical models \cite{Matsumoto2002,Humphreys2013,Tsangetal2016,Pezzeetal2017optimal,Rehaceketal2018optmeas,Yang2019partialcommutativity,Yangetal2019attaining,Len2021,Almeidaetal2021collective}.
Somewhat paradoxically, preceding methods assume knowledge on the target point $\param$ to be estimated in order to identify the optimal measurement \cite{Matsumoto2002,Humphreys2013,Pezzeetal2017optimal,Yang2019partialcommutativity}, rendering the parameter estimation task to, strictly speaking in the case, applicable only for local parameter estimation.

Yang \emph{et. al.} \cite{Yangetal2019attaining} presents a method to attain the best precision limit without any prior knowledge of the target point.
The method is composed of two steps: a rough expectation of the target-point by a tomography, and the precise estimation with the optimal measurement depending on the rough expectation.
The measurement setting has to be adaptively updated according to the previous measurements.
Thus the measurement still depends on the target-point.

Another drawback of proposed optimal measurements is the inclusion of projectors onto the target state \cite{Humphreys2013}.
Such a projector could be highly entangled for quantum metrology.
This is why optimal measurements are sometimes given up in multiparameter quantum metrology \cite{BaumgratzDatta2016}.

This article proposes the use of antiunitary symmetries to side-step the problems of incompatibility and unfeasible measurement in multiparameter quantum estimation.
Any quantum statistical model endowed with an antiunitary symmetry is compatible, which means that the QCRB is attainable in the asymptotic limit.
Furthermore, if the system consists only of pure states, the antiunitary symmetry guarantees a target-independent optimal measurement.
The construction of globally optimal measurements requires a step beyond the studies of compatibility conditions \cite{Matsumoto2002,Ragyetal2016,BaumgratzDatta2016,Pezzeetal2017optimal,Yang2019partialcommutativity,BelliardoGiovannetti2021}.
Our optimal measurement has a continuous degree of freedom, and one can choose any of them to achieve the global optimality.

With these advantages in mind, we present methods to implement antiunitary symmetries in multiparameter quantum metrology.
The so-designed pure-state models for quantum metrology have globally optimal measurements by construction.
Our methods do not sacrifice the attainable precisions when implementing antiunitary symmetries.

Though unrecognized, popular models such as N00N states \cite{LeeKokDowling2002GHZequivalentNOON,Giovanettietal2004science,Giovanettietal2011}, antiparallel spins \cite{GisinPopescu1999,Changetal2014} and 3D magnetometry \cite{BaumgratzDatta2016} enjoy antiunitary symmetry.
For parameter estimations of these models, we point out how to make the full use of their antiunitary symmetry.
For example, we obtain an optimal measurement for the 3D magnetometry which is only bipartite-entangled irrespective to the system size.

Besides practical implications, our result bridges two seemingly different concepts, namely, the incompatibility and the imaginarity.
The incompatibility characterizes quantumness of phase transitions in quantum statistical mechanics \cite{Carollo2018,Carolloetal2018fermion,Leonforte2019muc,Basconeetal2019muc,Basconeetal2019mucPRB,Carollo2019}.
The imaginarity is formulated as a meaningful resource in quantum information processing \cite{HickeyGour2018imaginarity,Wu2021imaginarityPRL,Wu2021imaginarityPRA}.
We refine the relation between these two concepts.

We start by reviewing theories of quantum multi-parameter estimation and antilinear operators in Sec.~\ref{sec:pre}.
Section \ref{sec:symmetry} presents the notion of antiunitary symmetry of quantum statistical models, with its direct link to compatible, optimal, and target-independent multiparameter estimations.
Then we propose methods to implement antiunitary symmetry in metrological settings in Sec.~\ref{sec:reduction}.
Section \ref{sec:discussions} offers discussions concerning a refinement of the link between antiunitary symmetry and compatibility.
We offer our conclusion at Sec.~\ref{sec:conclusion}.

\section{Preliminaries}\label{sec:pre}

\subsection{Quantum multi-parameter estimation theory}
Let us consider an estimation of parameters that determines density operators of finite dimensional Hilbert space $\hil$.
The density operator $\state$ is parametrized by $\param = (x_1, x_2,...,x_n)^\transpose \in \region \subset \Re^n$.
The set $\model$ is called a \emph{quantum statistical model} or simply a \emph{model}.
The task is to estimate parameters $\param$ from results of quantum measurements performed on multiple copies $\state^{\otimes \nq \times \nc}$.
The identical measurement on $\state^{\otimes \nq}$ is assumed to be performed independently $\nc$ times.
Each measurement on $\state^{\otimes \nq}$ can be entangled over the $\nq$ copies.
An estimation process is described by the pair of a positive operator-valued measures (POVMs) $\Pi = \{ \Pi_\omega \in \B(\hil^{\otimes \nq}) \}_{\omega \in \Omega}$ and an estimator $\estimator : \Omega^{\nc} \rightarrow \region$, where $\Omega$ is the set of measurement results ($\B$ stands of the space of bounded linear operators).

Precision of estimation is characterized by the following mean square error matrix:
\begin{align}
	\nonumber &\msem (\Pi,\estimator) \\
	\label{msem} &:= \sum_{{\bm \omega} \in \Omega^{\nc}} p({\bm \omega} | \param) \left( \estimator({\bm \omega}) - \param \right) \left( \estimator({\bm \omega}) - \param \right)^\transpose,
\end{align}
where $p({\bm \omega}|\param) = \prod_{k=1}^{\nc} \trace[ \Pi_{\omega_k} \state^{\otimes \nq} ]$ for ${\bm \omega} = (\omega_1,...,\omega_{\nc})^\transpose$.
For estimators satisfying
\begin{equation}
\label{finite_consistent}	\sum_{{\bm \omega} \in \Omega^{\nc}} p({\bm \omega} | \param) \estimator({\bm \omega}) = \param,
\end{equation}
the mean square error matrix is a covariance matrix around the target-point (true value) $\param$.
We consider locally unbiased estimators and consistent estimators in this article.
The estimator is called locally unbiased at $\param$ when \eqref{finite_consistent} and
\begin{equation}
	\sum_{{\bm \omega} \in \Omega^{\nc}} \check{x}_i ({\bm \omega}) \partial_j p({\bm \omega}|\param) = \delta_{ij},
\end{equation}
holds at the point $\param$ ($\partial_j$ stands for the partial derivative with respect to $x_j$).
The estimator is called consistent if \eqref{finite_consistent} is satisfied at any point $\param$ in the asymptotic limit, namely, if
\begin{equation}
\label{consistent}	\lim_{\nc \rightarrow \infty} \sum_{{\bm \omega} \in \Omega^{\nc}} p({\bm \omega} | \param) \estimator({\bm \omega}) = \param, ~(\forall \pspace)
\end{equation}
holds (to be more precise, consistency is defined on a sequence of estimators $\estimator_{\nc}: \Omega^{\nc} \rightarrow \region$ for $\nc \in \mathbb{N}$).
The maximum likelihood estimator
\begin{equation}
	\estimator = \mathrm{argmax} \left\{ \sum_{l=1}^{\nc} \ln p(\omega_l | \param) \middle| \pspace \right\},
\end{equation}
is consistent.
While consistent estimators assume sufficiently many rounds of measurements, the unbiasedness condition \eqref{consistent} holds globally over the parameter space $\region$.

One of the goals of quantum multiparameter estimation is to find the optimal estimator $\estimator$ and measurement $\Pi$ to minimize the covariance matrix $\msem (\Pi,\estimator)$.
Classical and quantum Cram\'{e}r-Rao bounds deal with the minimization over estimators and measurements, respectively.
Define the $n \times n$ classical Fisher information matrix (CFIM) for measurement $\Pi = \{ \Pi_\omega \in \B(\hil^{\otimes \nq}) \}_{\omega \in \Omega}$ by
\begin{align}
	\nonumber & [\fisher_C (\Pi,\param)]_{ij} \\
	&:= \sum_{\omega \in \Omega} \frac{\partial_i \trace[ \Pi_\omega \state^{\otimes \nq} ] \partial_j \trace[ \Pi_\omega \state^{\otimes \nq} ] }{\trace[ \Pi_\omega \state^{\otimes \nq} ]},
\end{align}
where $\partial_i$ represents the partial derivative with respect to $x_i$.
CFIM provides the classical Cram\'{e}r-Rao bound of the covariance matrix
\begin{equation}
\label{classicalCRB}	\nc \msem (\Pi,\estimator) \geq \fisher_C (\Pi,\param)^{-1}.
\end{equation}
At any point $\pspace$, there exists an $\param$-dependent locally unbiased estimator saturating the inequality for any $\nc$ (see e.g. Chapter 2 of \cite{AmariNagaoka2000} for an explicit construction of the estimator).
Note that the locally unbiased estimator at $\param$ is generally impossible to construct if the target-point $\param$ is unknown.
The maximum likelihood estimator saturates the same inequality globally in the asymptotic limit $\nc \rightarrow \infty$.

The remaining optimization is about the measurement.  
For general model $\model$, symmetric logarithmic derivatives (SLDs) $\sld[i] \in \B(\hil)$ ($i=1,...,n$) are Hermitian operators defined indirectly by
\begin{equation}
\label{sld}	2 \partial_i \state = \sld[i] \state + \state \sld[i].
\end{equation}
The solution of Eq.~\eqref{sld} is not unique but one is given by
\begin{equation}
\label{integralSLD}	\sld[i] = 2 \int_0^\infty e^{-t \state} \partial_i \state e^{-t \state} \dif t.
\end{equation}
SLDs define $n \times n$ quantum Fisher information matrix (QFIM) via \cite{Helstrom1967,Helstrom1968}
\begin{equation}
	[\fisher_Q(\param)]_{ij} := \Real \left[ \trace[\state \sld[i] \sld[j] ] \right],
\end{equation}
and provides the quantum Cram\'{e}r-Rao bound (QCRB) of the CFIM
\begin{equation}
\label{HCRB}	\fisher_C (\Pi,\param) \leq \nq \fisher_Q(\param),
\end{equation}
which holds for any measurement $\Pi$ \cite{Helstrom1976,Holevo1982}.
While SLDs are not necessarily determined uniquely by Eq.~\eqref{sld}, the QFIM is independent of the choice of SLDs.

For multiparameter estimation $(n \geq 2)$, the QCRB is not necessarily saturated by any measurement, unlike for single-parameter estimations.
This limitation is caused by the incompatibility between estimations of different parameters.
A necessary and sufficient condition for the saturation $\fisher_C (\Pi,\param) = \nq \fisher_Q(\param)$ in the limit $\nq \rightarrow \infty$ is the \emph{weak commutativity} \cite{Matsumoto2002,Ragyetal2016}:
\begin{equation}
\label{uhlmann} 	[\mathcal{U}(\param)]_{ij} := \frac{\Imag \left[\trace[\state \sld[i] \sld[j]] \right]}{2}  = 0. ~(\forall i,j)
\end{equation}
The matrix $\mathcal{U}$ is called as the mean Uhlmann curvature \cite{Carollo2018}, and is independent of the choice of SLDs.
In this article, compatible models refer to weakly commutative models.

If the saturation $\fisher_C (\Pi,\param) = \nq \fisher_Q(\param)$ occurs for $\nq=1$, then the multiple product of the same measurement on the copied model saturates it for any $\nq$, since the CFIM is additive with respect to the tensor product and separable measurements.
For $\nq=1$, a necessary condition for the saturation $\fisher_C (\Pi,\param) = \fisher_Q(\param)$ is the partial commutativity:
\begin{equation}
	\Pi_\param [\sld[i], \sld[j]] \Pi_\param =0,~(\forall i,j)
\end{equation}
where $\Pi_\param$ is the projector onto the support of $\state$ \cite{Yang2019partialcommutativity}.
A sufficient condition for the saturation of QCRB with $\nq=1$ is quasi-classicality represented by
\begin{equation}
[ \sld[i], \sld[j] ] = 0,~(\forall i,j),
\end{equation}
at the point $\pspace$ in question.
Quasi-classical models do not demand collective measurements for saturating the QCRBs.

All pure weakly commutative models are quasi-classical, and have measurements to saturate QCRB \eqref{HCRB} with $\nq=1$ \cite{Matsumoto2002}.
For a pure-state model $\state = \ketbra[\psi_\param]$, a necessary and sufficient condition for a POVM $\E$ on $\hil$ to saturate QCRB \eqref{HCRB} at $\pspace$ is \cite{Yang2019partialcommutativity},
\begin{align}
	\nonumber	& {\rm If} ~ \bra[\psi_\param] E \ket[\psi_\param] =0 ~ {\rm then}~ \exists \eta_{i,j} \in \Re,\\
		\label{Yang1} &  ~E \sld[i] \ket[\psi_\param] = \eta_{i,j} E \sld[j] \ket[\psi_\param], ~~(\forall i,~j),
\end{align}
and
\begin{align}
	\nonumber	& {\rm If} ~ \bra[\psi_\param] E \ket[\psi_\param] \neq 0 ~ {\rm then}~\exists \xi_i \in \Re,\\
		\label{Yang2} & E \sld[i] \ket[\psi_\param] = \xi_i E \ket[\psi_\param],~~(\forall i),
\end{align}
for any element $E$ of $\E$.
Note that condition \eqref{Yang1} should be interpreted so that when the vector $E \sld[j] \ket[\psi_\param]$ on the right-hand-side of the equation is zero, condition \eqref{Yang1} still holds with $\eta_{i,j}$ a real infinity \cite{Yang2021}.
More precisely, condition \eqref{Yang1} is equivalent to
\begin{align}
	\nonumber	& {\rm If} ~ \bra[\psi_\param] E \ket[\psi_\param] =0 ~ {\rm then}~ \exists \ket[\psi] \in \hil,~\eta_i \in \Re,\\
		\label{Yang3} & \eta_i \ket[\psi] = E \sld[i] \ket[\psi_\param],~~(\forall i).
\end{align}
While this interpretation of condition \eqref{Yang1} is not obvious in the first look at the statement of \cite{Yang2019partialcommutativity} Theorem 2, it can be deduced from its proof.

As in the case for single-parameter estimation, the optimal measurement $\Pi$ generally depend on the target point $\pspace$ to be estimated.
Explicit methods to construct the optimal measurements proposed so far typically generates parameter-dependent POVMs \cite{Humphreys2013,Yang2019partialcommutativity}.
To perform such a measurement, one needs a prerequisite knowledge on the point $\pspace$, or must start from adaptive measurements to find $\pspace$ in advance.

\subsection{Antiunitary operators}
We review basic properties of antiunitary operators on finite dimensional Hilbert spaces.
Differences between general antiunitary operators and conjugations are relevant in this article.
More advanced review on the topic can be found in \cite{Uhlmann2016antilinear}.

An operator $\antiu$ on $\hil$ is said to be antilinear if
\begin{equation}
	\antiu (c_1 \ket[\psi_1] + c_2 \ket[\psi_2]) = c_1^\ast \antiu \ket[\psi_1] + c_2^\ast \antiu \ket[\psi_2],
\end{equation}
holds for any complex coefficients $c_1,~c_2$ and vectors $\ket[\psi_1],~\ket[\psi_2]$.
The Hermitian adjoint $\antiu^\dagger$ of an antilinear operator is defined differently to linear ones by
\begin{equation}
	\bra[\psi_1] ( \antiu^\dagger \ket[\psi_2]) = \bra[\psi_2] ( \antiu \ket[\psi_1]),
\end{equation}
from which $(\antiu^\dagger)^\dagger = \antiu$, $(\antiu_1 \antiu_2)^\dagger = \antiu_2^\dagger \antiu_1^\dagger$, $(A \antiu)^\dagger = \antiu^\dagger A^\dagger$ and $(\antiu A)^\dagger = A^\dagger \antiu^\dagger$ follow for antilinear operators $\antiu_1,~\antiu_2$ and linear operator $A$.
Hermitian adjoints of antilinear operators are again antilinear.
Antilinear operator $\antiu$ is said to be Hermitian and unitary when $\antiu^\dagger = \antiu$ and $\antiu^\dagger = \antiu^{-1}$ hold, respectively.
Unitary antilinear operators are called antiunitaries.
Hermitian antiunitary operators are called conjugations.
Equivalently, conjugation $\conj$ is an antiunitary satisfying $\conj^2= \id_\hil$.
In this article, antiunitary operators are represented by $\antiu$, and conjugations are by $\conj$.

A conjugation is a complex conjugation in its reference-bases.
To see this, choose a basis $\{ \ket[\psi_i] \}_{i=1,...,d}$ of a $d$-dimensional Hilbert space $\hil$. 
Define a matrix representation $[ \antiu ]$ of antilinear operator $\antiu$ in this basis by
\begin{equation}
\label{matrix_repn1}	\antiu \ket[\psi_j] = \sum_{i=1}^d [ \antiu ]_{ij} \ket[\psi_i],
\end{equation}
from which
\begin{equation}
\label{matrix_repn2}	\antiu \sum_j c_j \ket[\psi_j] = \sum_{ij} c_j^\ast [ \antiu ]_{ij} \ket[\psi_i],
\end{equation}
follows.
The matrix $[ \antiu ]$ is unitary and symmetric when $\antiu$ is unitary and Hermitian, respectively.
A conjugation $\conj$ has a symmetric unitary matrix representation $[ \conj ]$, and thus decomposed as
\begin{equation}
\label{Autonne-Takagi}	[ \conj ] = V V^\transpose,
\end{equation}
with a unitary matrix $V$.
This is a direct consequence from Autonne$-$Takagi factorization.
Inserting Eq.~\eqref{Autonne-Takagi} to Eq.~\eqref{matrix_repn2}, we have
\begin{align}
\nonumber	\conj \sum_j c_j \ket[\psi_j] &= \sum_{ijk} c_j^\ast V_{ik} V^\transpose_{kj} \ket[\psi_i] \\
	= \sum_{k} & \left( \sum_j c_j V^\dagger_{kj} \right)^\ast   \left( \sum_{i} V_{ik}  \ket[\psi_i] \right).
\end{align}
The action of $\conj$ is thus a complex conjugation in the $V$-rotated basis $\ket[\psi_k'] := \sum_{i} V_{ik}  \ket[\psi_i]$.

The reference-basis of a conjugation is not unique.
If a basis $\{ \ket[\psi_j] \}_{j=1,...,\dim \hil}$ is a reference-basis of $\conj$, then another basis $\{ \ket[\psi'_j] \}_{j=1,...,\dim \hil}$ defined by
\begin{equation}
	\ket[\psi'_j] := R \ket[\psi_j],~(j=1,...,\dim \hil),
\end{equation}
with a real orthogonal matrix $R$ is also a reference-basis of $\conj$.

A general antiunitary operator $\antiu$ has a decomposition
\begin{equation}
\label{decompo}	\antiu = U \conj,
\end{equation}
into a (linear) unitary operator $U$ and a conjugation $\conj$.
$\antiu$ is a conjugation if $U$ is symmetric with respect to the reference basis that makes $\conj$ a complex conjugation.
Otherwise $\antiu$ is an antiunitary that is not a conjugation.

Spin-flip is an example of antiunitary operator that is not a conjugation.
Consider a 2-dimensional space $\hil$ with a basis $\{ \ket[\psi^+],~\ket[\psi^-] \}$.
Spin-flip $\antiu_f$ is defined by
\begin{equation}
\label{spin-flip}	\antiu_f = \sigma_Y \conj,
\end{equation}
where $\sigma_Y$ is the Pauli-$Y$ operator and $\conj$ is the complex conjugation in the same basis.
Obviously $\sigma_Y$ is not symmetric, and $\antiu_f^2 = - \id_2 \neq \id_2$.

In most part of this article antiunitary operators act in adjunction
\begin{equation}
	\antiu A \antiu^\dagger,
\end{equation}
on (linear) Hermitian operators $A$ such as quantum states.
The above operator is equal to
\begin{equation}
	U \conj A \conj U^\dagger = U A^\ast U^\dagger,
\end{equation}
from Eq.~\eqref{decompo}, where $A^\ast$ is the complex conjugation of $A$ in the reference basis of $\conj$.
The adjoint action of conjugation on a Hermitian operator is equivalent to transposition since $A^\ast = (A^\dagger)^\ast = A^\transpose$.
When $A$ is a quantum state, namely, a density operator, the transformation $A \mapsto \antiu A \antiu^\dagger$ cannot be a completely positive map.

Direct sums and tensor products of antiunitary operators are well-defined, and both result in antiunitary operators again.
The tensor product of antiunitary operators $\antiu_A$ on $\hil_A$ and $\antiu_B$ on $\hil_B$ is defined By
\begin{align}
\nonumber	\antiu_A \otimes \antiu_B \sum_j & c_j \ket[\psi^A_j] \otimes \ket[\psi^B_j]\\
\label{antiunitary_tensor} &:= \sum_j c_j^\ast \ket[\antiu_A \psi^A_j] \otimes \ket[\antiu_B \psi^B_j],
\end{align}
for vector $\sum_j c_j \ket[\psi^A_j] \otimes \ket[\psi^B_j]$ in the composite system $\hil_A \otimes \hil_B$.
For arbitrary two vectors in $\hil_A \otimes \hil_B$ decomposed by
\begin{align}
	\ket[\psi] &= \sum_j c_j \ket[\psi^A_j] \otimes \ket[\psi^B_j],\\
	\ket[\xi] &= \sum_k d_k \ket[\xi^A_k] \otimes \ket[\xi^B_k],
\end{align}
the inner product $\langle \antiu_A \otimes \antiu_B \xi | \antiu_A \otimes \antiu_B \psi \rangle$ is calculated as
\begin{align}
\nonumber	& \langle \antiu_A \otimes \antiu_B \xi | \antiu_A \otimes \antiu_B \psi \rangle \\
\nonumber	&= \sum_{j,k} d_k c^\ast_j \langle \antiu_A \xi^A_k | \antiu_A \psi^A_j \rangle \langle \antiu_B \xi^B_k | \antiu_B \psi^B_j \rangle \\
	&= \sum_{j,k} d_k c^\ast_j \langle \psi^A_k | \xi^A_j \rangle \langle \psi^B_k | \xi^B_j \rangle = \langle \psi | \xi \rangle,
\end{align}
by using Eq.~\eqref{antiunitary_tensor}.
Equation $\langle \antiu_A \otimes \antiu_B \xi | \antiu_A \otimes \antiu_B \psi \rangle = \langle \psi | \xi \rangle$ reveals that $\antiu_A \otimes \antiu_B$ is well-defined and is antiunitary.
We also have
\begin{equation}
	(U_A \antiu_A) \otimes (U_B \antiu_B) = (U_A \otimes U_B) (\antiu_A \otimes \antiu_B),
\end{equation}
where $U_A$ and $U_B$ are linear operators, and the tensor product in the middle is of linear operators.
Tensor products of pairs of an antiunitary operator and a linear operator is not well-defined.
There is no ``partial conjugation,'' in contrast to the well-defined partial transposition.

\section{Global antiunitary symmetry}\label{sec:symmetry}
This section presents a direct link between antiunitary symmetry and weak commutativity.
Theorem \ref{thm:global} states that quantum statistical models with ``global'' antiunitary symmetries are weakly commutative at all points in the parameter spaces.
Theorem \ref{thm:measurement} ensures the target-independent optimal measurements for pure-state models with global antiunitary symmetries.
These theorems underlie our methods to find globally optimal measurements.

\subsection{Global antiunitary symmetry and conjugation}
Let us define models with global antiunitary symmetry, and clarify the distinction between general antiunitaries and conjugations with respect to the definition.
\begin{definition}\label{def:global}
	A quantum statistical model $\model$ is said to have a \emph{global antiunitary symmetry} (GAS) when there is an antiunitary operator $\antiu$ such that
	\begin{equation}
\label{global}	\antiu \rho_\param \antiu^\dagger = \rho_\param,
	\end{equation}
	holds for any $\pspace$.
	Models with GASs are said to be \emph{imaginarity-free}.
\end{definition}
The antiunitary operator $\antiu$ in the definition does not depend on the parameter $\param$.
This is why we call the antiunitary symmetry in the sense of Def.~\ref{def:global} as being global.

As a trivial example, any classical model, whose states are diagonal in a fixed basis for all $\pspace$, i.e., when all the eigenvectors are independent of $\param$, is imaginarity-free.
The antiunitary operator is the complex conjugation in the diagonal basis.

There also are various non-classical imaginarity-free models.
This article presents GASs of N00N states \cite{LeeKokDowling2002GHZequivalentNOON,Giovanettietal2004science,Giovanettietal2011}, super-dense coding \cite{Fujiwara2001} and 3D magnetometry \cite{BaumgratzDatta2016} in Sec.~\ref{sec:examples}, and of antiparallel spins \cite{GisinPopescu1999,Changetal2014}, qubit inside a disc of Bloch sphere \cite{Suzuki2016} and embedding quantum simulators \cite{Casanovaetal2011eqs,Candiaetal2013,Zhangetal2015eqs,Chenetal2016eqs,Loredoetal2016eqs,Chengetal2017eqs} in Appx.~\ref{sec:models_appx}.

The GAS must be a conjugation for certain classes of models.
\begin{lemma}\label{lem:anti_vs_conj}
	A GAS, if exists, is given only by a conjugations if the model $\model$ satisfies either of the following two conditions:\\
	\noindent {\rm (a)} $\rho_\param$ is a pure state at all points in $\region$ and the pure states span the entire Hilbert space $\hil$,\\
	\noindent {\rm (b)} $\rho_\param$ is non-degenerate at some point in $\region$.
\end{lemma}
We use following two lemmas to prove lemma~\ref{lem:anti_vs_conj}.
\begin{lemma}\label{lem:rank1_antiunitary}
	For any antiunitary operator $\antiu$ and a vector $\ket[\psi]$ on a Hilbert space $\hil$, $\antiu \ketbra[\psi] \antiu^\dagger = \ketbra[\psi]$ implies $\antiu^2 \ket[\psi] = \ket[\psi]$.
\end{lemma}
\emph{proof})
$\antiu \ketbra[\psi] \antiu^\dagger = \ketbra[\psi]$ holds if and only if $\antiu \ket[\psi] = \epsilon \ket[\psi]$ with some unimodular $\epsilon$.
This implies $\antiu^2 \ket[\psi] = \epsilon^\ast \epsilon \ket[\psi] =\ket[\psi]$.
$\qed$
\begin{lemma}\label{lem:antiunitary_decomposition}
	Let $A$ be an Hermitian operator on Hilbert space $\hil$ with eigensubspaces $\hil_i$ ($\oplus_i \hil_i = \hil$), and $\antiu$ be an antiunitary operator on $\hil$ which satisfies $\antiu A \antiu^\dagger = A$.
	Then $\antiu$ has a decomposition
	\begin{equation}
	\label{antiunitary_decomposition}	\antiu = \oplus_i \antiu_i,
	\end{equation}
	where $\antiu_i$ is an antiunitary operator on $\hil_i$ for each $i$.
\end{lemma}
\emph{proof})
Let us denote the eigenvector of $A$ with eigenvalue $\lambda \in \Re$ by $\ket[\psi_\lambda]$.
Then we have
\begin{align}
\nonumber	\lambda \bra[\psi_\lambda] \antiu \ket[\psi_{\lambda'}] = \bra[\psi_\lambda] A \antiu \ket[\psi_{\lambda'}] &= \bra[\psi_\lambda] \antiu A \ket[\psi_{\lambda'}] \\
	\label{lambda}	=& \lambda' \bra[\psi_\lambda] \antiu \ket[\psi_{\lambda'}],
\end{align}
for any eigenvalues $\lambda,~\lambda'$, where the second equality follows from $\antiu A \antiu^\dagger = A$.
Equation \eqref{lambda} is possible only if when $\lambda \neq \lambda'$, $\antiu \ket[\psi_{\lambda'}]$ and $\ket[\psi_\lambda]$ are orthogonal to each other.
Equivalently, $\ket[\psi] \in \hil_i$ implies $\antiu \ket[\psi] \in \hil_i$.
Thus $\antiu$ is a sum of antiunitaries on the eigensubspaces of $A$.
$\qed$\\
\emph{proof of lemma \ref{lem:anti_vs_conj}})
(a) Let $\antiu$ be the GAS of a pure state model $\{ \ketbra[\psi_\param] | \pspace \}$.
We have $\antiu^2 \ket[\psi_\param] = \ket[\psi_\param]$ at all points from Lem.~\ref{lem:rank1_antiunitary}.
Because $\antiu^2$ is linear and $\hil$ is spanned by $\{ \ket[\psi_\param] | \pspace \}$, $\antiu^2$ must be identity.

(b) Let $\param_0 \in \region$ be the point where $\rho_{\param_0}$ becomes non-degenerate.
As $\rho_{\param_0}$ is Hermitian and satisfies $\antiu \rho_{\param_0} \antiu^\dagger = \rho_{\param_0}$, lemma \ref{lem:antiunitary_decomposition} reveals a decomposition of $\antiu$ given by Eq.~\eqref{antiunitary_decomposition}.
The dimension of eigensubspaces are all $1$ because of the non-degeneracy.
Then lemma \ref{lem:rank1_antiunitary} applies to each $\antiu_i$, and we have $\antiu_i^2 = \id_{\hil_i}$.
This implies $\antiu^2 = \id_\hil$, and completes the proof for the case (b).
$\qed$\\

Even if a model has a GAS which is not a conjugation, the same model has another symmetry provided by a conjugation:
\begin{theorem}\label{thm:anti_equal_conj}
	If a model $\model$ is imaginarity-free, there is a conjugation which is a GAS for $\model$.
\end{theorem}
This theorem remains valid even if the model is replaced by any set of Hermitian operators, where the definition of GAS is straightforwardly extended.
In the proof presented below, $\state$ can be a Hermitian operator and $X$ can be a discrete set.\\
\noindent \emph{proof})
Let $\model$ be a set of Hermitian operators in Hilbert space $\hil$ with a GAS $\antiu$.
We explicitly construct a conjugation replacing $\antiu$.

Let $\hil_i$ $(i \in I)$ be the disjoint subspaces of $\hil$ that are simultaneous eigensubspaces of all the states in $\model$.
The state is decomposed as
\begin{equation}
\label{universal_state_decomposition}	\state = \Pi' \state \Pi' \oplus  \bigoplus_{i \in I} \lambda_i(\param) \Pi_i,
\end{equation}
where $\Pi_i$ are projectors each onto $\hil_i$, $\Pi'$ is a projector onto the complement subspace
\begin{equation}
	\hil' := \hil \backslash (\oplus_{i \in I} \hil_i),
\end{equation}
and $\lambda_i(\param)$ are eigenvalues.

For arbitrary vector $\ket[\psi]$ orthogonal to $\hil_i$, there is a point $\param_\psi$ such that $\ket[\psi]$ is in the complement subspace of the eigensubspace of $\rho_{\param_\psi}$ with eigenvalue $\lambda_i(\param_\psi)$. 
By applying Lem.~\ref{lem:antiunitary_decomposition} to $\rho_{\param_\psi}$, we obtain a decomposition of $\antiu$ such that $\ket[\psi]$ belongs to the different sector from $\hil_i$.
As $\ket[\psi]$ is any vector orthogonal to $\hil_i$, we have a decomposition of the antiunitary
\begin{equation}
\label{eigensubspace_decomposition1}	\antiu = \antiu_i' \oplus \antiu_i,
\end{equation}
where antiunitary operators $\antiu_i$ and $\antiu_i'$ acts on $\hil_i$ and its complement, respectively.
Since the decomposition \eqref{eigensubspace_decomposition1} exists for each $i \in I$, we have
\begin{equation}
\label{eigensubspace_decomposition2}	\antiu = \antiu' \oplus \bigoplus_{i \in I} \antiu_i,
\end{equation}
where $\antiu'$ is an antiunitary operator on $\hil'$.

The defining equation $\antiu \state \antiu^\dagger = \state$ of the GAS is rewritten by the decompositions of the state \eqref{universal_state_decomposition} and the antiunitary \eqref{eigensubspace_decomposition2} to
\begin{equation}
\label{complement_symmetry}	\antiu' \state' \antiu'^\dagger = \state',
\end{equation}
for the truncated (unnormalized) state $\state' : = \Pi' \state \Pi'$, and
\begin{equation}
\label{subspace_symmetry}	\antiu_i \Pi_i \antiu_i^\dagger = \Pi_i. ~(\forall i \in I)
\end{equation}

Any antiunitary operator on $\hil_i$ satisfies Eq.~\eqref{subspace_symmetry} instead of $\antiu_i$.
Thus all $\antiu_i$ in $\antiu$ can be replaced by conjugations.

We now show that $\antiu'$ is a conjugation.
To do this, we consider the following step-by-step procedure to decompose $\hil'$ and $\antiu'$.
The procedure starts from $N=1$ with the trivial decomposition $\hil^{(1)}_1 = \hil'$ and $\antiu^{(1)}_1 = \antiu'$, and ends at $N=\dim \hil'$ with a decomposition of $\hil'$ into $1$-dimensional subspaces.

At the step $N$, a disjoint subspace decomposition $\hil^{(N)}_j$ ($j = 1,...,N$) of $\hil'$ and corresponding decomposition $\antiu' = \oplus_{j=1}^{N} \antiu^{(N)}_j$ of antiunitary is given.
Choose a subspace with dimension more than $2$, say $\hil^{(N)}_k$, and a unit vector $\ket[\psi] \in \hil^{(N)}_k$ therein.
By definition of the truncated state $\state'$, there is a point $\param$ such that $\ket[\psi]$ is not an eigenvector of $\rho_{\param}$.
From Lem.~\ref{lem:antiunitary_decomposition}, $\antiu'$ is decomposed as
\begin{equation}
\label{separate}	\antiu' = \antiu_{\psi}^a \oplus \antiu_{\psi}^b,
\end{equation}
where $\antiu_{\psi}^a$ and $\antiu_{\psi}^b$ are antiunitary operator on disjoint subspaces $\hil_{\psi}^a$, $\hil_{\psi}^b$ ($\hil' = \hil_{\psi}^a \oplus \hil_{\psi}^b$), respectively, such that $\ket[\psi]$ is not orthogonal to either $\hil_{\psi}^a$ or $\hil_{\psi}^b$.
For the two decompositions \eqref{separate} and $\antiu' = \oplus_{j=1}^N \antiu^{(N)}_j$ to be consistent, we necessarily have the subspace decomposition $\hil^{(N+1)}_j$ ($j = 1,...,N+1$) of $\hil'$ defined by
\begin{align}
	\hil^{(N+1)}_j &= \hil^{(N)}_j, ~ (j \neq k,~N+1)\\
	\hil^{(N+1)}_k &= \hil^{(N)}_k \wedge \hil_{\psi}^a,\\
	\hil^{(N+1)}_{N+1} &= \hil^{(N)}_k \wedge \hil_{\psi}^b,
\end{align}
and the corresponding decomposition $\antiu' = \oplus_{j = 1}^{N+1} \antiu^{(N+1)}_j$ of antiunitary.

At the end of the procedure, we obtain the decomposition of antiunitary
\begin{equation}
	\antiu' = \oplus_{j = 1}^{\dim \hil'} \antiu^{\dim \hil'}_j,
\end{equation}
into $1$-dimensional antiunitaries.
Since any $1$-dimensional antiunitary must be a conjugation, $\antiu^{\dim \hil'}_j$ is a conjugation for all $j=1,...,N$, and hence $\antiu'$ is, too.

Now that $\antiu'$ is a conjugation and $\antiu_i$ can be replaced by a conjugation $\conj_i$ on $\hil_i$ for all $i \in I$,
\begin{equation}
	\conj := \antiu' \oplus \bigoplus_{i \in I} \conj_i,
\end{equation}
is a conjugation on $\hil$ which satisfies Eqs.~\eqref{complement_symmetry} and \eqref{subspace_symmetry}.
Thus $\conj$ is a conjugation that gives the GAS for $\model$.
$\qed$\\

When a conjugation $\conj$ provides a GAS for model $\model$, there is a basis that represents $\state$ as a real matrix for all $\pspace$.
The model $\model$ is ``free from imaginarity'' in this sense.
Conversely, any model consisting only of real density matrices have a conjugation as a GAS.
Theorem \ref{thm:anti_equal_conj}, by identifying antiunitary-symmetric models with conjugation-symmetric models, justifies the term ``imaginarity-free'' to be used on models with GASs.

\subsubsection{Antiunitary symmetry defined on horizontal lift}
A lift of a quantum statistical model is its orbit in the associated fiber bundle.
For pure-state models, the defined GAS is equivalent to the antiunitary symmetry on the so-called horizontal lift of the model.
We briefly explain this equivalence without any prerequisite knowledge on geometry here.
The content of this section is independent on other parts of this article.
See \cite{Matsumoto1997,Matsumoto2005} for complete expositions.

For a pure-state model $\model$, a lift is identified with a parameterized unit vector $\ket[\psi_\param]$ in $\hil$ such that
\begin{equation}
	\ketbra[\psi_\param] = \state,
\end{equation}
for any $\pspace$.
The lift is said to be horizontal, if it satisfies
\begin{equation}
\label{horizontal_lift}	\Imag \left[ \langle \psi_\param | \partial_i \psi_\param \rangle \right] = 0,
\end{equation}
for any $i$ and $\pspace$.
The horizontal lift exists if and only if the model $\model$ is quasi-classical.

The global antiunitary symmetry (GAS) of a horizontal lift is defined as the antiunitary operator $\antiu$ such that
\begin{equation}
\label{GAS_lift}	\antiu \ket[\psi_\param] = \ket[\psi_\param],
\end{equation}
holds for any $\pspace$.
Now it is readily seen that $\antiu$ becomes a GAS of the model $\model$.
Conversely, if a model $\model$ has a GAS $\antiu$, then there is a lift satisfying \eqref{GAS_lift}.
This lift is horizontal since $\ket[\psi_\param]$ and $\ket[\partial_i \psi_\param]$ are in the same real subspace of $\hil$ invariant by the antiunitary $\antiu$.

The model has a GAS if its horizontal lift does, and they share the same antiunitary operator.

\subsection{Global antiunitary symmetry and weak commutativity}
The weak commutativity is ensured by a GAS.
\begin{lemma}\label{lem:symmetricSLD}
	If model $\model$ has a GAS $\antiu$, there is an SLD $\sld[j]$ satisfying $\antiu \sld[j] \antiu^\dagger = \sld[j]$ for any point $\pspace$ and for any parameter $x_j$.
\end{lemma}
\emph{proof})
Define the SLD by Eq.~\eqref{integralSLD}.
Then we have
\begin{align}
\nonumber	& \antiu \sld[j] \antiu^\dagger \\
\nonumber	&= 2 \int_0^\infty \antiu e^{-t \state} \antiu^\dagger \antiu \partial_j \state \antiu^\dagger \antiu e^{-t \state} \antiu^\dagger \dif t  \\
\nonumber	&= \int_0^\infty e^{-t \antiu \state \antiu^\dagger} \partial_j \state e^{-t \antiu \state \antiu^\dagger} \dif t \\
	&= \int_0^\infty e^{-t \state} \partial_j \state e^{-t \state } \dif t = \sld[j],
\end{align}
where the second equality follows from $\antiu \partial_j \state \antiu^\dagger = \partial_j \state$.
$\qed$
\begin{theorem}\label{thm:global}
	Any imaginarity-free model $\model$ is weakly commutative at any point in $\region$.
\end{theorem}
\emph{proof})
Let $\antiu$ be the GAS of model $\model$.
From Lem.~\ref{lem:symmetricSLD}, there are SLDs satisfying $\antiu \sld[j] \antiu^\dagger = \sld[j]$ for all $j$.
For these SLDs we have
\begin{align}
\nonumber	\trace[ \state \sld[i] \sld[j] ]^\ast &= \trace[ \antiu \state \antiu^\dagger \antiu \sld[i] \antiu^\dagger \antiu \sld[j] \antiu^\dagger ] \\
	&= \trace[ \state \sld[i] \sld[j]].
\end{align}
Therefore the Uhlmann curvature \eqref{uhlmann} is zero for any pairs of $(i,j)$.
$\qed$

According to Thm.~\ref{thm:global}, parameter encoding schemes are not crucial for compatible multiparameter estimations.
There are physical systems which provide only weakly commutative models, irrespective to the parameter encoding schemes.
For example, Majorana fermions have antiunitary particle-hole symmetry.
Thus any models constituting of Majorana fermions are weakly commutative.

Quantum mechanics in real Hilbert spaces lacks an uncertainty principle between observables of a specific form \cite{Stueckelberg1960real}.
If $A$ and $B$ are observables such that $A^\ast = A$ and $B^\ast = B$, the expectation value $\trace[ \rho [A,B] ]$ is zero for any real density operator $\rho = \rho^\ast$.
The Heisenberg uncertainty principle $\langle \Delta A \rangle \langle \Delta B \rangle \geq |\langle [A,B] \rangle|/2$ is trivialized.
Nevertheless there is an uncertainty principles of a different form \cite{Stueckelberg1960real,LahtiMaczynski1987real}.

We are not sure how Thm.~\ref{thm:global} relates to the absence of Heisenberg uncertainty principle.
The link between uncertainty principles and incompatibility of parameters should be clarified in more detail.
Such an attempt has started only recently \cite{LuWang2021heisenberg}.

Theorem \ref{thm:global} directly relates geometric phases and the phases of off-diagonal elements in density operators.
The mean Uhlmann curvature is the Uhlmann geometric phase (per unit area) accumulated by infinitesimal loops around the point \cite{Carollo2018}.
The Uhlmann geometric phase coincides to Berry phase for pure states \cite{Uhlmann1986holonomy}.
If the phases of off-diagonal elements of the density operator are all zero at the neighborhood of $\pspace$, the geometric phases accumulated by infinitesimal loops around the point is also zero.

Antiunitary symmetries are sometimes easier to check than the zero of mean Uhlmann curvature.
Consider thermal state of a parametrized Hamiltonian $H_\param$ presented by a canonical ensemble $e^{-H_\param} /Z$.
Here we assume that the temperature is contained in the Hamiltonian as a parameter.
Such a model is rather studied in quantum statistical mechanics, than in quantum metrology. 
Non-zero mean Uhlmann curvature characterizes quantum phase transitions of the thermal state \cite{Carollo2018,Carolloetal2018fermion,Leonforte2019muc,Basconeetal2019muc,Basconeetal2019mucPRB,Carollo2019}.
Lemma \ref{lem:antiunitary_decomposition} indicates that $e^{-H_\param} /Z$ is imaginarity-free if and only if $H_\param$ does.
That is, if the Hamiltonian endows with an antiunitary symmetry such as time-reverse, the thermal state has zero mean Uhlmann curvature.

One might wonder if QFIMs of imaginarity-free models have a particular form.
As we will see in Sec.~\ref{sec:antiparallel}, it is possible to construct an imaginarity-free model with QFIM $\fisher_Q(\param)$ from arbitrary model with QFIM $\fisher_Q(\param)/2$.
GASs do not pose any restriction on the form of QFIMs.
As a consequence, imaginarity-freeness does not imply the stricter notion of ``compatibility'' referring to the saturation of QCRBs with diagonal QFIMs \cite{Ragyetal2016}.

As a final remark, there are weakly commutative models without any GAS.
This implies that the converse of Thm.~\ref{thm:global} is not true in general: weak commutativity does not necessarily imply imaginarity-freeness.
Even some single-parameter models, such as off-equator phase estimation presented in Appx.~\ref{sec:phase_estimation}, do not have any GAS.

\subsection{Optimal target-independent measurements}
Any imaginarity-free pure-state model is accompanied by a set of globally optimal POVMs.
This means that the QCRB is saturated by the corresponding measurement at any target-point.

In this section, we set the number $\nq$ of copies in joint measurements to $1$, because we focus mainly on pure imaginarity-free models which are automatically quasi-classical \cite{Matsumoto2002}.

A set of optimal POVMs is known for antiparallel spins \cite{Changetal2014} (see Appx.~\ref{sec:antiparallel_spins} for details).
These POVM operators are first proposed in \cite{GisinPopescu1999}, and proven to maximize a fidelity measure \cite{Massar2000}.
The POVMs must be independent from the target point $\param$ in the parameter space, in order to be optimal for the fidelity measure.
Later the same set of operators is shown to saturate QCRB \cite{Changetal2014}, namely, it is optimal in terms of a precision bound as well.

The target-independent POVMs are fascinating because they are optimal all over the parameter space.
In general, around different regions in the parameter space, we require different POVMs to achieve the best precision limit in the region.
This is the case even for single-parameter estimations.
If the optimal POVMs are independent of the target point, the precise measurement can be carried out without any prerequisite knowledge on the source.

The following theorem presents target-independent optimal POVMs.
\begin{theorem}\label{thm:measurement}
	Let $\antiu$ be an antiunitary on $\hil$.
	Any POVM $\E_\antiu$ satisfying conditions
	\begin{align}
	\label{rank1}	{\rm rank} E = 1 ~&(\forall E \in \E_\antiu),\\
	\label{invariantPOVM}	\antiu E \antiu^\dagger = E ~&(\forall E \in \E_\antiu),
	\end{align}
	saturates the QCRB \eqref{HCRB} of a pure-state model $\model$ with $\nq=1$ everywhere in $\region$ if $\antiu$ is a GAS of $\model$.
\end{theorem}
Since CFIM is additive with respect to tensor products and product measurements, the $\nq$ product of antiunitary-invariant rank-1 measurements saturates the QCRB \eqref{HCRB} for any $\nq$.

For proving Thm.~\ref{thm:measurement}, we use the following lemma to restrict GASs to conjugations.
Recall that any imaginarity-free model has a conjugation as a GAS (Thm.~\ref{thm:anti_equal_conj}).
\begin{lemma}\label{lem:conjugation}
	There exists a POVM $\E_\antiu$ on $\hil$ satisfying Eqs.~\eqref{rank1} and \eqref{invariantPOVM} if and only if $\antiu$ is a conjugation.
\end{lemma}
\emph{proof})
If $\antiu$ is a conjugation, projectors onto vectors of $\antiu$'s reference-basis form a POVM with required properties.

Conversely, let $\E_\antiu = \{ q_\omega \ket[E_\omega] \bra[E_\omega] \}_{\omega \in \Omega}$ ($q_\omega \in (0,1]$) be a POVM satisfying conditions~\eqref{rank1} and \eqref{invariantPOVM}.
From Lem.~\ref{lem:rank1_antiunitary}, we have $\antiu^2 \ket[E_\omega] = \ket[E_\omega]$ for all $\omega$.
Because $\{ \ket[E_\omega] \}_{\omega \in \Omega}$ spans the entire Hilbert space $\hil$, we have $\antiu^2 = \id_\hil$.
In other words, $\antiu$ is a conjugation.
$\qed$\\
\emph{proof of Thm.~\ref{thm:measurement}})
From Lem.~\ref{lem:conjugation}, we can assume that the antiunitary operator is a conjugation $\conj$.

Let $\E_\conj$ be a POVM satisfying conditions \eqref{rank1} and \eqref{invariantPOVM}.
From Lem.~\ref{lem:symmetricSLD}, there are SLDs obeying the antiunitary symmetry $\conj \sld[j] \conj = \sld[j]$.
It suffices to show that conditions \eqref{Yang2} and \eqref{Yang3} (Theorems 1 and 2 of \cite{Yang2019partialcommutativity}) hold for these antiunitary-invariant SLDs and for any element in $\E$.

For this purpose, we choose a particular pure-state vector for $\model$.
Let $\ket[\psi_\param]$ be the continuously parametrized unit vector such that $\ketbra[\psi_\param] = \state$ and $\conj \ket[\psi_\param] = \ket[\psi_\param]$ hold everywhere.
Equivalently, $\{ \ket[\psi_\param] ~|~\pspace \}$ is defined to be the horizontal lift of $\model$.

Let $E$ be an element of POVM $\E$.
Take a reference-basis $\{ \ket[\Pi^E_k] \}_{k=1,...,\dim \hil}$ of $\conj$ so that
\begin{equation}
	\ket[\Pi^E_1] \bra[\Pi^E_1] = q_E E,~\conj \ket[\Pi^E_k] = \ket[\Pi^E_k],
\end{equation}
for $k=1,...,\dim \hil$, where $q_E \in (0,1]$ is a positive coefficient.
SLDs are represented by real matrices
\begin{equation}
\label{SLD_decomposition}	\sld[j] = \sum_{k,l=1}^{\dim \hil} \ell^j_{kl}(E,\param) \ket[\Pi^E_k] \bra[\Pi^E_l],
\end{equation}
in the basis $\{ \ket[\Pi^E_k] \}_{k=1,...,\dim \hil}$, where $\ell^j_{kl}(E,\param)$ are real coefficients.
Since $\{ \ket[\Pi^E_k] \}_{k=1,...,\dim \hil}$ spans the real Hilbert subspace of $\hil$ invariant under $\conj$, the vector $\ket[\psi_\param]$ of the horizontal lift have the decomposition
\begin{equation}
\label{lift_decomposition}	\ket[\psi_\param] = \sum_{k=1}^{\dim \hil} r_k(E,\param) \ket[\Pi^E_k],
\end{equation}
with real coefficients $r_k(E,\param)$.

Now we check the conditions \eqref{Yang2} and \eqref{Yang3} on $E$.
Equation \eqref{lift_decomposition} alone implies
\begin{equation}
	E \ket[\psi_\param] = q_E r_1(E,\param) \ket[\Pi^E_1],
\end{equation}
and Eqs.~\eqref{SLD_decomposition} and \eqref{lift_decomposition} together imply
\begin{equation}
	E \sld[j] \ket[\psi_\param] = q_E \left( \sum_{k=1}^{\dim \hil} \ell^j_{1k}(E,\param) r_k(E,\param) \right) \ket[\Pi^E_1].
\end{equation}
We find that $E \ket[\psi_\param]$ and $E \sld[j] \ket[\psi_\param]$ are all proportional to $\ket[\Pi^E_k]$ by real coefficients.
This implies that $E$ satisfies conditions \eqref{Yang2} and \eqref{Yang3} at any point $\pspace$.

Because $E$ is an arbitrary element in $\E$, conditions \eqref{Yang2} and \eqref{Yang3} are satisfied by all elements in $\E$ everywhere.
$\qed$

The optimality of antiunitary-invariant POVM is independent not only on the target point, but also on how the parameters are encoded in the state.
The optimal POVMs depend only on the GAS, which in turn depends only on the states $\model$ as a set.
Thus re-parametrization change neither the GAS nor the optimal POVMs.
One does not have to use different POVMs according to different parametrization.

The condition on POVMs indicated by Thm.~\ref{thm:measurement} is so loose that continuously many such POVMs exist.
To see this, let us consider a pure-state model whose GAS is a conjugation $\conj$.
The conjugation defines a $\dim \hil$-dimensional real Hilbert subspace of $\hil$ spanned by vectors satisfying $\conj \ket[\psi] = \ket[\psi]$.
You can take any vectors from this $\conj$-invariant real subspace to construct the optimal POVM by $E= \ket[\psi] \bra[\psi]$.
For example, the reference basis for representing a conjugation $\conj$ as a complex conjugation is not unique.
In Appx.~\ref{sec:antiparallel_spins}, we find two different sets of optimal POVMs for antiparallel spins from its GAS.
One of them is given in \cite{GisinPopescu1999} and the other one is new.
As another example, rank-1 projectors distributed randomly over the real subspace also saturates the QCRB \cite{Matsumoto2005}.

While any antiunitary-invariant measurements work equally well, there should be experimentally preferable ones, such as those requiring less entanglement.
Consider, for example, an estimation of parameters $\param$ encoded as a unitary evolution $e^{-i H_\param}$.
A typical method of quantum metrology utilizes multiple copies of the same system driven by $e^{-i \sum_{n=1}^N H^{(n)}_\param } = \otimes_{n=1}^N e^{-i H^{(n)}_\param } $.
If the $1$-particle Hamiltonian $H_\param$ has an antiunitary skew-symmetry $\antiu H_\param \antiu^\dagger = - H_\param$, the overall Hamiltonian has an antiunitary skew-symmetry under $\antiu^{\otimes N}$.
We show in Sec.~\ref{sec:optimized_initial} how to construct imaginarity-free models based on Hamiltonians' antiunitary skew-symmetries.
If you construct an imaginarity-free model based on the GAS $\antiu^{\otimes N}$, the POVM must be invariant under $\antiu^{\otimes N}$.
You would prefer the measurement in the product POVM, each being invariant under $\antiu$.

We apply the above reasoning to find the optimal measurement basis for 3D magnetometry \cite{BaumgratzDatta2016} in Sec.~\ref{sec:3Dmagnetometry}.
There the GAS provides a parameter-independent optimal measurement basis.
The entanglement required for the measurement is reduced from $N$-partite to $2$-partite.
The search of reasonable measurements is possible thanks to the freedom of the optimal measurement for imaginarity-free models.

We close this section by extending the globally optimal POVMs to globally optimal estimation strategies.
Even if a target-independent measurement saturates the QCRB everywhere, the optimal locally unbiased estimator could still depend on the target-point.

Maximum likelihood estimators are independent on the target-point to be estimated.
The only condition for the maximum likelihood estimator to work globally is the identifiability of the parameter.
More precisely, the probability distribution $p(\Omega|\param)$ generated by the measurement must be in one-to-one correspondence with the point $\param$.
Projectors onto a single reference-basis for $\conj$ will not produce sufficiently informatic probability distributions to identify the parameters in general.

Fortunately, the condition on the optimal POVMs described in Thm.~\ref{thm:measurement} is loose enough to admit the POVMs to identify the parameters.
Under the restrictions \eqref{rank1} and \eqref{invariantPOVM}, it is possible to construct a POVM $\{ E_\omega \}_{\omega \in \Omega}$ such that the map $\rho \mapsto (\trace[E_\omega \rho] )_{\omega \in \Omega}$ from $\conj$-invariant density operators $\{ \rho \in \B(\hil)~|~\rho \geq 0, ~\trace[\rho]=1,~\conj \rho \conj =\rho \}$ to $\Re^\Omega$ is one-to-one.
An example of such \emph{real}-informationally-complete POVMs associated to $\conj$ is the rank-1 projectors distributed randomly over the real subspace defined by $\conj$ \cite{Matsumoto2005}.
Because the space of real density operators associated to $\conj$ contains $\model$ if $\conj$ is its GAS, the probability distribution generated by $\{ E_\omega \}_{\omega \in \Omega}$ identifies the parameter.
In this way we obtain the following corollary:
\begin{corollary}\label{cor:global_estimation}
	Let $\model$ be an imaginarity-free pure-state model.
	Suppose that the model $\model$ is identifiable in that the map $\param \mapsto \state$ is invertible.
	Then there is a POVM $\Pi = \{ \Pi_k \}_{k \in \Omega}$ on $\hil$ with which the maximum likelihood estimator $\estimator:\Omega^{\nc} \rightarrow \region$ satisfies
	\begin{equation}
		\lim_{\nc \rightarrow \infty} \nc \msem (\Pi,\estimator) = \mathcal{F}_Q (\param)^{-1},
	\end{equation}
	at any point $\pspace$.
\end{corollary}

We can globally saturate QCRB by a fixed measurement and a maximum likelihood estimator without any prior knowledge on the target-point, if the pure-state model has a GAS.
In practice, there is no need to stick to the real-informationally-complete POVMs.
One can employ POVMs that identify the state from the model $\model$ only.

\subsection{Examples of imaginarity-free models}\label{sec:examples}
\begin{figure}[tb]
	\centering
	\includegraphics[width=8cm]{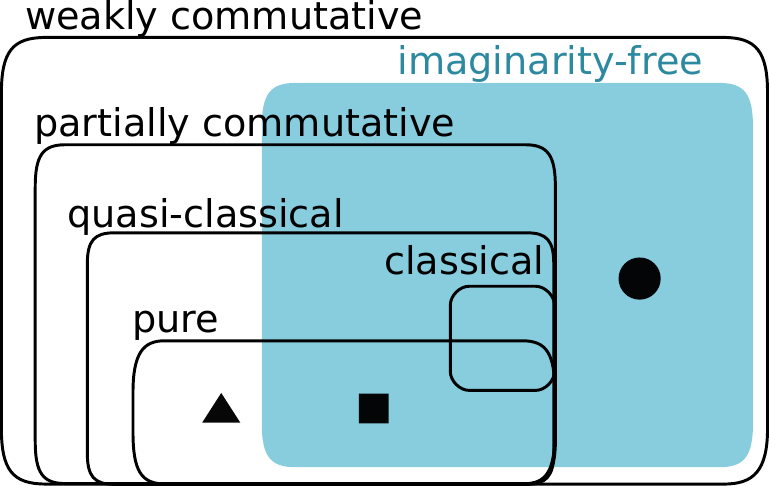}
	\caption{Classifications of weakly commutative models. A model is said to be classical when its states are simultaneously diagonalizable. Example models provided in this article are classified to either of the marked regions. The triangle contains off-equator phase estimation of a spin (Appx.~\ref{sec:phase_estimation}). The square contains N00N states (Sec.~\ref{sec:N00N}), super-dense coding (Sec.~\ref{sec:superdense}), 3D magnetometry (Sec.~\ref{sec:3Dmagnetometry}), and antiparallel spins (Appx.~\ref{sec:antiparallel_spins}). The circle contains depolarized antiparallel spins (Appx.~\ref{sec:antiparallel_spins}) and qubit inside a disc of Bloch sphere (Appx.~\ref{sec:suzuki}).}
	\label{fig:classification}
\end{figure}

In this section, we introduce three examples of imaginarity-free models.
These are N00N states (Sec.~\ref{sec:N00N}), super-dense coding (Sec.~\ref{sec:superdense}), 3D magnetometry (Sec.~\ref{sec:3Dmagnetometry}).
The Venn diagram in FIG.~\ref{fig:classification} summarizes known classifications of weakly commutative models together with imaginarity-free models.
The examples presented in this section all correspond to the square symbol in FIG.~\ref{fig:classification}.
To ensure the existences of models at each region of the diagram, we present some other examples in Appendix~\ref{sec:models_appx}.

In this section again, we set the number $\nq$ of copies in joint measurements to $1$.

\subsubsection{N00N states}\label{sec:N00N}
The first example is a single-parameter model.
The N00N state of optical interferometry is a milestone of quantum metrology and is still under theoretical and experimental studies.
See, e.g. \cite{LeeKokDowling2002GHZequivalentNOON,Giovanettietal2004science,Giovanettietal2011} for theoretical reviews and \cite{Polinoetal2020} for a review including experiments.

The pure-state model with a phase parameter $\phi$ is given by
\begin{equation}
\label{N00N}	\ket[\psi_\phi] := \frac{\ket[N,0] + e^{-i N \phi} \ket[0,N]}{\sqrt{2}},
\end{equation}
where $\ket[N_1,N_2]$ represents a two-mode state having $N_1$ and $N_2$ photons in mode 1 and mode 2, respectively.
We can readily see that the N00N state distributes on the equator of Bloch sphere with poles $\ket[N,0]$ and $\ket[0,N]$, and thus can be represented by real density matrices in a suitable basis.
Any antiunitary operator $\antiu$ satisfying
\begin{align}
\label{GASN00N}	\antiu c \ket[0,N] = c^\ast \ket[N,0],&~ \antiu c \ket[N,0] = c^\ast \ket[0,N],\\
\nonumber & \hspace{2cm} (\forall c \in \mathbb{C})
\end{align}
is a GAS for N00N state model \eqref{N00N}.
From the GAS \eqref{GASN00N} we recover a target-independent optimal measurement basis including vectors
\begin{equation}
	\frac{\ket[N,0] \pm \ket[0,N]}{\sqrt{2}}.
\end{equation}
The measurement on this basis is carried out by a combination of a beam-splitter and a photon-number-counting measurement.

Note that not all the pure-state single-parameter models are imaginarity-free in general.
A counterexample is provided in Appx.~\ref{sec:phase_estimation} by an off-equator phase estimation of a spin.
Even for single-parameter models, the globally optimal measurement is not guaranteed if the model does not have a GAS.

\subsubsection{Super-dense coding}\label{sec:superdense}
Fujiwara \cite{Fujiwara2001} considered an estimation of SU$(2)$ unitary $U$ represented on $2$-dimensional space.
The model is given by
\begin{equation}
\label{superdense}	\left\{ \rho_U = (U \otimes \id) \rho (U \otimes \id)^\dagger \middle| U \in \mathrm{SU}(2) \right\},
\end{equation}
where $\rho$ is a $2$-qubit state.
The unitary is represented by $3$-parameters.
We consider pure input state $\rho = \ketbra[\psi_r]$ defined by $ \ket[\psi_r] = \sqrt{r} \ket[0,0] + \sqrt{1-r} \ket[1,1]$ with a constant $r \in [0,1]$.
This is a parameter estimation version of super-dense coding \cite{Bennettetal1992}.
The model is weakly commutative if and only if $r=1/2$, namely, when $\rho$ is maximally entangled \cite{Fujiwara2001}.

Model $\{ \rho_U | U \in \mathrm{SU}(2) \}$ is indeed imaginarity-free for maximally entangled $\rho$.
For conjugation $\conj$ in the reference basis $\{ \ket[0],~\ket[1] \}$ we have
\begin{align}
\nonumber	(\conj \otimes \conj) (U \otimes \id) & \ket[\psi_{1/2}] = (\conj U \conj \otimes \id) \ket[\psi_{1/2}] \\
\label{dense2}	&= -( \sigma_Y U \sigma_Y \otimes \id) \ket[\psi_{1/2}] \\
\label{dense3}	&= (\sigma_Y \otimes \sigma_Y)( U \otimes \id) \ket[\psi_{1/2}]
\end{align}
where we have used identities $(\sigma_Y \otimes \id) \ket[\psi_{1/2}] = (\id \otimes \sigma_Y^\transpose) \ket[\psi_{1/2}]$ and $\conj U \conj = -\sigma_Y U \sigma_Y$ holding for any SU$(2)$ unitary $U$.
Thus the model is symmetric under the product conjugation $\conj \otimes \conj$ followed by $\sigma_Y \otimes \sigma_Y$.

With the aid of GAS we can construct the target-independent optimal measurement with $\nq=1$ which is not addressed in \cite{Fujiwara2001}.
Since Eq.~\eqref{dense3} holds for $U=i \sigma_k$ ($k=X,Y,Z$, $\sigma_k$ represents the Pauli-$k$ operator), projectors onto the Bell measurement basis $\{ \ket[\psi_{1/2}],~(i \sigma_k \otimes \id) \ket[\psi_{1/2}] ~|~ k=X,Y,Z \}$ forms a rank-1 antiunitary-invariant optimal POVM.
The measurement used by the receiver of original super-dense coding scheme \cite{Bennettetal1992} is an optimal measurement for the parameter-estimation version.

If the input state is not maximally entangled, namely, if $r \neq 1/2$, there is no unitary $V$ on $\mathbb{C}^2$ such that $(\sigma_Y \otimes \id) \ket[\psi_r] = (\id \otimes V) \ket[\psi_r]$ holds.
Therefore it is impossible to transform Eq.~\eqref{dense2} to Eq.~\eqref{dense3} by replacing $\sigma_Y \otimes \sigma_Y$ to $\sigma_Y \otimes V$.
Indeed, there is no GAS when $r \neq 1/2$ because the model \eqref{superdense} is not weakly commutative for these $r$ \cite{Fujiwara2001}.

Note that we did not require the explicit expression of parametrized unitaries to find a GAS.
Theorem \ref{thm:global} enables the shortcut in deducing weak commutativity.

\subsubsection{3D magnetometry}\label{sec:3Dmagnetometry}
3D magnetometry has an $N$ spin system interacting with the external magnetic field \cite{BaumgratzDatta2016}.
The Hamiltonian is
\begin{equation}
H_{\bm \varphi} := \sum_{i=1}^N {\bm \varphi} \cdot {\bm \sigma}_i,
\end{equation}
where ${\bm \varphi} = (\varphi_X , \varphi_Y, \varphi_Z) \in \Re^3$ is proportional to the magnetic field, and ${\bm \sigma}_i$ is the vector of Pauli matrices for $i$th spin.
The parameter to be estimated is ${\bm \varphi}$.
The $N$-spin system undergoes unitary transformation $e^{-iH_{\bm \varphi}}$.
The model of 3D magnetometry is
\begin{equation}
\label{3D_magnetometry}	\left\{ e^{-iH_{\bm \varphi}} \ketbra[\Psi] e^{iH_{\bm \varphi}} \middle| {\bm \varphi} \in \Re^3  \right\},
\end{equation}
where $\ket[\Psi]$ is the initial state to be optimized.

Inspired by the single-parameter phase estimation, in \cite{BaumgratzDatta2016} they considered
\begin{equation}
	\label{initial_state} \ket[\Psi] := \frac{\ket[\Psi_X^{GHZ}] + e^{i \delta_Y} \ket[\Psi_Y^{GHZ}] + e^{i \delta_Z} \ket[\Psi_Z^{GHZ}]}{\mathcal{N}},
\end{equation}
as the initial state.
Here $\mathcal{N}$ is the normalization factor, $\delta_Y, \delta_Z \in [0,2\pi]$ are constants to be chosen for making the model weakly commutative.
State $\ket[\Psi_k^{GHZ}] =(\ket[\psi_k^+]^{\otimes N} + \ket[\psi_k^-]^{\otimes N})/\sqrt{2}$ is a $N$-partite GHZ state with eigenstates of Pauli operators $\sigma_k \ket[\psi_k^\pm] = \pm \ket[\psi_k^\pm]$.
For later convenience, we choose the phases of $\ket[\psi_k^\pm]$ to have
\begin{align}
	\ket[\psi_X^+] = \frac{1}{\sqrt{2}} \left( \begin{array}{c}
		1\\
		1
	\end{array} \right),~&
	\ket[\psi_X^-] = \frac{1}{\sqrt{2}} \left( \begin{array}{c}
		1\\
		-1
	\end{array} \right),\\
	\ket[\psi_Y^+] = \frac{1}{\sqrt{2}} \left( \begin{array}{c}
		1\\
		i
	\end{array} \right),~&
	\ket[\psi_Y^-] = \frac{1}{\sqrt{2}} \left( \begin{array}{c}
		i\\
		1
	\end{array} \right),\\
	\ket[\psi_Z^+] = \left( \begin{array}{c}
		1\\
		0
	\end{array} \right),~&
	\ket[\psi_Z^-] = \left( \begin{array}{c}
		0\\
		1
	\end{array} \right).
\end{align}

We consider the symmetry under spin-flip on $N$ systems,
\begin{equation}
\label{N_spin-flip}	\antiu_f^{\otimes N},
\end{equation}
where $\antiu_f = \sigma_Y \conj$ and $\conj$'s reference basis is $\{ \ket[\psi^+_Z],~\ket[\psi^-_Z] \}$.
Since the spin-flip changes the sign of Pauli operators ${\bm \sigma} \mapsto - {\bm \sigma}$, the unitary evolution $e^{-iH_{\bm \varphi}}$ is invariant under spin-flip.
Thus the spin-flip is a GAS of the model if the initial state $\ket[\Psi]$ is also invariant under the spin-flip.

Now we derive the values of $\delta_Y$ and $\delta_Z$ to make the model imaginarity-free.
From our choice of $\ket[\psi_k^\pm]$ follow
\begin{align}
\label{spin_vector}	\antiu_f \ket[\psi_k^+] = i \ket[\psi_k^-],~& \antiu_f \ket[\psi_k^-] = -i \ket[\psi_k^+],\\
\nonumber &(k=X,Y,Z)
\end{align}
and hence
\begin{align}
	\antiu_f^{\otimes N} \ket[\Psi_k^{GHZ}] = \left\{ \begin{array}{lr}
		\frac{\ket[\psi_k^+]^{\otimes N} - \ket[\psi_k^-]^{\otimes N}}{(-i)^N\sqrt{2}} & (N:\mathrm{odd}), \\
		(-1)^{N/2} \ket[\Psi_k^{GHZ}] & (N:\mathrm{even}),
	\end{array} \right.
\end{align}
for any integer $N$.
Therefore $\antiu_f^{\otimes N} \ket[\Psi] \propto \ket[\Psi]$ is the case when $\delta_Y=0,~\pi$ and $\delta_Z = 0,~\pi$ for $N=4m$, and $\delta_Y=\pi/2,~3\pi/2$ and $\delta_Z = \pi/2,~3\pi/2$ for $N=4m+2$ with some integer $m$.
The model $e^{-iH_{\bm \varphi}} \ket[\Psi]$ has a GAS and thus is weakly commutative for these $\delta_Y$ and $\delta_Z$.
We are not sure if the difference between our solutions on $(\delta_Y,\delta_Z)$ with those found in \cite{BaumgratzDatta2016} is entirely due to the choice of phases since vectors $\ket[\psi_k^\pm]$ are not explicitly indicated in \cite{BaumgratzDatta2016}.

Baumgratz and Datta \cite{BaumgratzDatta2016} found an optimal measurement basis including the target state $e^{-iH_{\bm \varphi}} \ketbra[\Psi] e^{iH_{\bm \varphi}}$, which is $N$-partite entangled.
They also considered projectors
\begin{equation}
	\Pi_{k,\pm} = \frac{\id \pm \sigma_k^{\otimes N}}{6},~(k=1,2,3)
\end{equation}
and numerically exhibited that these POVMs are close to the optimal.
The projector is invariant under the GAS \eqref{N_spin-flip} since
\begin{equation}
	\antiu_f^{\otimes N} \Pi_{k,\pm} \left( \antiu_f^{\otimes N} \right)^\dagger = \frac{\id \pm (-1)^N \sigma_k^{\otimes N}}{6} = \Pi_{k,\pm},
\end{equation}
holds for all $k=1,2,3$ and even $N$.
That the high-rank projectors $\Pi_{k,\pm}$ almost achieves the ultimate precision limit is still non-trivial.

We derive different measurement bases which are guaranteed to be optimal from the GAS.
The GAS $\antiu_f^{\otimes N}$ is a tensor product of $N/2$ conjugations $\antiu_f^{\otimes 2}$.
Equation \eqref{spin_vector} implies that for $k=X,Y,Z$, the basis
\begin{equation}
\label{2-partite_basis}	\left\{ \frac{\ket[\psi_k^-,\psi_k^-] \pm \ket[\psi_k^+,\psi_k^+]}{-\sqrt{2} i},~ \frac{\ket[\psi_k^-,\psi_k^+] \pm \ket[\psi_k^+,\psi_k^-]}{\sqrt{2}}   \right\},
\end{equation}
is a reference basis for the bipartite conjugation $\antiu_f^{\otimes 2}$,
and thus $\antiu_f^{\otimes N}$ has the $N/2$-product of \eqref{2-partite_basis} as a reference basis.
This implies that the bipartite measurements in the basis \eqref{2-partite_basis}, performed independently on $N/2$ copies, is optimal.

\section{Methods to implement global antiunitary symmetries}\label{sec:reduction}
In this section we consider methods to construct imaginarity-free models.
The benefit of implementing the GASs is twofold.
First, the weak commutativity of the model is ensured at any point (Thm.~\ref{thm:global}).
Second, the models have parameter-independent optimal POVMs (Thm.~\ref{thm:measurement} and Cor.~\ref{cor:global_estimation}), when the models are pure.

We focus on quantum metrology, and thus assume some ability to modify the physical systems to adjust the models.
If the model $\model$ cannot be adjusted, the attainable precision does not improve even if we implement antiunitary symmetry on $\state$ by pre-measurement operations.
The assumed ability is to replace the initial states before interactions.

Even if you can adjust the models, you may still wonder if the restriction to GASs is worth taking.
GASs inevitably restrict available models.
There may be a model without GAS but still offers better precisions for all the parameters.
In short, you may wonder if compatible models offer maximum precisions.
Unfortunately, previous methods to design compatible models for quantum metrology did not consider optimizations on the attainable precision \cite{Matsumoto2002,Ragyetal2016,BelliardoGiovannetti2021}.

This question has almost been ignored since it was difficult to calculate the optimal precision bounds for incompatible models.
To propose a metrology scheme, you have to evaluate the precisions attained by the scheme.
If the model is compatible, it suffices to calculate the QCRB.
Otherwise you have to calculate the so-called Holevo Cram\'{e}r-Rao bound \cite{Holevo1976,Holevo2011,GutaKahn2006,HayashiMatsumoto2008,KahnGuta2009,Yamagata2013,Yangetal2019attaining}, whose closed-form expression is missing.
This has been one of the reasons for researchers to seek for compatible models.
However, methods to calculate the Holevo Cram\'{e}r-Rao bound has been developed \cite{Suzuki2016,Matsumoto2002,Bradshawetal2017,Bradshawetal2018,Gorecki2020optimalprobeserror,Yamagata2021}, and a semi-definite programming is available now \cite{Albarellietal2019}.
The complexity is further reduced when you require only a lower bound of the Holevo Cram\'{e}r-Rao bound \cite{Sidhuetal2021}.
The final blow on the QCRB is its modest advantage: the Holevo Cram\'{e}r-Rao bound secures at least half of the QCRB even in the worst case \cite{Carollo2019}.
This encourages the pursuit for better precisions in the realm of incompatible models.

Our answer to the question on the simultaneous realizability of best precisions and GASs is twofold.

In Sec.~\ref{sec:optimized_initial}, we consider a particular generator of unitary transformation.
The optimal initial state for the phase estimation inevitably leads to imaginarity-free models.
The GAS coexists with the optimal initial state, and provides optimal global measurements at the same time.

In Sec.~\ref{sec:antiparallel}, we propose a different method to implement antiunitary symmetry by generalizing antiparallel spins.
The method implements GASs without deteriorating the precision limit of a given model, under some symmetry conditions.

\subsection{Antiunitary skew-symmetric Hamiltonians}\label{sec:optimized_initial}
The first method is inspired by 3D magnetometry.
Consider a model whose state given by
\begin{equation}
\label{skew-symmetric_model}	e^{-iH_\param} \rho_{\rm ini} e^{iH_\param},
\end{equation}
where $H_\param$ is a parametrized Hamiltonian.
The Hamiltonian is assumed to have a skew-symmetry
\begin{equation}
\label{skew-symmetry}	\antiu H_\param \antiu^\dagger = - H_\param,
\end{equation}
under an antiunitary $\antiu$.
If there is a state $\ket[\psi_{\rm ini}]$ satisfying
\begin{equation}
\label{antiunitary-invariant}	\antiu \ket[\psi_{\rm ini}] = \epsilon \ket[\psi_{\rm ini}],
\end{equation}
with some unimodular $\epsilon$, $\antiu$ becomes a GAS of the model \eqref{skew-symmetric_model} by setting the initial state to $\rho_{\rm ini} = \ketbra[\psi_{\rm ini}]$.
The existence of antiunitary skew-symmetry \eqref{skew-symmetry} is the key for implementing the GAS.

Now we question the optimality of the antiunitary-invariant initial state \eqref{antiunitary-invariant} for estimating $\param$.
The N00N state reviewed in Sec.~\ref{sec:N00N} is antiunitary-invariant and, at the same time, the best initial state for the phase estimation on interferometry.
We show in the following that the simultaneous realization of the optimal initial state and the GAS is not a mere coincidence.

Optimal initial states are known for particular single-parameter estimations.
Consider a quantum system driven by a traceless generator $H$.
Now we are given a task to estimate the single parameter $t$ from the state
\begin{equation}
\label{phase_estimation}	e^{-iHt} \rho_{\rm ini} e^{iHt},
\end{equation}
where the initial state $\rho_{ini}$ is of our choice.
The best initial state that maximizes the quantum Fisher information for $t$ of the state \eqref{phase_estimation} is given by \cite{GrossCaves2020onefrommany}
\begin{equation}
	\label{generalGHZ}	\rho_{\rm ini} = \ketbra[\psi_{\rm ini}],~~ \ket[\psi_{\rm ini}] = \frac{\ket[\lambda_+] + e^{i\phi} \ket[\lambda_-]}{\sqrt{2}},
\end{equation}
where $\phi \in [0,2\pi]$ is arbitrary, and $\ket[\lambda_+]$ and $\ket[\lambda_-]$ are eigenstates of $H$ with maximum and minimum eigenvalues $\lambda_+$ and $\lambda_-$, respectively.
The optimal state maximizes the quantum Fisher information by maximizing the uncertainty in expecting $H$ \cite{BraunsteinCaves1994,Braunsteinetal1996}.

We can readily see that the resulting model
\begin{equation}
\label{maximumQFIM}	e^{-iHt} \ket[\psi_{\rm ini}] = \frac{e^{-i \lambda_+ t} \ket[\lambda_+] + e^{i(\phi- \lambda_-) t} \ket[\lambda_-]}{\sqrt{2}},
\end{equation}
is imaginarity-free.
The state lies on a great circle of the Bloch sphere of the qubit system spanned by $\ket[\lambda_+]$ and $\ket[\lambda_-]$.
Therefore the state is represented with real coefficients in a suitable basis.

When the Hamiltonian has a specific form, a GAS of the model \eqref{maximumQFIM} is derived from the antiunitary skew-symmetry of the Hamiltonian.
Suppose that the spectrum of the generator $H$ distributes symmetrically about the origin.
We further assume that the eigensubspaces $\hil_\lambda$ and $\hil_{-\lambda}$ belonging to eigenvalues $\lambda$ and $-\lambda$, respectively, have the same dimension.
Then an operator $\antiu$ on $\hil$ defined by
\begin{align}
	&\antiu \sum_{\lambda} \sum_{d=1}^{\dim \hil_\lambda} c_{\lambda,d} \ket[\lambda,d] = \sum_{\lambda} \sum_{d=1}^{\dim \hil_\lambda} c_{\lambda,d}^\ast \ket[- \lambda,d],\\
\nonumber	&\hspace{1cm} \left( \ket[\lambda_+,1] := \ket[\lambda_+],~\ket[\lambda_-,1] := \ket[\lambda_-] \right)
\end{align}
is an antiunitary.
This antiunitary is a symmetry of the initial state.
We have $\antiu \ket[\psi_{\rm ini}] = e^{- i\phi} \ket[\psi_{\rm ini}]$ and hence
\begin{equation}
\label{invariantGHZ}	\antiu \rho_{\rm ini} \antiu^\dagger = \rho_{\rm ini},
\end{equation}
by definition.
Moreover, the skew-symmetry
\begin{equation}
\label{skew-symmetricH}	\antiu H \antiu^\dagger = - H,
\end{equation}
of the generator under the action of $\antiu$ follows from the definition.

Combining Eqs.~\eqref{invariantGHZ} and \eqref{skew-symmetricH}, we arrive at the equality
\begin{equation}
	\antiu e^{-iHt} \rho_{\rm ini} e^{iHt} \antiu^\dagger = e^{-iHt} \rho_{\rm ini} e^{iHt},
\end{equation}
meaning that $\antiu$ is a GAS of the model \eqref{phase_estimation} for any optimal initial state.
Its optimal global measurements are lead by Thm.~\ref{thm:measurement} and Cor.~\ref{cor:global_estimation}.

The optimization of initial state presented above underlies single-parameter metrology schemes including the N00N state reviewed in Sec.~\ref{sec:N00N}.

In summary, if the spectrum and the eigensubspaces of generator $H$ distribute symmetrically about the origin, the model \eqref{phase_estimation} with any optimal initial state \eqref{generalGHZ} is imaginarity-free.
The resulting model with optimal initial state provides the target-independent optimal measurement to attain the best QCRB.
For a special generator of this kind, the optimization of initial state naturally leads to imaginarity-free models.

This observation dispels a doubt that GASs spoil advantages of quantum statistical models.
Imaginarity-free models look close to classical ones in that the freedom of imaginary numbers disappears, that all the parameters are compatible, and that global optimal measurements exist.
The classical behavior can, however, coexist with the most effective metrological schemes offered by quantum models.
In this sense, the imaginarity-free estimations benefits from both of classical and quantum estimations.

\subsection{Antiparallel model}\label{sec:antiparallel}
\begin{figure*}[tb]
	\centering
	\includegraphics[width=16cm]{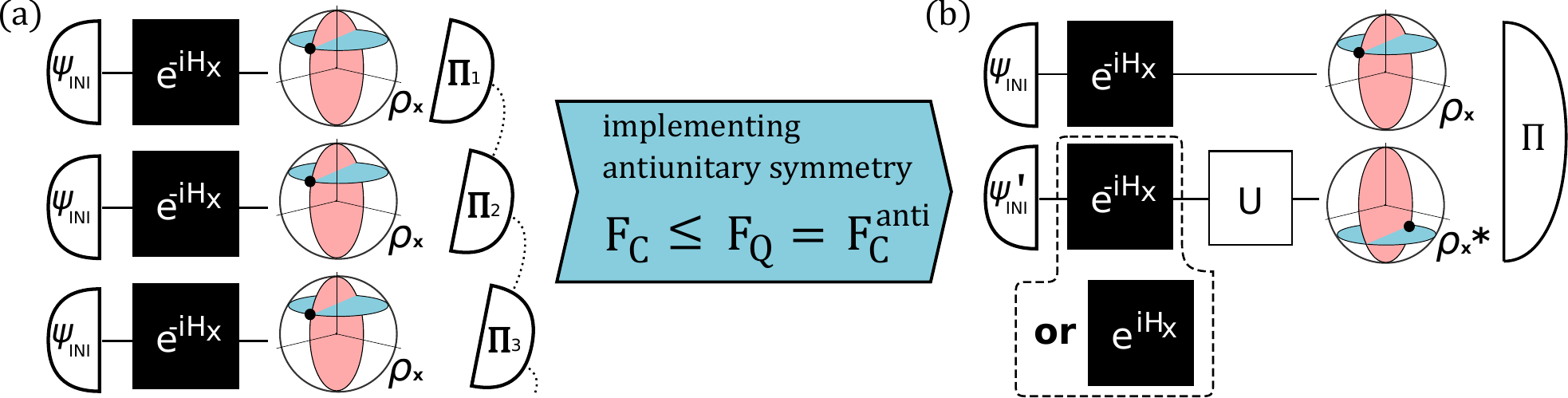}
	\caption{Methods and advantages to implement GASs by antiparallel models. The task is to estimate the values of parameters encoded in a Hamiltonian $H_\param$ driving the system. The initial states you prepare go through the parameter-dependent unitary evolution $e^{-iH_\param}$. They are output as the parametrized states $\state$, on which you perform measurements to estimate the parameters. (a) If the parametrized states do not form a weakly commutative model, QCRB cannot be saturated. Even if the model is weakly commutative, optimal measurements may depend on the output state, and require adaptive measurements $\Pi_1,\Pi_2,...$ in general. (b) If the model is imaginarity-free, the weak commutativity is guaranteed. If the outputs are pure states, there is a target-independent optimal measurement $\Pi$. Sometime it is possible to produce an antiparallel model, with implemented GAS. This is the case at least when the Hamiltonian has an antiunitary skew-symmetry $\antiu H_\param \antiu = - H_\param$ (Sec.~\ref{sec:skew_symmetry}), and when the inverse state evolution $e^{iH_\param}$ is available (Sec.~\ref{sec:inverse}).}
	\label{fig:implementing}
\end{figure*}

Our second method requires the state $\state$ of the original model and its conjugation $\state^\ast$.
This is a non-trivial requirement on the physical system since there is no completely-positive map to transform given state $\state$ to its conjugation $\state^\ast$.
We consider how to produce mutually conjugate state pairs later.
See FIG.~\ref{fig:implementing} for a schematic representation of our second proposal.

We propose to use, if possible, half-conjugated copies $\state \otimes \state^\ast$ instead of the normal copies $\state \otimes \state$.
\begin{definition}
	Let $\model$ be a quantum statistical model.
	\emph{Antiparallel model} for $\model$ is defined by
	\begin{equation}
	\label{antiparallel_model}	\left\{ \rho_\param \otimes \rho_\param^\ast \middle| \pspace \right\},
	\end{equation}
	where $\cdot^\ast$ represents a complex conjugation in a fixed basis.
\end{definition}
The antiparallel model has a GAS
\begin{equation}
\label{symmetry_antiparallel}	\swap \conj^{\otimes 2},
\end{equation}
\emph{independently} to the original model $\model$, where $S$ represents the swap operator $S \ket[\psi,\psi'] = \ket[\psi',\psi]$.
Thus the antiparallel model is weakly commutative at any point in the parameter space (Thm.~\ref{thm:global}) \emph{independently} to the original model $\model$.

The antiparallel model does not deteriorate the attainable precision.
QFIMs of the original model $\model$ ($\fisher_Q (\param)$), of the antiparallel model \eqref{antiparallel_model} ($\fisher_Q^\mathrm{anti} (\param)$), and of ``parallel model'' $\{ \state \otimes \state | \pspace \}$ ($\fisher_Q^\mathrm{para} (\param)$) are related by
\begin{equation}
\label{doubling_original}	\fisher_Q^\mathrm{para} (\param) = 2 \fisher_Q (\param) = \fisher_Q^\mathrm{anti} (\param)
\end{equation}
at any point $\pspace$.
The second equality is obtained directly from SLDs for the antiparallel model
\begin{equation}
\sld[i]^{\rm anti} := \sld[i] \otimes \id + \id \otimes \sld[i]^\ast,~(i=1,...,n),
\end{equation}
where $\sld[i]$ are SLDs of the original model.
Since the antiparallel model is weakly commutative, the QCRB \eqref{doubling_original} is asymptotically ($\nq \rightarrow \infty$) attainable on the antiparallel model.
The same bound is not attainable on the original incompatible model and the parallel model.

A reference basis to represent the GAS \eqref{symmetry_antiparallel} as a complex conjugation is
\begin{equation}
\label{antiparallel_measurement}	\left\{ \ket[i,i], \frac{\ket[i,j] + \ket[j,i]}{\sqrt{2}}, \frac{\ket[i,j] - \ket[j,i]}{\sqrt{2}i} \middle| i \neq j \right\},
\end{equation}
where $\ket[i,j]$ are the product basis for the complex conjugation.
When the model is pure, the corresponding projectors form an optimal set of POVMs to achieve QCRB \eqref{doubling_original} with $\nq=1$ \emph{independently} to the original model $\model$ (Thm.~\ref{thm:measurement}).

A notable theoretical aspect of ``antiparallelization'' is its model-independence.
According to the precision bound \eqref{doubling_original}, antiparallel state pair $\state \otimes \state^\ast$ always surpasses the parallel one $\state \otimes \state$ with respect to the attainable precision limit.
Mutually conjugated state pairs have revealed their intrinsic utility both theoretically \cite{GisinPopescu1999,Massar2000,CerfIblisdir2001,BraunsteinGhoshSeverini2007,Kato2009,Miyazaki2020strongly,Changetal2014} and experimentally \cite{Tangetal2020}.
Since these works all focus on restricted models such as spins, the extendible advantages of mutually conjugated state pairs has been kept unknown.
The reduction of incompatibility manifests a model-independent benefit of conjugate state pairs.

A potential intuitive reason why the conjugated state pair offers better mean square errors than the mere pair is that the former distributes over larger operator-space compared to the latter.
Volumes of spin ensembles characterized by entropy or code dimension are considered to be positively correlated to the fidelity in estimating the direction of spins from the sources of ensembles \cite{GisinPopescu1999,Baganetal2000,JoszaSchlienz2000,Miyazaki2020strongly}.
We can define a state ensemble from a quantum statistical model, by, for example, assuming uniform distribution over the parameter space.
Note, however, that mean square errors are defined locally at each point unlike the ensemble volumes and the fidelity.
Therefore mean square error should be compared to certain local quantity such as rate of volume growth around each point, rather than to the ensemble volume itself.
We leave the investigation on the ensemble volume of general antiparallel models as a future work.

Note that the antiparallel model is not the only method to implement a GAS independently of the original model.
Appendix \ref{sec:EQS} introduces a different kind of compatible models which are inspired by embedding quantum simulators \cite{Casanovaetal2011eqs,Candiaetal2013,Zhangetal2015eqs,Chenetal2016eqs,Loredoetal2016eqs,Chengetal2017eqs}.

There are instances where the mutually conjugate states $\state$ and $\state^\ast$ are available at the same time.
Here we consider quantum metrological schemes to estimate parameters encoded in a Hamiltonian $H_\param$.
Tasks of quantum metrology abstract to estimations of parameters from states $e^{-iH_\param} \ket[\psi]$ with the initial state $\ket[\psi]$ of your choice.
The antiparallel models additionally require the conjugated states $\conj e^{-iH_\param} \ket[\psi] = (\conj e^{-iH_\param} \conj) \ket[\conj \psi]$.
If it is possible to produce arbitrary initial states, the remaining task is to realize conjugate state evolution $\conj e^{-iH_\param} \conj = e^{i \conj H_\param \conj}$.

In the following subsections we introduce two major instances where the conjugate state evolution is available.

\subsubsection{Hamiltonians' skew-symmetry}\label{sec:skew_symmetry}
Let us assume that a Hamiltonian $H_\param$ flips its sign
\begin{equation}
	\antiu H_\param \antiu^\dagger= -H_\param,
\end{equation}
under an antiunitary $\antiu = U \conj$. 
In this case, the conjugated state evolution $e^{i \conj H_\param \conj}$ can be decomposed to
\begin{equation}
	e^{i \conj H_\param \conj} = U^\dagger e^{i \antiu H_\param \antiu^\dagger} U = U^\dagger e^{-iH_\param} U.
\end{equation}
Namely, the conjugate state evolution is unitarily equivalent to the original evolution $e^{-iH_\param}$.
Since the original evolution is available, the conjugated state
\begin{equation}
	U^\dagger e^{-iH_\param} U \ket[\conj \psi],
\end{equation}
can be produced by implementing unitary transformations $U$ and $U^\dagger$ before and after the evolution.

Note that if there is a state $\ket[\psi]$ such that $\antiu \ket[\psi] = \epsilon \ket[\psi]$ with some unimodular $\epsilon$, the model $e^{-iH_\param} \ket[\psi]$ itself obtains the GAS $\antiu$ because $\antiu e^{-iH_\param} \ket[\psi] = \epsilon e^{-iH_\param} \ket[\psi]$ holds (see Sec.~\ref{sec:optimized_initial} for details).
This is how 3D magnetometry and models in Sec.~\ref{sec:optimized_initial} acquire their GASs.

\subsubsection{Inverse transformation}\label{sec:inverse}
Suppose that the inverse transformation $e^{iH_\param}$ is available in addition to $e^{-iH_\param}$.
Let $\conj$ be any conjugation that leaves $H_\param$ invariant.
For example, one may take the eigenbasis of $H_\param$ as the reference basis for conjugation.
Then for any initial state $\ket[\psi]$,
\begin{equation}
	e^{iH_\param} \ket[\conj \psi],
\end{equation}
is the conjugate state of $e^{-iH_\param} \ket[\psi]$.

\section{Discussions}\label{sec:discussions}
\subsection{Antiunitary invariant POVMs on mixed states}\label{sec:mixed}
The antiunitary-invariant POVMs do not necessarily saturate QCRBs for mixed models.
This is partly because multipartite measurements ($\nq \geq 2$) are necessary for saturating QCRBs in general.
Any single-copy measurement cannot saturate the QCRB for such a model.
Theorem \ref{thm:measurement} does not generalizes to models with mixed states.

Now you may wonder what if the imaginarity-free mixed-state model is quasi-classical.
Any quasi-classical model is accompanied by a (potentially target-dependent) single-copy measurement to saturate the QCRB.
Does our antiunitary-invariant POVM saturate QCRBs for quasi-classical mixed models?

The answer is ``no.''
Even if a quasi-classical mixed state model has a GAS, the antiunitary-invariant POVM does not necessarily form the optimal POVM.
A counterexample is given in Appx.~\ref{sec:depolarized_antiparallel} by the off-equator estimation on depolarized antiparallel spins.
Our antiunitary-invariant POVM is not optimal to this quasi-classical mixed model.

The antiunitary-invariant POVMs do not seem to have any special property in the mixed state regime.
In Appx.~\ref{sec:depolarized_antiparallel}, we demonstrate that different antiunitary-invariant POVMs give different CFIMs for the off-equator phase estimation on depolarized antiparallel spins.
This implies that the antiunitary-invariant POVMs do not share anything in terms of the measurement precisions at the mixed state regime.
Their coincidence is only at the pure state limit.

\subsection{Local antiunitary symmetry}\label{sec:localantiunitary}
We have seen that the converse of Thm.~\ref{thm:global} is not true. There are weakly commutative models without any GAS.
To refine Thm.~\ref{thm:global}, we compare weak commutativity and antiunitary symmetry at each point in the parameter spaces.

We are motivated from the following geometric perspective.
Consider application of Thm.~\ref{thm:global} to subspaces of the whole parameter space. 
If $\region_1$ is a subspace of the parameter space $\region$ and if there is a GAS $\antiu_1 \rho_\param \antiu_1^\dagger = \rho_\param$ inside $\region_1$, then the model $\model$ is weakly commutative at all points in $\region_1$.
If there are such ``small'' GASs $\antiu_i \rho_\param \antiu_i^\dagger = \rho_\param$ for covering subspaces $\region_i$ $(i \in I,~\region = \cup_{i \in I} \region_i)$, then the model $\model$ is weakly commutative at all points in $\region$.
Note that $\antiu_i$s do not necessarily coincide.

Now a question is, whether any weakly commutative models have such covering subspaces endowed with antiunitary symmetries.
While global weak commutativity does not imply GAS, it may imply patchworks of the small antiunitary symmetries.
If so, weak commutativity and the existence of antiunitary symmetries should be equivalent at sufficiently small regions.
In the light of Thm.~\ref{thm:measurement}, such a ``local'' antiunitary symmetry might be useful for constructing target-\emph{dependent} optimal POVMs.

The answer to the question is positive for pure-state models and negative in general.
We show that all weakly commutative pure-state models have local antiunitary symmetries.
However, this result does not extend to mixed-state models.
The weakly commutativity does not imply local antiunitary symmetries.

We define the local antiunitary symmetry (LAS) as a straightforward limitation of the GAS to neighborhoods of a point.
\begin{definition}\label{def:local_antiunitary}
	A model $\model$ is said to have a LAS at $\pspace$, if there is an antiunitary operator $\antiu$ such that $\antiu \state \antiu^\dagger = \state$ and $\antiu \partial_i \state \antiu^\dagger = \partial_i \state$ hold for all $i$.
\end{definition}
To see how LAS relates to the GAS, consider the first-order approximation of state $\rho_{\param + \Delta \param}$ around the point $\pspace$:
\begin{equation}
	\rho_{\param + \Delta \param} \approx \state + \sum_{i=1}^n \Delta x_i \partial_i \state,
\end{equation}
where $\Delta x_i$ are small real numbers.
Then the LAS $\antiu$ satisfies
\begin{equation}
\label{local_antiunitary}	\antiu \rho_{\param + \Delta \param} \antiu^\dagger 	\approx \rho_{\param + \Delta \param},
\end{equation}
in the first-order approximation.
Thus if there is a LAS at $\pspace$, the states in a small neighborhood around $\param$ share an approximately good GAS.

LAS does imply weak commutativity at each point.
Furthermore, these two are equivalent for pure-state models.
\begin{theorem}\label{thm:local_antiunitary}
	If a model $\model$ has a LAS at $\pspace$, it is weakly commutative at $\pspace$.
	Pure models have LASs at points where the models are weakly commutative.
\end{theorem}
The proof of Thm.~\ref{thm:local_antiunitary} is delegated to Appx.~\ref{sec:proof_LAS}.
The off-equator phase estimation on spin, for example, have LASs at all points, despite having no GAS.
This implies that the piecewise antiunitary symmetries do not necessarily joint together to form a single antiunitary symmetry.

The LAS accompanies the optimal POVMs to saturate QCRB for pure-state models (see Appx.~\ref{sec:optimal_local_measurement} for details).
Similarly to the case of GAS (Thm.~\ref{thm:measurement}), they are the rank-$1$ POVM invariant under the LAS.
Since the LAS may depend on the target point, the optimal POVMs are also target-dependent in general.

Regarding mixed states, there are weakly commutative models without any LAS.
We can further say that there are quasi-classical models without any LAS when $\dim \hil \geq 3$, but not when $\dim \hil =2$.
Examples are presented in Appx.~\ref{sec:matsumotosan}.
The structure of compatibility is more involved than to be explained by antiunitary symmetries.

\subsection{Quantifying antiunitary asymmetry}\label{sec:imaginarity_measure}
In this section, we try quantifying antiunitary asymmetry of quantum statistical models.
Theorems \ref{thm:global} and \ref{thm:local_antiunitary} both states that incompatibility, a unique property of quantum mechanics, is absent for models with antiunitary symmetry.
Conversely, a model behaves quantum mechanically when it does not have antiunitary symmetry.

There will be many functions measuring antiunitary asymmetry of a quantum statistical model.
For a model $\model$ at a point $\pspace$ define
\begin{align}
	\nonumber & M (\state) \\
	\label{asymmetry_measure} &:= \min_{\substack{\antiu \in \mathcal{U}^\ast(\hil) \\ \antiu \state \antiu^\dagger = \state}} \frac{1}{n} \trace \left[ \sum_{i=1}^n  ( \partial_i \state - \antiu \partial_i \state \antiu^\dagger )^2 \right],
\end{align}
where $\mathcal{U}^\ast(\hil)$ denotes the set of antiunitary operators on $\hil$.
This measure is faithful in that it takes zero if and only if there is a \emph{local} antiunitary symmetry at $\pspace$.
It is also invariant under isometric coordinate transformations $\param \mapsto R \param$ ($R$: isometric matrix, see Appx.~\ref{sec:imaginarity_measure} for details).

We are inspired by imaginarity measures formulated in the resource theory to define the asymmetry measure \eqref{asymmetry_measure}.
Originally, the imaginarity of a single quantum state is measured with respect to a fixed basis $\{ \ket[k] \}_{k=1,...,d}$ \cite{HickeyGour2018imaginarity,Wu2021imaginarityPRL,Wu2021imaginarityPRA}, or equivalently to a fixed conjugation $\conj$.
The formalisms resemble resource theories of coherence \cite{BaumgratzPlenio2014coherence,ChitambarGour2016coherence,WinterDong2016coherence}, in that they both require a fixed reference basis.

In contrast, the weak commutativity is implied by any conjugation.
Such a basis-independent definition of imaginarity has not been studied yet.
However, there is a basis-independent definition of coherence termed ``set coherence'' \cite{DesignolleBrunner2021setcoherence}.
In Appx.~\ref{sec:imaginarity} we define ``set imaginarity'' by following the procedure for set coherence, and extend it to asymmetry measures for quantum statistical models.

The faithful measure of antiunitary asymmetry characterizes quantumness of the model, similarly to an incompatibility measure.
Carollo \emph{et. al.} \cite{Carollo2019} introduced a faithful measure of incompatibility.
As being faithful, it takes zero only if the mean Uhlmann curvature is zero.
The incompatibility measure is used to detect quantumness of phase transitions through the incompatibility of parametrized thermal states \cite{Carollo2019}.
Belliardo and Giovannetti \cite{BelliardoGiovannetti2021} introduced another faithful measure of incompatibility as a ratio between attainable precision and the precision based on SLD QFIMs.

Faithful asymmetry measures could be more sensitive to systems' quantumness than the incompatibility is.
As is presented in Appx.~\ref{sec:matsumotosan}, there are models with zero incompatibility but with positive antiunitary asymmetry.
Furthermore, imaginarity can be defined on single-parameter models, while the incompatibility requires more than $2$ parameters.
Thus measures of antiunitary asymmetry could be more useful than the incompatibility measure for characterizing quantum phase transitions.
We hope to study this in more detail on a different occasion.

\section{Conclusion}\label{sec:conclusion}
We introduced a class of quantum statistical models endowed with symmetries under antiunitary operators.
A model of this class consists only of real density matrices.
The null phase of density matrices implies the null mean Uhlmann curvature, enabling the compatible multiparameter estimation. 
A variety of existing models have antiunitary symmetries.

If a pure-state model has an antiunitary symmetry, any rank-$1$ antiunitary-invariant POVM saturates the QCRB.
Once you find the antiunitary symmetry, you discover an optimal measurement at the same time.
The antiunitary-invariant measurement is optimal independently of the target point to be estimated.
To perform such a measurement, you require neither prerequisite knowledge on the target, nor the adaptive measurement to guess the target.
By utilizing the freedom of the optimal measurements, we derived an antiunitary-invariant measurement for 3D magnetometry, which requires only bipartite entanglement irrespective to the number of spins.
The antiunitary-invariant measurement is, however, not necessarily optimal for mixed-state models. 

We proposed methods to implement antiunitary symmetries in several settings of quantum metrology.

For phase estimations of particular unitary evolutions, all the optimal initial states lead to imaginarity-free models.
Antiunitary symmetries can coexist with optimal metrological schemes in this setting.

Mutually conjugate state pair $\state \otimes \state^\ast$ has better precision bounds than $\state \otimes \state$ does, independently of the model $\state$.
The conjugate pair has an antiunitary symmetry independent of the original model $\state$.
We consider how to produce the conjugated state $\state^\ast$ when a parametrized unitary evolution generates $\state$.
This is possible especially if the Hamiltonian has an antiunitary skew-symmetry, or if the inverse state evolution is also available.

Finally, we studied antiunitary symmetries varying along the target points of a model.
A pure-state model is compatible if and only if it has the target-dependent antiunitary symmetry.
Based on the study, we introduced a function to measure antiunitary asymmetry of quantum statistical models at given target points.
The measure is faithful and invariant under isometric coordinate transformations on the parameter space.

Our results have room for extension.
Despite no mention here, infinite dimensional Hilbert spaces have unique uncertainty relations.
Target-independent optimal measurements are, including their existence, still unknown for mixed state models.
Extending our results in these untouched regions will help develop various metrology schemes such as superresolution \cite{Tsangetal2016,Chrostowskietal2017,Rehaceketal2017,Rehaceketal2018optmeas,YuPrasad2018,Napolietal2019}.

We are currently relying on physicists' intuition to discover the antiunitary symmetry of quantum statistical models.
While intuition sometimes works perfectly on physics-born models, a general algorithm to find the antiunitary symmetry (if it exists) would be much more effective.
Such an algorithm will be useful not only for quantum metrology, but also for understanding quantum phase transitions via asymmetry measures.

\section*{Acknowledgement}
JM would like to thank M. T. Quintino, M. Ara\'{u}jo and J. Sidhu for valuable discussions.

\onecolumn\newpage
\twocolumn
\appendix

\section{Additional examples of imaginarity-free models}\label{sec:models_appx}
\subsection{Antiparallel spins}
Some pure quasi-classical models cease to be quasi-classical when they pass depolarizing channels.
We demonstrate this phenomena with antiparallel spins in \ref{sec:antiparallel_spins}.
In contrast, depolarizing channels preserve the existence and non-existence of GAS.
For $\delta \in (0,1)$, the depolarized model
\begin{equation}
\left\{ (1- \delta) \state + \delta \frac{\id}{\dim \hil} \middle| \pspace \right\},
\end{equation}
is imaginarity-free if and only if the original model $\model$ is.
In fact, $\antiu \state \antiu^\dagger = \state$ and $\antiu \id \antiu^\dagger = \id$ implies $\antiu (p \state + q \id ) \antiu^\dagger = p \state + q \id$ for any real numbers $p$ and $q$.
Depolarizing channels break quasi-classicality while preserving GASs.

\subsubsection{Spin}\label{sec:spin}
\begin{figure}[tbh]
	\centering
	\includegraphics[width=8cm]{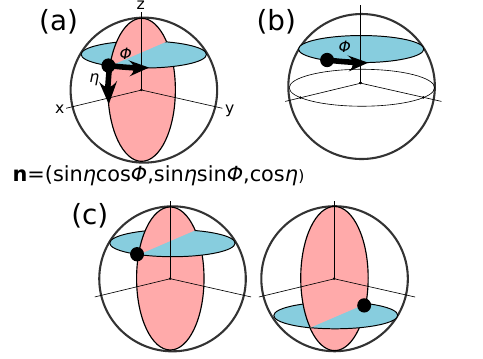}
	\caption{Bloch sphere representations of the spin model and its variants. (a) A (single) spin model. The state is represented by the Bloch vector, which is parametrized by the spherical coordinates (with unit radius). (b) The off-equator phase estimation on spin model. The latitude $\eta$ is regarded as a constant. (c) The antiparallel spins. The Bloch vectors are directing opposite to each other.}
	\label{fig:bloch}
\end{figure}

Before proceeding to the weakly commutative models, we review a model that is itself not weakly commutative.
For spherical coordinates $\eta \in [0, \pi]$ and $\phi \in [0,2\pi]$ define a qubit state on $2$-dimensional space $\hil_2$
\begin{align}
\label{polar_model}	&\rho_{(\eta,\phi)} = \frac{\id_2 + \vecn_{(\eta,\phi)} \cdot {\bm \sigma} }{2},\\
\nonumber & \left( \vecn_{(\eta,\phi)} = (\sin \eta \cos \phi, \sin \eta \sin \phi, \cos \eta) \right)
\end{align}
where ${\bm \sigma} = (\sigma_x, \sigma_y, \sigma_z)$ is the vector of unitary Pauli matrices.
The Bloch sphere representation of a spin is given in FIG.~\ref{fig:bloch} (a).
The mean Uhlmann curvature of model
\begin{equation}
\label{spin}	\left\{ \rho_{(\eta,\phi)} \middle| \eta \in [0, \pi],~ \phi \in [0,2\pi] \right\},
\end{equation}
is given by
\begin{align}
	\label{miu_qubit_polar}	\mathcal{U} = \left[ \begin{array}{cc}
		\mathcal{U}_{\eta \eta} & \mathcal{U}_{\eta \phi} \\
		\mathcal{U}_{\phi \eta} & \mathcal{U}_{\phi \phi} \\
	\end{array} \right] = \frac{1}{2}
	\left[ \begin{array}{cc}
		0 & \sin \eta \\
		- \sin \eta & 0 \\
	\end{array} \right].
\end{align}
Thus \eqref{spin} is weakly commutative only at the north and south poles $\eta=0, \pi$.
We call the state $\rho_{(\eta,\phi)}$ and the model \eqref{spin} as a spin and a spin model, respectively.
The spin model turns weakly commutative with some modifications presented in FIG.~\ref{fig:bloch} (b) and (c).

\subsubsection{Off-equator phase estimation}\label{sec:phase_estimation}
Fixing the latitude $\eta$ of the spin to a constant $c$ results in a single-parameter model of phase $\phi \in [0, 2 \pi]$.
States of the single-parameter model
\begin{equation}
\label{single_circle}	\left\{ \rho_{(c,\phi)} \middle| \phi \in [0,2\pi] \right\},
\end{equation}
form a circle in the Bloch sphere with a constant latitude $c \in (0,\pi)$ (a schematic representation is in FIG.~\ref{fig:bloch} (b)).

The model does not have any GAS except at the equator $c = \pi/2$.
At the equator, the spin-flip followed by $\pi$-rotation along the $z$-axis becomes a GAS.
Outside the equator, the circular orbits are passed to the other hemisphere by the spin-flip.
There is no unitary to return the whole circles to the original hemispheres \cite{BraunsteinGhoshSeverini2007}.

\subsubsection{Antiparallel spins}\label{sec:antiparallel_spins}
Antiparallel spins are two spins directing the opposite to each other.
A schematic representation is given in FIG.~\ref{fig:bloch} (c).
The antiparallel spin model is represented by
\begin{equation}
\label{antiparallel_spins}	\left\{ \rho_{(\eta,\phi)} \otimes \antiu_f \rho_{(\eta,\phi)} \antiu_f^\dagger \middle| \eta \in [0, \pi],~ \phi \in [0,2\pi] \right\},
\end{equation}
since the spin-flip operator $\antiu_f$ flips a spin to its opposite direction.

Gisin and Popescu \cite{GisinPopescu1999} showed that the spin direction $\vecn_{(\eta,\phi)}$ can be better guessed from the antiparallel spins than from parallel spins.
Later Chang \emph{et. al.} \cite{Changetal2014} demonstrated the advantage of antiparallel spins for the estimation of spherical coordinates $(\eta,\phi)$.
The weak commutativity of antiparallel spins is shown there.
Since the model consists of pure states, it must be quasi-classical \cite{Matsumoto2002}.

The antiparallel spins has a GAS
\begin{equation}
\label{symmetry_antiparallel_spins}	\swap \antiu_f^{\otimes 2},
\end{equation}
where $\antiu_f^{\otimes 2}$ is the tensor product of spin-flips, and $\swap$ is the operator which swaps the sides of bipartite system $\hil_2 \otimes \hil_2$.
The antiunitary operator $\antiu_f^{\otimes 2}$ flips the directions of spins on both sides, and then unitary $\swap$ swaps the spins to recover the original state.
The GAS \eqref{symmetry_antiparallel_spins} is a conjugation despite $\antiu_f$ is not.

Depolarized antiparallel spins
\begin{equation}
\label{depolarized_antiparallel}	(1- \delta) \rho_{(\eta,\phi)} \otimes \antiu_f \rho_{(\eta,\phi)} \antiu_f^\dagger + \delta \frac{\id_4}{4}.
\end{equation}
are no longer quasi-classical, nor partially commutative (see Appx.~\ref{sec:depolarized_antiparallel} for details).
Nevertheless, the GAS \eqref{symmetry_antiparallel_spins} is preserved by the depolarized channel.

Let us consider the optimal POVMs for antiparallel spins.
Vectors
\begin{equation}
\label{gisin_basis}	\ket[\Pi_k] = \frac{\sqrt{3}+1}{2 \sqrt{2}} \ket[\vecn_k,-\vecn_k] + \frac{\sqrt{3}-1}{2 \sqrt{2}} \ket[-\vecn_k,\vecn_k],
\end{equation}
with properly chosen four Bloch vectors $\vecn_k$ ($k=1,...,4$) are known to form an optimal measurement basis for antiparallel spins \cite{Changetal2014}.
Here, $\ket[\vecn]$ represents a state vector with Bloch vector $\vecn$ (see \cite{GisinPopescu1999,Changetal2014} for details on the phase factors).
This set of vectors indeed satisfy
\begin{equation}
	\swap \antiu_f^{\otimes 2} \ket[\Pi_k] = - \ket[\Pi_k],
\end{equation}
and thus $\{ i \ket[\Pi_k] \}_{k=1,...,4}$ is a reference basis that represents $\swap \antiu_f^{\otimes 2}$ as a mere complex conjugation.

There are different reference bases for the same conjugation $\swap \antiu_f^{\otimes 2}$.
An example is
\begin{equation}
\label{antiparallel_basis}	\left\{  i \ket[0,1],~ i \ket[1,0],~\frac{\ket[0,0]+\ket[1,1]}{\sqrt{2}},~\frac{\ket[0,0]-\ket[1,1]}{\sqrt{2}i} \right\},
\end{equation}
where $\{ \ket[0,0],~\ket[0,1],~\ket[1,0],~\ket[1,1] \}$ now represents the computational basis.
Later we see that our basis \eqref{antiparallel_basis} sometimes outperforms \eqref{gisin_basis} in terms of the precision for depolarized antiparallel spins (Sec.~\ref{sec:mixed}).
This is unexpected because the basis \eqref{gisin_basis} provided by Gisin and Popescu \cite{GisinPopescu1999} looks much specialized to the estimation of $\vecn_{(\eta,\phi)}$.
In fact, each component of \eqref{gisin_basis} is the optimal covariant measurement vector to maximize a fidelity measure for this specific problem \cite{Massar2000}.

\subsubsection{Classical and quantum Fisher information matrices for antiparallel spins}\label{sec:depolarized_antiparallel}
We first show that the depolarized antiparallel spins is not quasi-classical.
Then we present the deviation $\Delta \phi^2$ of single-parameter $\phi$ stemming from POVM measurements on depolarized antiparallel spins.
The latter corresponds to the precision of the off-equator phase estimation on the depolarized antiparallel spins.

The state of depolarized antiparallel spins is defined by Eq.~\eqref{depolarized_antiparallel}.
A pair of SLDs for depolarized antiparallel spins is given by
\begin{equation}
	L^{\rm anti}_x = (1-\delta) \left( L_x \otimes \id_2 + \id_2 \otimes \antiu_f L_x \antiu_f^\dagger \right),
\end{equation}
for $x=\eta,\phi$, where $L_x$ is the SLD for single spin, given by
\begin{align}
	L_\eta &= \left[ \begin{array}{cc}
		- \sin \eta & e^{- i \phi} \cos \eta \\
		e^{i \phi} \cos \eta & \sin \eta \\
	\end{array} \right], \\
	L_\phi &= \left[ \begin{array}{cc}
		0 & - i e^{- i \phi} \sin \eta \\
		i e^{i \phi} \sin \eta & 0 \\
	\end{array} \right].
\end{align}
The SLD operators for the depolarized antiparallel spins is unique since the state is of full-rank.
The commutator $[L^{\rm anti}_\eta , L^{\rm anti}_\phi]$ of SLD operators reduces to
\begin{equation}
	(1-\delta)^2 \left( [L_\eta , L_\phi] \otimes \id_2 + \id_2 \otimes \antiu_f [L_\eta , L_\phi] \antiu_f \right).
\end{equation}
This never becomes zero since $[L_\eta , L_\phi]$ is not proportional to $\id_2$.
Thus the depolarized antiparallel spins is not quasi-classical.
It is not partially commutative as being of full-rank. 

The deviation $\Delta \phi^2$ can be calculated as the inverse of the classical Fisher information about $\phi$. 
We here provide a full analysis on the two CFIMs for reference-basis measurements \eqref{gisin_basis} and \eqref{antiparallel_basis}.
Note that for general depolarized state $(1-\delta) \rho_\param + \delta \id/\dim \hil$ and a projector $\Pi_k$, probability to have the corresponding result $k$ is
\begin{align}
& \trace \left[ \left( (1-\delta) \rho_\param + \frac{\delta \id}{\dim \hil} \right) \Pi_k \right] \\
&= (1-\delta) \trace[ \rho_\param \Pi_k] + \frac{\delta}{\dim \hil}.
\end{align}
The $i,j$-component of the CFIM is
\begin{equation}
[\fisher_C (\param)]_{ij} = (1-\delta) \sum_{k} \frac{ \partial_i \trace[ \rho_\param \Pi_k] \partial_j \trace[ \rho_\param \Pi_k]}{\trace[ \rho_\param \Pi_k] + \frac{\delta}{4(1-\delta)}}.
\end{equation}
The CFIM is calculated from the probabilities $\trace[ \rho_\param \Pi_k]$ to obtain the result $k$ by the measurement on the original state.

Let us first consider the basis \eqref{gisin_basis} given by Gisin and Popescu \cite{GisinPopescu1999}.
The probabilities to obtain result $k$ ($k=1,...,4$) can be found, for example, in the appendix of \cite{Changetal2014}:
\begin{align}
	& \bra[\Pi_k] \rho_{(\eta,\phi)} \otimes \antiu_f \rho_{(\eta,\phi)} \antiu_f^\dagger \ket[\Pi_k] = \frac{1}{3} A_k(\eta,\phi)^2, \\
\nonumber	& A_k(\eta,\phi) := \sin \eta \cos \left( \phi - \frac{(k-1)2\pi}{3}\right) \\
	&\hspace{3cm} - \frac{\sqrt{2}}{4} \cos \eta + \frac{\sqrt{6}}{4}.
\end{align}
The corresponding elements of CFIM for the depolarized antiparallel spins are given by
\begin{align}
\nonumber	[\fisher_C]_{ij} =& \frac{4(1-\delta)}{3} \\
\times &\sum_{k=1}^4 \frac{A_k(\eta,\phi)^2 B_k^i(\eta,\phi) B_k^j(\eta,\phi)}{A_k(\eta,\phi)^2 + \frac{3\delta}{4(1-\delta)}},\\
\nonumber	B_k^\eta(\eta,\phi) :=& \cos \eta \cos \left( \phi - \frac{(k-1)2\pi}{3}\right) \\
	& \hspace{2.5cm} - \frac{\sqrt{2}}{4} \sin \eta,\\
	B_k^\phi(\eta,\phi) :=& - \sin \eta \sin \left( \phi - \frac{(k-1)2\pi}{3}\right).
\end{align}
The deviation $\Delta \phi^2$ for this measurement is given by $[\fisher_C (\param)]_{\phi \phi}^{-1}$.

If our measurement \eqref{antiparallel_basis} is applied on the pure antiparallel spins, the probabilities are
\begin{align}
	&\bra[0,1] \rho_{(\eta,\phi)} \otimes \antiu_f \rho_{(\eta,\phi)} \antiu_f^\dagger \ket[0,1] = \frac{(1+\cos \eta)^2}{4},\\
	&\bra[1,0] \rho_{(\eta,\phi)} \otimes \antiu_f \rho_{(\eta,\phi)} \antiu_f^\dagger \ket[1,0] = \frac{(1 - \cos \eta)^2}{4},\\
\nonumber	& \frac{(\bra[0,0] \pm \bra[1,1]) \rho_{(\eta,\phi)} \otimes \antiu_f \rho_{(\eta,\phi)} \antiu_f^\dagger (\ket[0,0] \pm \ket[1,1])}{2} \\
	&= \frac{\sin^2 \eta (1 \mp \cos 2\phi)}{4}.
\end{align}
The corresponding elements of CFIM are given by
\begin{widetext}
\begin{align}
\nonumber		\left[ \fisher_C \right]_{\eta \eta} =& (1-\delta) \left\{ \frac{\sin^2 \eta (1+\cos \eta)^2}{(1+\cos \eta)^2 + \frac{\delta}{1- \delta}} + \frac{\sin^2 \eta (1-\cos \eta)^2}{(1-\cos \eta)^2 + \frac{\delta}{1- \delta}} \right\} \\
		&\hspace{1cm} + (1-\delta) \left\{ \frac{\sin^2 \eta \cos^2 \eta (1+\cos 2\phi)^2}{\sin^2 \eta (1+\cos 2 \phi) +\frac{\delta}{1-\delta}} + \frac{\sin^2 \eta \cos^2 \eta (1-\cos 2\phi)^2}{\sin^2 \eta (1-\cos 2 \phi) +\frac{\delta}{1-\delta}} \right\},\\
		\left[ \fisher_C \right]_{\eta \phi} =& (1-\delta) \left\{ \frac{\sin^3 \eta \cos \eta \sin 2 \phi (1-\cos 2\phi)}{\sin^2 \eta (1-\cos 2 \phi) +\frac{\delta}{1-\delta}} - \frac{\sin^3 \eta \cos \eta \sin 2\phi (1+\cos 2\phi)}{\sin^2 \eta (1+\cos 2 \phi) +\frac{\delta}{1-\delta}} \right\},\\
		\left[ \fisher_C \right]_{\phi \phi} =& (1-\delta) \left\{ \frac{\sin^4 \eta \cos^2 2 \phi }{\sin^2 \eta (1+\cos 2\phi) +\frac{\delta}{1-\delta}} + \frac{\sin^4 \eta \cos^2 2 \phi}{\sin^2 \eta (1-\cos 2 \phi) +\frac{\delta}{1-\delta}} \right\}.
\end{align}
\end{widetext}
The deviation $\Delta \phi^2$ is again given by the inverse of $[\fisher_C (\param)]_{\phi \phi}$.

The QFIM of pure antiparallel model is
\begin{equation}
\fisher_Q^{\rm anti} = 2 \left[\begin{array}{cc}
1 & 0 \\
0 & \sin^2 \eta
\end{array}\right] = 2 \fisher_Q^{\rm spin},
\end{equation}
where $\fisher_Q^{\rm spin}$ is the QFIM of single spin.
While this relation can be derived by a direct calculation, it is expected by Eq.~\eqref{doubling_original}.
The depolarization deteriorates QFIM to
\begin{equation}
\fisher_Q^{\rm anti} = 2(1-\delta)^2 \left[\begin{array}{cc}
1 & 0 \\
0 & \sin^2 \eta
\end{array}\right].
\end{equation}
Again the minimum of deviation $\Delta \phi^2$ is the inverse $[\fisher_Q^{\rm anti}]_{\phi \phi} = 1/ 2(1-\delta)^2 \sin^2 \eta$.
This bound is the best attained by the (potentially target-dependent) optimal measurements each on a single copy of the depolarized antiparallel spin pairs.

\begin{figure}[bt]
	\centering
	\includegraphics[width=8.3cm]{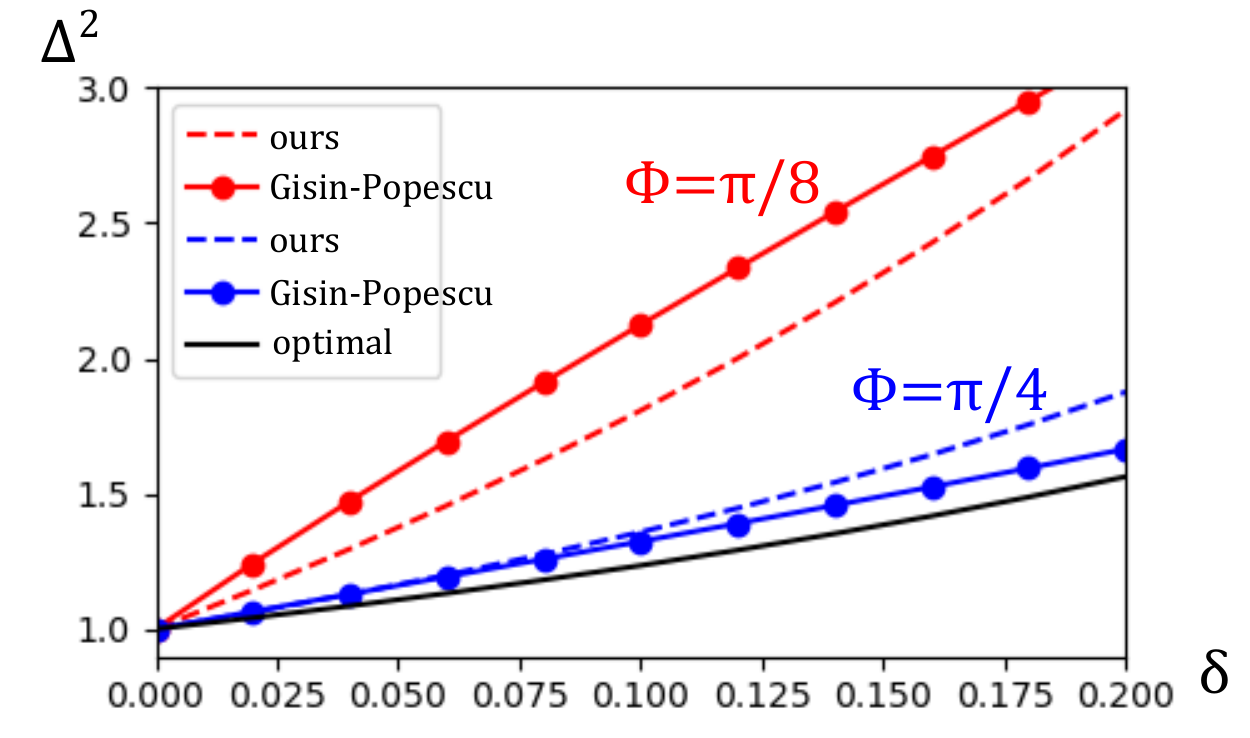}
	\caption{Performances of the reference-basis measurements for the off-equator phase estimation on depolarized antiparallel spins. The deviation $\Delta \phi^2 = [\fisher_C]_{\phi \phi}^{-1}$ stemming from measurements in the bases \eqref{gisin_basis} (dotted solid lines) and \eqref{antiparallel_basis} (dashed lines) are represented as functions of $\delta$ and compared to the optimal bound $[\fisher_Q]_{\phi \phi}^{-1}$ (black solid line). The target points are $(\eta,\phi) = (3\pi/4,\pi/8)$ (red lines) and $(3\pi/4,\pi/4)$ (blue lines), respectively. These measurements are optimal only at the pure limit $\delta  =0$. See Appendix \ref{sec:depolarized_antiparallel} for analytic expressions of $\Delta \phi^2$.}
	\label{fig:antiphase}
\end{figure}

We take the latitude $\eta$ as a constant and consider the single-parameter estimation of phase $\phi$, which is trivially quasi-classical.
This is the antiparallel version of off-equator phase estimation.
The deviation $\Delta \phi^2$ based on the measurements in reference bases \eqref{gisin_basis} and \eqref{antiparallel_basis} is compared to the optimal deviation in FIG.~\ref{fig:antiphase}.
The deviations are equal to the optimal value only at the pure limit ($\delta=0$), and otherwise lager.
As the coloured lines in FIG.~\ref{fig:antiphase} show, there is a separation of CFIMs calculated from two measurement bases \eqref{gisin_basis} and \eqref{antiparallel_basis}.

\subsection{Qubit inside a disc}\label{sec:suzuki}
Suzuki \cite{Suzuki2016} gave an weakly commutative model of a qubit
\begin{align}
\label{disc}	\left\{ \rho_\param = \frac{\id_2 + \vecn_\param \cdot {\bm \sigma} }{2} \middle| \begin{array}{l}\vecn_\param = f_1(\param) \vecn_1 + f_2(\param) \vecn_2, \\ \pspace \end{array} \right\},
\end{align}
where $f_1$ and $f_2$ are arbitrary real functions of the parameter $\pspace$ such that $|\vecn_\param | \leq 1$, and $\vecn_1$ and $\vecn_2$ are arbitrary (not necessarily orthogonal) unit vectors on $3$-dimensional real space.
In the Bloch sphere representation, the qubit is in a disc spanned by two vectors $\vecn_1$ and $\vecn_2$.
The disc cross the origin $(0,0,0)$ and cuts the sphere into halves.
This model is not quasi-classical \cite{Suzuki2016}.

The qubit inside a disc \eqref{disc} is imaginarity-free.
The state transforms under the spin-flip $\antiu_f$ as
\begin{equation}
	\frac{\id_2 + \vecn_\param \cdot {\bm \sigma} }{2} \mapsto \frac{\id_2 - \vecn_\param \cdot {\bm \sigma} }{2}.
\end{equation}
Mutually opposite vectors $\vecn_\param$ and $-\vecn_\param$ are always in the same plane which includes the origin $(0,0,0)$.
Thus the spin-flip followed by the unitary $\pi$-rotation along this plane takes any state $\rho_\param$ to itself.

It is crucial for the disc to cross the origin.
Otherwise the model does not have a GAS.

\subsection{Embedding quantum simulators}\label{sec:EQS}
Here we introduce a class of model inspired by embedding quantum simulators \cite{Casanovaetal2011eqs,Candiaetal2013,Zhangetal2015eqs,Chenetal2016eqs,Loredoetal2016eqs,Chengetal2017eqs}.
\emph{EQS model} for pure state model $\{ \ketbra[\psi_\param] | \pspace \}$ is defined by
\begin{equation}
\label{eqs_model}	\left\{ \eqsmix := \ketbra[\Psi_\param^\mathrm{EQS}] \in \B(\hil \otimes \hil_2) \middle| \pspace \right\},
\end{equation}
where
\begin{equation}
\eqspure := \frac{ \ket[\psi_\param] \otimes \ket[0] + (\conj \ket[\psi_\param]) \otimes \ket[1]}{\sqrt{2}},
\end{equation}
with some conjugation $\conj$ on $\hil$, where $\hil_2$ is a two dimensional ancillary system with an orthonormal basis $\{ \ket[0],\ket[1] \}$.
The transformation $\ket[\psi_\param] \mapsto \eqspure$ is not realized as a completely positive map.
Extra resources are required to produce the state $\eqspure$.

The global phase subtlety matters when modifying to EQS models.
The modifications to EQS models result in different models from $\ket[\psi_\param]$ and from $\ket[e^{i\phi(\param)} \psi_\param]$ where $\phi(\param)$ is a parameter-dependent global phase.
While these two original models are equivalent, their EQS models demonstrates different precision bounds.

Embedding quantum simulators typically utilize the encoding of state $\ket[\psi_\param]$ in the form
\begin{equation}
\label{eqs_original}	\frac{\ket[\Real \psi_\param] \otimes \ket[0] + \ket[\Imag \psi_\param] \otimes \ket[1]}{\sqrt{2}}
\end{equation}
instead of $\eqspure$ \cite{Casanovaetal2011eqs}.
With \eqref{eqs_original} we arrive at the same precision bound since it is unitarily equivalent to $\eqspure$.

A GAS of EQS model \eqref{eqs_model} is
\begin{equation}
\label{antiunitary_symmetry_EQS}	(\id_\hil \otimes \sigma_X) (\conj \otimes \conj_2),
\end{equation}
where $\conj \otimes \conj_2$ is the conjugation in the product basis $\ket[i] \otimes \ket[j]$ such that $\conj \ket[i] = \ket[i]$ ($i=1,...,\dim \hil, ~j=0,1$).
A reference basis to represent \eqref{antiunitary_symmetry_EQS} as a complex conjugation is
\begin{equation}
\left\{ \ket[i] \otimes \frac{\ket[0] + \ket[1]}{\sqrt{2}},~\ket[i] \otimes \frac{\ket[0] - \ket[1]}{\sqrt{2}i} \middle| ~i=1,...,\dim \hil \right\}.
\end{equation}
Thus a product measurement saturates QCRB.

QFIMs of EQS models are sometimes improved from the original models.
QFIM of an EQS model relates to that of the original model by
\begin{equation}
\label{sldeqs}	\fisher_Q^\mathrm{EQS} (\param)= \fisher_Q (\param) + {\bm X} {\bm X}^\transpose,
\end{equation}
where
\begin{equation}
{\bm X}^\transpose = 2 i \left(\langle \partial_1 \psi_\param | \psi_\param \rangle ~ ... ~ \langle \partial_n \psi_\param| \psi_\param \rangle \right),
\end{equation}
is a real vector.
It is futile to modify EQS models again to their EQS models to obtain even better precision bound.
The second modifications are equivalent to mere unitary transformations.

\section{Local antiunitary symmetry}\label{sec:LAS}
This section technically supports the discussion of Sec.~\ref{sec:localantiunitary} on LAS by providing proofs.
We prove Thm.~\ref{thm:local_antiunitary} in Sec.~\ref{sec:proof_LAS}.
Section \ref{sec:optimal_local_measurement} derives optimal local measurements from LAS.
Section \ref{sec:matsumotosan} gives examples of quasi-classical models without any antiunitary symmetry.

Following lemma is frequently employed throughout this section.
\begin{lemma}\label{lem:LAS}
	A model $\model$ has a LAS $\antiu$ at $\pspace$ if and only if (1) $\antiu \rho \antiu^\dagger = \state$ holds and (2) there are set of SLD operators satisfying
	\begin{equation}
	\label{LAS_lemma}	\antiu \sld[j] \antiu^\dagger = \sld[j], ~(j=1,...,m)
	\end{equation}
	at $\param$.
\end{lemma}
\emph{proof})
Suppose $\model$ has a LAS $\antiu$ at $\pspace$.
A set of SLD operators is defined by Eq.~\eqref{integralSLD}.
These SLD operators satisfy Eq.~\eqref{LAS_lemma} because $\antiu \state \antiu^\dagger = \state$ and $\antiu \partial \state \antiu^\dagger = \partial \state$ holds by the assumption.

Conversely, if $\state$ and $\sld[j]$ are $\antiu$-invariant, the defining equation \eqref{sld} implies that $\antiu$ is a LAS of $\model$ at $\pspace$.
$\qed$\\
Note that if the density operator is of full-rank, then the SLDs are in one-to-one correspondence with the derivatives $\partial_j \state$.

\subsection{Proof of Theorem \ref{thm:local_antiunitary}}\label{sec:proof_LAS}
We start by proving that any model $\model$ with a LAS $\antiu$ at $\pspace$ is weakly commutative at $\param$.
From Lem.~\ref{lem:LAS}, there are SLD operators $\sld[j]$ satisfying Eq.~\eqref{LAS_lemma}.
Then we have
\begin{align}
\nonumber	&2 \Imag \left[ \trace[ \state \sld[i] \sld[j] ] \right] \\
\nonumber	&= \trace[ \state \sld[i] \sld[j] ] - \trace[ \state \sld[i] \sld[j] ]^\ast \\
\nonumber	&= \trace[ \state \sld[i] \sld[j] ] - \trace[ \antiu \state \sld[i] \sld[j] \antiu^\dagger] \\
	&=0,
\end{align}
because $\sld[i]$, $\sld[j]$ and $\state$ are all invariant under $\antiu$.
This reveals the weak commutativity of $\model$ at $\param$.

Conversely, assume that a pure-state model $\{ \ketbra[\psi_\param] | \pspace \}$ is weakly commutative at point $\pspace$.
Since all pure weakly commutative models are quasi-classical \cite{Matsumoto2002}, there is a set of SLDs satisfying
\begin{equation}
	[ \sld[i], \sld[j] ] = 0,
\end{equation}
for any pairs of $i,j$.
Let $\{ \ket[\Pi_k] \}_{k=1,...,d}$ be a basis of $\hil$ in which the SLDs are diagonalized
\begin{equation}
	\sld[i] = \sum_{k=1}^d a^{(i)}_k \ketbra[\Pi_k],
\end{equation}
where $a^{(i)}_k$ are the real eigenvalues.
We represent the state vector
\begin{equation}
	\ket[\psi_\param] = \sum_{k=1}^d b_k e^{i \phi_k} \ket[\Pi_k],
\end{equation}
with the real coefficients $b_k$ and phase $\phi_k$ ($k=1,...,d$).

Define a new basis vectors $\ket[\Pi_k']$ by
\begin{equation}
	\ket[\Pi_k'] := e^{i \phi_k} \ket[\Pi_k],
\end{equation}
for $k=1,...,d$.
The state $\ketbra[\psi_\param]$ is represented by a real matrix in this basis since $\ket[\psi_\param] = \sum_{k=1}^d b_k \ket[\Pi_k']$.
The SLDs are still diagonalized by the new basis.
The derivative $\partial_i (\ketbra[\psi_\param])$ is also real in this basis since
\begin{align}
\nonumber	2 \partial_i (\ketbra[\psi_\param]) &= \sld[i] \ketbra[\psi_\param] + \ketbra[\psi_\param] \sld[i] \\
	&= \sum_{k,l=1}^d (a^{(i)}_k + a^{(i)}_l) b_k b_l \ket[\Pi_k'] \bra[\Pi_l'],
\end{align}
holds for all $i$.
The complex conjugation in the basis $\{ \ket[\Pi_k'] \}_{k=1,...,d}$ is a LAS of $\{ \ketbra[\psi_\param] | \pspace \}$ at $\pspace$.

\subsection{Optimal measurements from local antiunitary symmetry}\label{sec:optimal_local_measurement}
Here we prove that rank-$1$ POVM that is invariant under LASs forms optimal measurement for local multiparameter estimation.
\begin{theorem}\label{thm:local_measurement}
	If a POVM $\E_\antiu$ satisfies conditions \eqref{rank1} and \eqref{invariantPOVM}, and if $\antiu$ is a LAS of $\model$ at $\pspace$, $\E_\antiu$ saturates the QCRB \eqref{HCRB} of model $\model$ at $\pspace$.
\end{theorem}
\emph{proof})
From Thm.~\ref{thm:measurement}, the POVM $\E_\antiu$ satisfying conditions \eqref{rank1} and \eqref{invariantPOVM} exists only if $\antiu$ is a conjugation.
Therefore we assume that the antiunitary $\antiu$ is a conjugation $\conj$.

From Lem.~\ref{lem:LAS} there is a set of SLD operators $\{ \sld[j] \}_{j=1,...,n}$ satisfying $\conj \sld[j] \conj = \sld[j]$.
The remaining procedure of proof is the same to the proof of Thm.~\ref{thm:measurement}, except that all the statements are about the specific point $\param$. 
$\qed$\\

\subsection{Quasi-classical models without local antiunitary symmetries}\label{sec:matsumotosan}
We construct quasi-classical models without any LASs.
According to Thm.~\ref{thm:local_antiunitary}, these examples are inevitably mixed-state models.
We give two kinds of examples in Sec.~\ref{sec:miyazaki} and Sec.~\ref{sec:matsumoto}.
We also show that any quasi-classical qubit model has a LAS in Sec.~\ref{sec:qubit}.

The following lemma is used in exhibiting the non-existence of antiunitary symmetries for both of the examples.
\begin{lemma}\label{lem:diagonal_decomposition}
	Let $D$ be a real non-degenerate $d \times d$ diagonal matrix.
	If $U^\dagger D U$ remains real for a unitary $U$, then $U$ is decomposed to
	\begin{equation}
		\label{phase-orthogonal-decomposition}	U = \Phi R,
	\end{equation}
	where $R$ is an orthogonal matrix and
	\begin{equation}
		\Phi = {\rm diag}(e^{i\phi_1},...,e^{i \phi_d}),
	\end{equation}
	with some set of phases $\phi_j \in [0,2\pi]$.
\end{lemma}
\emph{proof})
Let us denote the real matrix $U^\dagger D U$ by $Q$.
Since $Q$ is diagonalized to $D = U Q U^\dagger$ by the inverse unitary $U^\dagger$, $U^\dagger$'s columns must be unit eigenvectors of $Q$.
Since the real Hermitian matrix $Q$ is symmetric, its eigenvectors $\vect[r]_j$ ($j=1,...,d$) can be taken real.
Therefore $U^\dagger$ is decomposed as
\begin{equation}
\label{Udecomposition}	U^\dagger = (e^{-i\phi_1} \vect[r]_1~e^{-i\phi_2} \vect[r]_2~...~e^{-i \phi_d} \vect[r]_d) = R^\dagger \Phi^\dagger,
\end{equation}
where $\phi_j$ are phases and $R^\dagger = (\vect[r]_1~...~\vect[r]_d)$ is an orthogonal matrix.
The lemma follows by taking the Hermitian adjoint on both sides of Eq.~\ref{Udecomposition}.
$\qed$\\
The action of diagonal phase matrix $\Phi$ as a similarity transformation is presented by
\begin{equation}
	\label{phase-action}    [ \Phi^\dagger A \Phi ]_{jk} = e^{i (\phi_k - \phi_j)} [A]_{jk},
\end{equation}
for arbitrary matrix $A$.

\subsubsection{Single-parameter model}\label{sec:miyazaki}
We consider a qutrit model whose state is
\begin{align}
\label{qutrit}	\rho_x = \left[
	\begin{array}{ccc}
		a& 0& 0 \\
		0& b& 0 \\
		0& 0& c \\ 
	\end{array} \right] + x \left[
	\begin{array}{ccc}
		0& e^{i \omega_{12}}& e^{i \omega_{13}} \\
        e^{-i \omega_{12}}& 0& e^{i \omega_{23}} \\
        e^{-i \omega_{13}}& e^{-i \omega_{23}}& 0\\
	\end{array} \right],
\end{align}
where $a,b,c$ are different positive real constants that sum to $1$, $\omega_{jk}$ are constant phases, and $x$ is the parameter.
Operator $\rho_x$ is a quantum state for sufficiently small $|x|$.
At $x=0$ we have
\begin{align}
	\rho_0 &= \left[
	\begin{array}{ccc}
		a& 0& 0 \\
		0& b& 0 \\
		0& 0& c \\ 
	\end{array} \right],\\
    \partial_x \rho_x |_{x=0} &= \left[
	\begin{array}{ccc}
		0& e^{i \omega_{12}}& e^{i \omega_{13}} \\
        e^{-i \omega_{12}}& 0& e^{i \omega_{23}} \\
        e^{-i \omega_{13}}& e^{-i \omega_{23}}& 0\\
	\end{array} \right].
\end{align}
We now prove that the model \eqref{qutrit} does not have any LAS at $x=0$ for a proper choice of $\omega_{jk}$.

From Thm.~\ref{thm:anti_equal_conj}, we can assume that the LAS is given by a conjugation.
Therefore, it is sufficient if we prove that $\rho_0$ and $\partial_x \rho_x |_{x=0}$ cannot be simultaneously decomplexified by a unitary transformation.

Let $U$ be any unitary that makes $U^\dagger \rho_0 U$ real.
Then $U$ is decomposed as Eq.~\eqref{phase-orthogonal-decomposition} into a phase matrix $\Phi = {\rm diag}(e^{i \phi_1},e^{i\phi_2},e^{i\phi_3})$ and an orthogonal matrix $R$ (Lem.~\ref{lem:diagonal_decomposition}).
For $j < k$ we have
\begin{equation}
    [ \Phi^\dagger \partial_x \rho_x |_{x=0} \Phi ]_{jk} = e^{i (\phi_k - \phi_j + \omega_{jk})},
\end{equation}
from Eq.~\eqref{phase-action}.
The components of $\Phi^\dagger \partial_x \rho_x |_{x=0} \Phi$ cannot be simultaneously real when
\begin{equation}
    \omega_{12} + \omega_{23} \neq \omega_{13},
\end{equation}
by any choice of the phases $(\phi_1,\phi_2,\phi_3)$.
Since $R$ is a real orthogonal matrix,
\begin{equation}
    R^\dagger \Phi^\dagger \partial_x \rho_x |_{x=0} \Phi R,
\end{equation}
cannot be real for any choice of the phases $(\phi_1,\phi_2,\phi_3)$ and $R$.
This completes the proof of non-existence of LAS at $x=0$.

\subsubsection{Multiparameter model}\label{sec:matsumoto}
Let $\hil$ be a $d$-dimensional space and $\{ \ket[f_j] | j=1,...,d \}$ be a basis.
Let $\rho$ be a full-rank density operator on $\hil$.
Define operators $L_j$ ($j=1,...,d$) by
\begin{equation}
	L_j := \ketbra[f_j] - \bra[f_j] \rho \ket[f_j] \id, ~(j=1,...,d)
\end{equation}
so that $\trace[\rho L_j] = 0 $ holds for any $j$, and $[L_j,L_k]=0$ holds for any pair.
The operators can be identified to SLDs of a model $\model$ with $d-1$-parameters $(x_1,...x_{d-1})$ such that
\begin{align}
	\rho_{(0,...,0)} &= \rho,\\
	 2 \partial_j \state |_{\mathbf{x}=(0,...,0)} &= L_j \rho + \rho L_j.\\
\nonumber	 &(j=1,...,d-1)
\end{align}
Because $\rho$ is full-rank, $L_j$ are the unique set of SLDs for model $\model$ at $\param = (0,...,0)$.

From Lem.~\ref{lem:LAS} and the uniqueness of SLDs, $\model$ has a LAS if and only if there is an antiunitary $\antiu$ such that
\begin{equation}
\label{LAS_1}	\antiu \rho \antiu^\dagger = \rho ,~\antiu L_j \antiu^\dagger = L_j,~(j=1,...,d-1)
\end{equation}
holds.
From Thm.~\eqref{thm:anti_equal_conj}, we can assume that $\antiu$ is a conjugation.
Since $\ketbra[f_j] = L_j + \bra[f_j] \rho \ket[f_j] \id$ and $\conj \id \conj = \id$ hold for any conjugation, Eq.~\eqref{LAS_1} holds for some conjugation if and only if $\rho$ and $\ketbra[f_j]$ ($j=1,...,d$) are simultaneously decomplexified in a basis of $\hil$.

Now, take $\{ \ket[f_j] | j=1,...,d \}$ as the reference-basis for matrix representations.
By applying Lem.~\ref{lem:diagonal_decomposition} to a non-degenerate real diagonal matrix $\sum_j r_j \ketbra[f_j]$ ($r_j \in \Real$, ~$r_j \neq r_k$ for $j \neq k$), any unitary matrix $U$ that keeps $\sum_j r_j \ketbra[f_j]$ real can be decomposed as $U = \Phi R$ to the diagonal phase unitary $\Phi = \mathrm{diag}(e^{i \phi_1},...,e^{i \phi_d})$ and an orthogonal matrix $R$.
The action of $\Phi$ on $\rho$ is given by Eq.~\eqref{phase-action}:
\begin{equation}
\label{matsumoto0}	[ \Phi^\dagger \rho \Phi ]_{jk} = e^{i (\phi_k - \phi_j)} [\rho]_{jk},
\end{equation}
for all $j$ and $k$.
The density matrix $\rho$ is decomplexified by $U$ if and only if $\Phi^\dagger \rho \Phi$ is real.

If $d \geq 3$, consider a density matrix such that
\begin{align}
\label{matsumoto1}	& \Imag [\rho_{12}] = - \Imag [\rho_{21}] \neq 0, \\
\label{matsumoto2}	& 0 \neq \rho_{jk} \in \Re, ~(\mathrm{otherwise})
\end{align}
hold.
Equations~\eqref{matsumoto2} and \eqref{matsumoto0} imply that the phase differences must make $e^{i (\phi_k - \phi_j)}$ real for any $j$ and $k$, if $\Phi^\dagger \rho \Phi$ is real.
For such a choice of phase differences, however, $[ \Phi^\dagger \rho \Phi ]_{12} = e^{i (\phi_2 - \phi_1)} [\rho]_{12}$ cannot be real from Eq.~\eqref{matsumoto1}.
Therefore, the density matrix $\rho$ satisfying Eqs.~\eqref{matsumoto1} and \eqref{matsumoto2} cannot be decomplexified by $U$.
The model $\model$ does not have any LAS at $\param = 0$ if we choose $\rho$ satisfying Eqs.~\eqref{matsumoto1} and \eqref{matsumoto2}.

\subsubsection{Qubit model}\label{sec:qubit}
If $d =2$, any quasi-classical model have a LAS.
For any qubit model, the maximum size of the set of mutually independent and commuting SLDs is $1$ \cite{Suzuki2019}.
Let $\sld[1] = r_1 \ketbra[1] + r_2 \ketbra[2]$ ($r_1 \neq r_2$) be the spectrum decomposition of the unique SLD $\sld[1]$.
Again we take the basis $\{ \ket[f_1],~ \ket[f_2] \}$ as the reference for matrix representation.
Choose two phases $\phi_1$ and $\phi_2$ so that
\begin{equation}
	e^{i (\phi_2 - \phi_1)} [\rho]_{12} = [ \Phi^\dagger \rho \Phi ]_{12} \in \Re,
\end{equation}
holds.
Then the matrices $\state$ and $\sld[1]$ are simultaneously decomplexified by the phase matrix $\Phi = \mathrm{diag}(e^{i \phi_1}, e^{i \phi_2})$.

\section{Set imaginarity, set antiunitary asymmetry, and asymmetry measures for quantum statistical models}\label{sec:imaginarity}
In this section, we consider measures of imaginarity and antiunitary asymmetry for a set of quantum states.
The purpose is to later introduce corresponding measures for quantum statistical models.
We employ measures of imaginarity from a resource theoretic framework reviewed below.
Then the measures are extended to fit to our purpose. 

\subsection{Resource theory of imaginarity}
Resource theories of imaginarity \cite{HickeyGour2018imaginarity,Wu2021imaginarityPRL,Wu2021imaginarityPRA} start from specifying a reference basis $\{ \ket[k] \}_{k=1,...,d}$.
Then imaginarity-free states are determined as real density matrices in the basis.
To fit to our discussion, we represent the set of imaginarity-free states as
\begin{equation}
\label{imaginarity_free}	F_\conj = \left\{ \rho ~ \mid ~ \conj \rho \conj = \rho \right\}.
\end{equation}
where $\conj$ is the complex conjugation in the reference basis $\{ \ket[k] \}_{k=1,...,d}$.
Measures of imaginarity quantify how much a given state differs from imaginarity-free states.
The measures must satisfy monotonicity under imaginarity-free operations (operations that does not affect imaginarity-free states \cite{HickeyGour2018imaginarity}), and return zero for imaginarity-free states.
Hickey and Gour proposed the distance measures
\begin{equation}
\label{distance_measures}	M^C_\conj (\rho) := \min_{\tau \in F_\conj} C(\rho,\tau), ~ (C:{\rm contractive~ metric})
\end{equation}
and the robustness of imaginarity \cite{HickeyGour2018imaginarity}.
If we choose the trace distance, the distance measure is expressed in a closed-form
\begin{equation}
\label{trace_distance}	M^1_\conj (\rho) := \min_{\tau \in F_\conj} || \rho - \tau ||_1 = \frac{|| \rho - \conj \rho \conj ||_1}{2},
\end{equation}
and is faithful, meaning that they are strictly positive outside the imaginarity-free states.
The robustness later turned out to be equal to $M^1_\conj$ \cite{Wu2021imaginarityPRL,Wu2021imaginarityPRA}.

\subsection{Set imaginarity}
The imaginarity thus defined is measured in terms of a fixed basis $\{ \ket[k] \}_{k=1,...,d}$, or equivalently a fixed conjugation $\conj$.
In contrast, the weak commutativity is implied by any conjugation.
In our sense, a model $\model$ should be called ``imaginarity-free'' if the states are simultaneously real for some reference basis which is not fixed in advance.
We are looking at a basis-independent definition of imaginarity.

Such a definition of imaginarity has not been studied yet.
However, there is a basis-independent definition of coherence \cite{BaumgratzPlenio2014coherence,ChitambarGour2016coherence,WinterDong2016coherence} termed ``set coherence'' \cite{DesignolleBrunner2021setcoherence}.
We follow the procedure to define ``set imaginarity,'' and ``set antiunitary asymmetry.''

A set of $n$ quantum states is presented by $\rho^{(n)} = \{ \rho_1,...,\rho_n \}$.
The free set is given by
\begin{equation}
F^{(n)} = \left\{ \rho^{(n)} \middle| \exists \conj, \rho^{(n)} \subset F_\conj \right\},
\end{equation}
for each $n$.
It is a union $F^{(n)} = \cup_{\conj} F^{(n)}_\conj$  of free sets for fixed conjugations defined by
\begin{equation}
F^{(n)}_\conj =  \left\{ \rho^{(n)} \middle|  \rho^{(n)} \subset F_\conj \right\}.
\end{equation}
Completely analogously to set coherence \cite{DesignolleBrunner2021setcoherence}, we may define max-distance
\begin{equation}
\label{max-distance}	M^C_{\max}(\rho^{(n)}) := \min_{\conj} \max_j M^C_\conj (\rho_j),
\end{equation}
and mean-distance
\begin{equation}
\label{mean-distance}	M^C_{\rm mean} (\rho^{(n)}) = \min_{\conj} \frac{1}{n} \sum_{j=1}^n M^C_\conj (\rho_j),
\end{equation}
as measures of set imaginarity.
These measures are faithful when we choose the trace-distance because $M^1_\conj$ is faithful for the basis-dependent imaginarity.
We do not touch operational meanings of the measures thus defined.
A measure of set coherence inherits an operational meaning from basis-dependent coherence \cite{DesignolleBrunner2021setcoherence}.

\subsection{Antiunitary asymmetry}
The procedure to define (basis-dependent) imaginarity generalizes straightforwardly to antiunitary asymmetry.
The representation \eqref{imaginarity_free} of imaginarity-free states extends the states with general antiunitary symmetry
\begin{equation}
F_\antiu = \left\{ \rho ~ \middle| ~ \antiu \rho \antiu^\dagger = \rho \right\},
\end{equation}
by replacing the conjugation $\conj$ to an antiunitary operator $\antiu$.
The distant measures are generalized accordingly.

The set imaginarity and ``set antiunitary asymmetry'' are equivalent when defined by distance measures.
Theorem \ref{thm:anti_equal_conj} implies that for any antiunitary operator $\antiu$, there is a conjugation $\conj$ such that
\begin{equation}
\label{conj_includes_anti1}	F_\antiu \subset F_\conj,
\end{equation}
and hence
\begin{equation}
\label{conj_includes_anti2}	M^C_\conj (\rho) \leq M^C_\antiu (\rho),~(\forall \rho)
\end{equation}
hold for any distance measure.
The set of free states for the set antiunitary asymmetry
\begin{equation}
\left\{ \rho^{(n)} \middle| \exists \antiu, \rho^{(n)} \subset F_\antiu \right\},
\end{equation}
is equal to the free set for imaginarity $F^{(n)}$ by Eq.~\eqref{conj_includes_anti1}.
The max- and mean-distances
\begin{equation}
\min_{\antiu} \max_j M^C_\antiu (\rho_j), ~\min_{\antiu} \frac{1}{n} \sum_{j=1}^n M^C_\antiu (\rho_j),
\end{equation}
are equals respectively to $M^C_{\max}(\rho^{(n)})$ and $M^C_{\rm mean}(\rho^{(n)})$ by Eq.~\eqref{conj_includes_anti2}.
The difference between imaginarity and antiunitary asymmetry only appears at the basis-dependent definitions for distance measures.

\subsection{Antiunitary asymmetry of quantum statistical models}
There are many ways to define antiunitary asymmetry of a quantum statistical model $\state$ at each point $\pspace$.
We require the measure to be faithful: the measure must be zero only if there is a LAS at the point.
The max-distance \eqref{max-distance} inspires us to define
\begin{equation}
\label{max_imaginarity}	M^1_{\max} (\param) := \min_\conj \lim_{ \Delta x \rightarrow 0} \frac{\max_i M^1_\conj (\state + \Delta x \partial_i \state)}{\Delta x},
\end{equation}
and call it max-imaginarity.
Since the limit diverges unless $\conj \state \conj = \state$, we have
\begin{equation}
M^1_{\max} (\param) = \min_{\conj, \conj \state \conj = \state} \max_i \frac{|| \partial_i \state - \conj \partial_i \state \conj ||_1}{2}.
\end{equation}
In a similar way we define mean-imaginarity as
\begin{align}
	\label{mean_imaginarity}	M^1_{\rm mean} (\param) &:= \min_\conj \lim_{ \Delta x \rightarrow 0} \frac{1}{n} \sum_{i=1}^n \frac{ M^1_\conj (\state + \Delta x \partial_i \state)}{\Delta x} \\
\nonumber	&= \min_{\conj, \conj \state \conj = \state} \frac{1}{n} \sum_{i=1}^n \frac{|| \partial_i \state - \conj \partial_i \state \conj ||_1}{2},
\end{align}
inspired by the mean-distance \eqref{mean-distance}.
The imaginarity and the antiunitary asymmetry remains equivalent for these definitions.

Our asymmetry measure \eqref{asymmetry_measure} is inspired by the mean-imaginarity \eqref{mean_imaginarity}.
We are not sure if the measure is equivalent to
\begin{align}
\label{imaginarity_measure}	\min_{\conj,~ \conj \state \conj = \state} \frac{1}{n} \trace \left[ \sum_{i=1}^n  ( \partial_i \state - \conj \partial_i \state \conj )^2 \right],
\end{align}
since we cease to use the trace distance.

Our asymmetry measure\eqref{asymmetry_measure} has an advantage of being invariant under isometric coordinate transformations.
Since the Jacobian of an isometric coordinate transformation is an orthogonal matrix $R$, we have
\begin{align}
\nonumber	&\sum_{i=1}^n ( \partial'_i \state - \antiu \partial'_i \state \antiu^\dagger )^2 \\
\nonumber	=& \sum_{i,j,k=1}^n R_{ij} (\partial_j \state - \antiu \partial_j \state \antiu^\dagger) \\
\nonumber	& \hspace{2.5cm} \times R_{ik} (\partial_k \state - \antiu \partial_k \state \antiu^\dagger)\\
\nonumber	=& \sum_{j,k=1}^n \delta_{jk} (\partial_j \state - \antiu \partial_j \state \antiu^\dagger) (\partial_k \state - \antiu \partial_k \state \antiu^\dagger)\\
	=&  \sum_{j=1}^n  (\partial_j \state - \antiu \partial_j \state \antiu^\dagger)^2,
\end{align}
where $\partial'$ are the derivative with respected to the transformed coordinates.

I am not sure if there is a faithful asymmetry measure that is invariant under general normal coordinate transformations.
The LASs, if they exist, are preserved by these transformations.
Namely, a model $\model$ has a LAS at $\pspace$ if and only if
\begin{equation}
	\left\{ \rho_{f^{-1}({\bm y})} \mid {\bm y} \in f(X) \right\},
\end{equation}
has a LAS at ${\bm y} = f(\param)$, where $f$ is any normal coordinate transformation.
Methods to extend this invariance of antiunitary symmetry to asymmetry measures remain open here.


\begin{thebibliography}{99}%
	\makeatletter
	\providecommand \@ifxundefined [1]{%
	 \@ifx{#1\undefined}
	}%
	\providecommand \@ifnum [1]{%
	 \ifnum #1\expandafter \@firstoftwo
	 \else \expandafter \@secondoftwo
	 \fi
	}%
	\providecommand \@ifx [1]{%
	 \ifx #1\expandafter \@firstoftwo
	 \else \expandafter \@secondoftwo
	 \fi
	}%
	\providecommand \natexlab [1]{#1}%
	\providecommand \enquote  [1]{``#1''}%
	\providecommand \bibnamefont  [1]{#1}%
	\providecommand \bibfnamefont [1]{#1}%
	\providecommand \citenamefont [1]{#1}%
	\providecommand \href@noop [0]{\@secondoftwo}%
	\providecommand \href [0]{\begingroup \@sanitize@url \@href}%
	\providecommand \@href[1]{\@@startlink{#1}\@@href}%
	\providecommand \@@href[1]{\endgroup#1\@@endlink}%
	\providecommand \@sanitize@url [0]{\catcode `\\12\catcode `\$12\catcode
	  `\&12\catcode `\#12\catcode `\^12\catcode `\_12\catcode `\%12\relax}%
	\providecommand \@@startlink[1]{}%
	\providecommand \@@endlink[0]{}%
	\providecommand \url  [0]{\begingroup\@sanitize@url \@url }%
	\providecommand \@url [1]{\endgroup\@href {#1}{\urlprefix }}%
	\providecommand \urlprefix  [0]{URL }%
	\providecommand \Eprint [0]{\href }%
	\providecommand \doibase [0]{http://dx.doi.org/}%
	\providecommand \selectlanguage [0]{\@gobble}%
	\providecommand \bibinfo  [0]{\@secondoftwo}%
	\providecommand \bibfield  [0]{\@secondoftwo}%
	\providecommand \translation [1]{[#1]}%
	\providecommand \BibitemOpen [0]{}%
	\providecommand \bibitemStop [0]{}%
	\providecommand \bibitemNoStop [0]{.\EOS\space}%
	\providecommand \EOS [0]{\spacefactor3000\relax}%
	\providecommand \BibitemShut  [1]{\csname bibitem#1\endcsname}%
	\let\auto@bib@innerbib\@empty
	\bibitem [{\citenamefont {Giovannetti}\ \emph {et~al.}(2006)\citenamefont
	  {Giovannetti}, \citenamefont {Lloyd},\ and\ \citenamefont
	  {Maccone}}]{Giovanettietal2006}%
	  \BibitemOpen
	  \bibfield  {author} {\bibinfo {author} {\bibfnamefont {V.}~\bibnamefont
	  {Giovannetti}}, \bibinfo {author} {\bibfnamefont {S.}~\bibnamefont {Lloyd}},
	  \ and\ \bibinfo {author} {\bibfnamefont {L.}~\bibnamefont {Maccone}},\ }\href
	  {\doibase 10.1103/PhysRevLett.96.010401} {\bibfield  {journal} {\bibinfo
	  {journal} {Phys. Rev. Lett.}\ }\textbf {\bibinfo {volume} {96}},\ \bibinfo
	  {pages} {010401} (\bibinfo {year} {2006})}\BibitemShut {NoStop}%
	\bibitem [{\citenamefont {Giovannetti}\ \emph {et~al.}(2011)\citenamefont
	  {Giovannetti}, \citenamefont {Lloyd},\ and\ \citenamefont
	  {Maccone}}]{Giovanettietal2011}%
	  \BibitemOpen
	  \bibfield  {author} {\bibinfo {author} {\bibfnamefont {V.}~\bibnamefont
	  {Giovannetti}}, \bibinfo {author} {\bibfnamefont {S.}~\bibnamefont {Lloyd}},
	  \ and\ \bibinfo {author} {\bibfnamefont {L.}~\bibnamefont {Maccone}},\ }\href
	  {\doibase 10.1038/nphoton.2011.35} {\bibfield  {journal} {\bibinfo  {journal}
	  {Nat. Photon.}\ }\textbf {\bibinfo {volume} {5}},\ \bibinfo {pages} {222}
	  (\bibinfo {year} {2011})}\BibitemShut {NoStop}%
	\bibitem [{\citenamefont {Demkowicz-Dobrza\'{n}ski}\ \emph
	  {et~al.}(2015)\citenamefont {Demkowicz-Dobrza\'{n}ski}, \citenamefont
	  {Jarzyna},\ and\ \citenamefont {Kołodyński}}]{Demkowiczetal2015}%
	  \BibitemOpen
	  \bibfield  {author} {\bibinfo {author} {\bibfnamefont {R.}~\bibnamefont
	  {Demkowicz-Dobrza\'{n}ski}}, \bibinfo {author} {\bibfnamefont
	  {M.}~\bibnamefont {Jarzyna}}, \ and\ \bibinfo {author} {\bibfnamefont
	  {J.}~\bibnamefont {Kołodyński}},\ }\href {\doibase 10.1016/bs.po.2015.02.003} {\bibfield  {journal} {\bibinfo  {journal} {Prog.
	  Opt.}\ }\textbf {\bibinfo {volume} {60}},\ \bibinfo {pages} {345 } (\bibinfo
	  {year} {2015})}\BibitemShut {NoStop}%
	\bibitem [{\citenamefont {Degen}\ \emph {et~al.}(2017)\citenamefont {Degen},
	  \citenamefont {Reinhard},\ and\ \citenamefont {Cappellaro}}]{Degenetal2017}%
	  \BibitemOpen
	  \bibfield  {author} {\bibinfo {author} {\bibfnamefont {C.~L.}\ \bibnamefont
	  {Degen}}, \bibinfo {author} {\bibfnamefont {F.}~\bibnamefont {Reinhard}}, \
	  and\ \bibinfo {author} {\bibfnamefont {P.}~\bibnamefont {Cappellaro}},\
	  }\href {\doibase 10.1103/RevModPhys.89.035002} {\bibfield  {journal}
	  {\bibinfo  {journal} {Rev. Mod. Phys.}\ }\textbf {\bibinfo {volume} {89}},\
	  \bibinfo {pages} {035002} (\bibinfo {year} {2017})}\BibitemShut {NoStop}%
	\bibitem [{\citenamefont {Pezz\`e}\ \emph {et~al.}(2018)\citenamefont
	  {Pezz\`e}, \citenamefont {Smerzi}, \citenamefont {Oberthaler}, \citenamefont
	  {Schmied},\ and\ \citenamefont {Treutlein}}]{Pezzeetal2018}%
	  \BibitemOpen
	  \bibfield  {author} {\bibinfo {author} {\bibfnamefont {L.}~\bibnamefont
	  {Pezz\`e}}, \bibinfo {author} {\bibfnamefont {A.}~\bibnamefont {Smerzi}},
	  \bibinfo {author} {\bibfnamefont {M.~K.}\ \bibnamefont {Oberthaler}},
	  \bibinfo {author} {\bibfnamefont {R.}~\bibnamefont {Schmied}}, \ and\
	  \bibinfo {author} {\bibfnamefont {P.}~\bibnamefont {Treutlein}},\ }\href
	  {\doibase 10.1103/RevModPhys.90.035005} {\bibfield  {journal} {\bibinfo
	  {journal} {Rev. Mod. Phys.}\ }\textbf {\bibinfo {volume} {90}},\ \bibinfo
	  {pages} {035005} (\bibinfo {year} {2018})}\BibitemShut {NoStop}%
	\bibitem [{\citenamefont {Braun}\ \emph {et~al.}(2018)\citenamefont {Braun},
	  \citenamefont {Adesso}, \citenamefont {Benatti}, \citenamefont {Floreanini},
	  \citenamefont {Marzolino}, \citenamefont {Mitchell},\ and\ \citenamefont
	  {Pirandola}}]{Braunetal2018}%
	  \BibitemOpen
	  \bibfield  {author} {\bibinfo {author} {\bibfnamefont {D.}~\bibnamefont
	  {Braun}}, \bibinfo {author} {\bibfnamefont {G.}~\bibnamefont {Adesso}},
	  \bibinfo {author} {\bibfnamefont {F.}~\bibnamefont {Benatti}}, \bibinfo
	  {author} {\bibfnamefont {R.}~\bibnamefont {Floreanini}}, \bibinfo {author}
	  {\bibfnamefont {U.}~\bibnamefont {Marzolino}}, \bibinfo {author}
	  {\bibfnamefont {M.~W.}\ \bibnamefont {Mitchell}}, \ and\ \bibinfo {author}
	  {\bibfnamefont {S.}~\bibnamefont {Pirandola}},\ }\href {\doibase 10.1103/RevModPhys.90.035006} {\bibfield  {journal} {\bibinfo  {journal}
	  {Rev. Mod. Phys.}\ }\textbf {\bibinfo {volume} {90}},\ \bibinfo {pages}
	  {035006} (\bibinfo {year} {2018})}\BibitemShut {NoStop}%
	\bibitem [{\citenamefont {Pirandola}\ \emph {et~al.}(2018)\citenamefont
	  {Pirandola}, \citenamefont {Bardhan}, \citenamefont {Gehring}, \citenamefont
	  {Weedbrook},\ and\ \citenamefont {Lloyd}}]{Pirandolaetal2018}%
	  \BibitemOpen
	  \bibfield  {author} {\bibinfo {author} {\bibfnamefont {S.}~\bibnamefont
	  {Pirandola}}, \bibinfo {author} {\bibfnamefont {B.~R.}\ \bibnamefont
	  {Bardhan}}, \bibinfo {author} {\bibfnamefont {T.}~\bibnamefont {Gehring}},
	  \bibinfo {author} {\bibfnamefont {C.}~\bibnamefont {Weedbrook}}, \ and\
	  \bibinfo {author} {\bibfnamefont {S.}~\bibnamefont {Lloyd}},\ }\href
	  {\doibase 10.1038/s41566-018-0301-6} {\bibfield  {journal} {\bibinfo
	  {journal} {Nat. Photonics}\ }\textbf {\bibinfo {volume} {12}},\ \bibinfo
	  {pages} {724} (\bibinfo {year} {2018})}\BibitemShut {NoStop}%
	\bibitem [{\citenamefont {Berchera}\ and\ \citenamefont
	  {Degiovanni}(2019)}]{BercheraDegiovanni2019}%
	  \BibitemOpen
	  \bibfield  {author} {\bibinfo {author} {\bibfnamefont {I.~R.}\ \bibnamefont
	  {Berchera}}\ and\ \bibinfo {author} {\bibfnamefont {I.~P.}\ \bibnamefont
	  {Degiovanni}},\ }\href {\doibase 10.1088/1681-7575/aaf7b2} {\bibfield
	  {journal} {\bibinfo  {journal} {Metrologia}\ }\textbf {\bibinfo {volume}
	  {56}},\ \bibinfo {pages} {024001} (\bibinfo {year} {2019})}\BibitemShut
	  {NoStop}%
	\bibitem [{\citenamefont {Helstrom}(1967)}]{Helstrom1967}%
	  \BibitemOpen
	  \bibfield  {author} {\bibinfo {author} {\bibfnamefont {C.}~\bibnamefont
	  {Helstrom}},\ }\href {\doibase 10.1016/0375-9601(67)90366-0} {\bibfield
	  {journal} {\bibinfo  {journal} {Phys. Lett. A}\ }\textbf {\bibinfo {volume}
	  {25}},\ \bibinfo {pages} {101 } (\bibinfo {year} {1967})}\BibitemShut
	  {NoStop}%
	\bibitem [{\citenamefont {{Helstrom}}(1968)}]{Helstrom1968}%
	  \BibitemOpen
	  \bibfield  {author} {\bibinfo {author} {\bibfnamefont {C.}~\bibnamefont
	  {{Helstrom}}},\ }\href {\doibase 10.1109/TIT.1968.1054108} {\bibfield
	  {journal} {\bibinfo  {journal} {IEEE Trans. Inf. Theory}\ }\textbf {\bibinfo
	  {volume} {14}},\ \bibinfo {pages} {234} (\bibinfo {year} {1968})}\BibitemShut
	  {NoStop}%
	\bibitem [{\citenamefont {Holevo}(2011)}]{Holevo2011}%
	  \BibitemOpen
	  \bibfield  {author} {\bibinfo {author} {\bibfnamefont {A.~S.}\ \bibnamefont
	  {Holevo}},\ }\href {\doibase 10.1007/978-88-7642-378-9} {\emph {\bibinfo
	  {title} {Probabilistic and statistical aspects of quantum theory}}},\
	  Vol.~\bibinfo {volume} {1}\ (\bibinfo  {publisher} {Springer Science \&
	  Business Media},\ \bibinfo {year} {2011})\BibitemShut {NoStop}%
	\bibitem [{\citenamefont {Hayashi}(2017)}]{Hayashi2017quantum}%
	  \BibitemOpen
	  \bibfield  {author} {\bibinfo {author} {\bibfnamefont {M.}~\bibnamefont
	  {Hayashi}},\ }\href {\doibase 10.1007/978-3-662-49725-8} {\emph {\bibinfo
	  {title} {Quantum Information Theory}}},\ Graduate Texts in Physics, Springer\
	  (\bibinfo  {publisher} {Springer},\ \bibinfo {year} {2017})\BibitemShut
	  {NoStop}%
	\bibitem [{\citenamefont {Szczykulska}\ \emph {et~al.}(2016)\citenamefont
	  {Szczykulska}, \citenamefont {Baumgratz},\ and\ \citenamefont
	  {Datta}}]{Szczykulskaetal2016}%
	  \BibitemOpen
	  \bibfield  {author} {\bibinfo {author} {\bibfnamefont {M.}~\bibnamefont
	  {Szczykulska}}, \bibinfo {author} {\bibfnamefont {T.}~\bibnamefont
	  {Baumgratz}}, \ and\ \bibinfo {author} {\bibfnamefont {A.}~\bibnamefont
	  {Datta}},\ }\href {\doibase 10.1080/23746149.2016.1230476} {\bibfield
	  {journal} {\bibinfo  {journal} {Adv. Phys. X}\ }\textbf {\bibinfo {volume}
	  {1}},\ \bibinfo {pages} {621} (\bibinfo {year} {2016})}\BibitemShut {NoStop}%
	\bibitem [{\citenamefont {Albarelli}\ \emph {et~al.}(2019)\citenamefont
	  {Albarelli}, \citenamefont {Friel},\ and\ \citenamefont
	  {Datta}}]{Albarellietal2019}%
	  \BibitemOpen
	  \bibfield  {author} {\bibinfo {author} {\bibfnamefont {F.}~\bibnamefont
	  {Albarelli}}, \bibinfo {author} {\bibfnamefont {J.~F.}\ \bibnamefont
	  {Friel}}, \ and\ \bibinfo {author} {\bibfnamefont {A.}~\bibnamefont
	  {Datta}},\ }\href {\doibase 10.1103/PhysRevLett.123.200503} {\bibfield
	  {journal} {\bibinfo  {journal} {Phys. Rev. Lett.}\ }\textbf {\bibinfo
	  {volume} {123}},\ \bibinfo {pages} {200503} (\bibinfo {year}
	  {2019})}\BibitemShut {NoStop}%
	\bibitem [{\citenamefont {Baumgratz}\ and\ \citenamefont
	  {Datta}(2016)}]{BaumgratzDatta2016}%
	  \BibitemOpen
	  \bibfield  {author} {\bibinfo {author} {\bibfnamefont {T.}~\bibnamefont
	  {Baumgratz}}\ and\ \bibinfo {author} {\bibfnamefont {A.}~\bibnamefont
	  {Datta}},\ }\href {\doibase 10.1103/PhysRevLett.116.030801} {\bibfield
	  {journal} {\bibinfo  {journal} {Phys. Rev. Lett.}\ }\textbf {\bibinfo
	  {volume} {116}},\ \bibinfo {pages} {030801} (\bibinfo {year}
	  {2016})}\BibitemShut {NoStop}%
	\bibitem [{\citenamefont {Humphreys}\ \emph {et~al.}(2013)\citenamefont
	  {Humphreys}, \citenamefont {Barbieri}, \citenamefont {Datta},\ and\
	  \citenamefont {Walmsley}}]{Humphreys2013}%
	  \BibitemOpen
	  \bibfield  {author} {\bibinfo {author} {\bibfnamefont {P.~C.}\ \bibnamefont
	  {Humphreys}}, \bibinfo {author} {\bibfnamefont {M.}~\bibnamefont {Barbieri}},
	  \bibinfo {author} {\bibfnamefont {A.}~\bibnamefont {Datta}}, \ and\ \bibinfo
	  {author} {\bibfnamefont {I.~A.}\ \bibnamefont {Walmsley}},\ }\href {\doibase 10.1103/PhysRevLett.111.070403} {\bibfield  {journal} {\bibinfo  {journal}
	  {Phys. Rev. Lett.}\ }\textbf {\bibinfo {volume} {111}},\ \bibinfo {pages}
	  {070403} (\bibinfo {year} {2013})}\BibitemShut {NoStop}%
	\bibitem [{\citenamefont {Gagatsos}\ \emph {et~al.}(2016)\citenamefont
	  {Gagatsos}, \citenamefont {Branford},\ and\ \citenamefont
	  {Datta}}]{Gagatsosetal2016}%
	  \BibitemOpen
	  \bibfield  {author} {\bibinfo {author} {\bibfnamefont {C.~N.}\ \bibnamefont
	  {Gagatsos}}, \bibinfo {author} {\bibfnamefont {D.}~\bibnamefont {Branford}},
	  \ and\ \bibinfo {author} {\bibfnamefont {A.}~\bibnamefont {Datta}},\ }\href
	  {\doibase 10.1103/PhysRevA.94.042342} {\bibfield  {journal} {\bibinfo
	  {journal} {Phys. Rev. A}\ }\textbf {\bibinfo {volume} {94}},\ \bibinfo
	  {pages} {042342} (\bibinfo {year} {2016})}\BibitemShut {NoStop}%
	\bibitem [{\citenamefont {Vidrighin}\ \emph {et~al.}(2014)\citenamefont
	  {Vidrighin}, \citenamefont {Donati}, \citenamefont {Genoni}, \citenamefont
	  {Jin}, \citenamefont {Kolthammer}, \citenamefont {Kim}, \citenamefont
	  {Datta}, \citenamefont {Barbieri},\ and\ \citenamefont
	  {Walmsley}}]{Vidrighin2014joint}%
	  \BibitemOpen
	  \bibfield  {author} {\bibinfo {author} {\bibfnamefont {M.~D.}\ \bibnamefont
	  {Vidrighin}}, \bibinfo {author} {\bibfnamefont {G.}~\bibnamefont {Donati}},
	  \bibinfo {author} {\bibfnamefont {M.~G.}\ \bibnamefont {Genoni}}, \bibinfo
	  {author} {\bibfnamefont {X.-M.}\ \bibnamefont {Jin}}, \bibinfo {author}
	  {\bibfnamefont {W.~S.}\ \bibnamefont {Kolthammer}}, \bibinfo {author}
	  {\bibfnamefont {M.}~\bibnamefont {Kim}}, \bibinfo {author} {\bibfnamefont
	  {A.}~\bibnamefont {Datta}}, \bibinfo {author} {\bibfnamefont
	  {M.}~\bibnamefont {Barbieri}}, \ and\ \bibinfo {author} {\bibfnamefont
	  {I.~A.}\ \bibnamefont {Walmsley}},\ }\href {\doibase 10.1038/ncomms4532}
	  {\bibfield  {journal} {\bibinfo  {journal} {Nat. Commun.}\ }\textbf {\bibinfo
	  {volume} {5}},\ \bibinfo {pages} {1} (\bibinfo {year} {2014})}\BibitemShut
	  {NoStop}%
	\bibitem [{\citenamefont {Roccia}\ \emph {et~al.}(2018)\citenamefont {Roccia},
	  \citenamefont {Cimini}, \citenamefont {Sbroscia}, \citenamefont {Gianani},
	  \citenamefont {Ruggiero}, \citenamefont {Mancino}, \citenamefont {Genoni},
	  \citenamefont {Ricci},\ and\ \citenamefont {Barbieri}}]{Rocciaetal2018}%
	  \BibitemOpen
	  \bibfield  {author} {\bibinfo {author} {\bibfnamefont {E.}~\bibnamefont
	  {Roccia}}, \bibinfo {author} {\bibfnamefont {V.}~\bibnamefont {Cimini}},
	  \bibinfo {author} {\bibfnamefont {M.}~\bibnamefont {Sbroscia}}, \bibinfo
	  {author} {\bibfnamefont {I.}~\bibnamefont {Gianani}}, \bibinfo {author}
	  {\bibfnamefont {L.}~\bibnamefont {Ruggiero}}, \bibinfo {author}
	  {\bibfnamefont {L.}~\bibnamefont {Mancino}}, \bibinfo {author} {\bibfnamefont
	  {M.~G.}\ \bibnamefont {Genoni}}, \bibinfo {author} {\bibfnamefont {M.~A.}\
	  \bibnamefont {Ricci}}, \ and\ \bibinfo {author} {\bibfnamefont
	  {M.}~\bibnamefont {Barbieri}},\ }\href {\doibase 10.1364/OPTICA.5.001171}
	  {\bibfield  {journal} {\bibinfo  {journal} {Optica}\ }\textbf {\bibinfo
	  {volume} {5}},\ \bibinfo {pages} {1171} (\bibinfo {year} {2018})}\BibitemShut
	  {NoStop}%
	\bibitem [{\citenamefont {Parniak}\ \emph {et~al.}(2018)\citenamefont
	  {Parniak}, \citenamefont {Bor\'owka}, \citenamefont {Boroszko}, \citenamefont
	  {Wasilewski}, \citenamefont {Banaszek},\ and\ \citenamefont
	  {Demkowicz-Dobrza\'{n}ski}}]{Parniaketal2018}%
	  \BibitemOpen
	  \bibfield  {author} {\bibinfo {author} {\bibfnamefont {M.}~\bibnamefont
	  {Parniak}}, \bibinfo {author} {\bibfnamefont {S.}~\bibnamefont {Bor\'owka}},
	  \bibinfo {author} {\bibfnamefont {K.}~\bibnamefont {Boroszko}}, \bibinfo
	  {author} {\bibfnamefont {W.}~\bibnamefont {Wasilewski}}, \bibinfo {author}
	  {\bibfnamefont {K.}~\bibnamefont {Banaszek}}, \ and\ \bibinfo {author}
	  {\bibfnamefont {R.}~\bibnamefont {Demkowicz-Dobrza\'{n}ski}},\ }\href
	  {\doibase 10.1103/PhysRevLett.121.250503} {\bibfield  {journal} {\bibinfo
	  {journal} {Phys. Rev. Lett.}\ }\textbf {\bibinfo {volume} {121}},\ \bibinfo
	  {pages} {250503} (\bibinfo {year} {2018})}\BibitemShut {NoStop}%
	\bibitem [{\citenamefont {Polino}\ \emph {et~al.}(2019)\citenamefont {Polino},
	  \citenamefont {Riva}, \citenamefont {Valeri}, \citenamefont {Silvestri},
	  \citenamefont {Corrielli}, \citenamefont {Crespi}, \citenamefont {Spagnolo},
	  \citenamefont {Osellame},\ and\ \citenamefont {Sciarrino}}]{Polinoetal2019}%
	  \BibitemOpen
	  \bibfield  {author} {\bibinfo {author} {\bibfnamefont {E.}~\bibnamefont
	  {Polino}}, \bibinfo {author} {\bibfnamefont {M.}~\bibnamefont {Riva}},
	  \bibinfo {author} {\bibfnamefont {M.}~\bibnamefont {Valeri}}, \bibinfo
	  {author} {\bibfnamefont {R.}~\bibnamefont {Silvestri}}, \bibinfo {author}
	  {\bibfnamefont {G.}~\bibnamefont {Corrielli}}, \bibinfo {author}
	  {\bibfnamefont {A.}~\bibnamefont {Crespi}}, \bibinfo {author} {\bibfnamefont
	  {N.}~\bibnamefont {Spagnolo}}, \bibinfo {author} {\bibfnamefont
	  {R.}~\bibnamefont {Osellame}}, \ and\ \bibinfo {author} {\bibfnamefont
	  {F.}~\bibnamefont {Sciarrino}},\ }\href {\doibase 10.1364/OPTICA.6.000288}
	  {\bibfield  {journal} {\bibinfo  {journal} {Optica}\ }\textbf {\bibinfo
	  {volume} {6}},\ \bibinfo {pages} {288} (\bibinfo {year} {2019})}\BibitemShut
	  {NoStop}%
	\bibitem [{\citenamefont {Helstrom}(1976)}]{Helstrom1976}%
	  \BibitemOpen
	  \bibfield  {author} {\bibinfo {author} {\bibfnamefont {C.~W.}\ \bibnamefont
	  {Helstrom}},\ }\href@noop {} {\emph {\bibinfo {title} {Quantum Detection and
	  Estimation Theory}}}\ (\bibinfo  {publisher} {Academic Press},\ \bibinfo
	  {year} {1976})\BibitemShut {NoStop}%
	\bibitem [{\citenamefont {Holevo}(1982)}]{Holevo1982}%
	  \BibitemOpen
	  \bibfield  {author} {\bibinfo {author} {\bibfnamefont {A.~S.}\ \bibnamefont
	  {Holevo}},\ }\href {\doibase 10.1007/978-88-7642-378-9} {\emph {\bibinfo
	  {title} {Probabilistic and Statistical Aspects of Quantum Theory}}}\
	  (\bibinfo  {publisher} {North Holland},\ \bibinfo {year} {1982})\BibitemShut
	  {NoStop}%
	\bibitem [{\citenamefont {Crowley}\ \emph {et~al.}(2014)\citenamefont
	  {Crowley}, \citenamefont {Datta}, \citenamefont {Barbieri},\ and\
	  \citenamefont {Walmsley}}]{Crowleyetal2014}%
	  \BibitemOpen
	  \bibfield  {author} {\bibinfo {author} {\bibfnamefont {P.~J.~D.}\
	  \bibnamefont {Crowley}}, \bibinfo {author} {\bibfnamefont {A.}~\bibnamefont
	  {Datta}}, \bibinfo {author} {\bibfnamefont {M.}~\bibnamefont {Barbieri}}, \
	  and\ \bibinfo {author} {\bibfnamefont {I.~A.}\ \bibnamefont {Walmsley}},\
	  }\href {\doibase 10.1103/PhysRevA.89.023845} {\bibfield  {journal} {\bibinfo
	  {journal} {Phys. Rev. A}\ }\textbf {\bibinfo {volume} {89}},\ \bibinfo
	  {pages} {023845} (\bibinfo {year} {2014})}\BibitemShut {NoStop}%
	\bibitem [{\citenamefont {Matsumoto}(2002)}]{Matsumoto2002}%
	  \BibitemOpen
	  \bibfield  {author} {\bibinfo {author} {\bibfnamefont {K.}~\bibnamefont
	  {Matsumoto}},\ }\href {\doibase 10.1088/0305-4470/35/13/307} {\bibfield
	  {journal} {\bibinfo  {journal} {J. Phys. A: Math. Gen.}\ }\textbf {\bibinfo
	  {volume} {35}},\ \bibinfo {pages} {3111} (\bibinfo {year}
	  {2002})}\BibitemShut {NoStop}%
	\bibitem [{\citenamefont {Vaneph}\ \emph {et~al.}(2013)\citenamefont {Vaneph},
	  \citenamefont {Tufarelli},\ and\ \citenamefont {Genoni}}]{Vanephetal2013}%
	  \BibitemOpen
	  \bibfield  {author} {\bibinfo {author} {\bibfnamefont {C.}~\bibnamefont
	  {Vaneph}}, \bibinfo {author} {\bibfnamefont {T.}~\bibnamefont {Tufarelli}}, \
	  and\ \bibinfo {author} {\bibfnamefont {M.~G.}\ \bibnamefont {Genoni}},\
	  }\href {\doibase 10.2478/qmetro-2013-0003} {\bibfield  {journal} {\bibinfo
	  {journal} {Quantum Meas. Quantum Metrol.}\ }\textbf {\bibinfo {volume} {1}},\
	  \bibinfo {pages} {12 } (\bibinfo {year} {25 Jun. 2013})}\BibitemShut
	  {NoStop}%
	\bibitem [{\citenamefont {Suzuki}(2016)}]{Suzuki2016}%
	  \BibitemOpen
	  \bibfield  {author} {\bibinfo {author} {\bibfnamefont {J.}~\bibnamefont
	  {Suzuki}},\ }\href {\doibase 10.1063/1.4945086} {\bibfield  {journal}
	  {\bibinfo  {journal} {J. Math. Phys.}\ }\textbf {\bibinfo {volume} {57}},\
	  \bibinfo {pages} {042201} (\bibinfo {year} {2016})}\BibitemShut {NoStop}%
	\bibitem [{\citenamefont {Ragy}\ \emph {et~al.}(2016)\citenamefont {Ragy},
	  \citenamefont {Jarzyna},\ and\ \citenamefont
	  {Demkowicz-Dobrza\'{n}ski}}]{Ragyetal2016}%
	  \BibitemOpen
	  \bibfield  {author} {\bibinfo {author} {\bibfnamefont {S.}~\bibnamefont
	  {Ragy}}, \bibinfo {author} {\bibfnamefont {M.}~\bibnamefont {Jarzyna}}, \
	  and\ \bibinfo {author} {\bibfnamefont {R.}~\bibnamefont
	  {Demkowicz-Dobrza\'{n}ski}},\ }\href {\doibase 10.1103/PhysRevA.94.052108}
	  {\bibfield  {journal} {\bibinfo  {journal} {Phys. Rev. A}\ }\textbf {\bibinfo
	  {volume} {94}},\ \bibinfo {pages} {052108} (\bibinfo {year}
	  {2016})}\BibitemShut {NoStop}%
	\bibitem [{\citenamefont {Pezz\`e}\ \emph {et~al.}(2017)\citenamefont
	  {Pezz\`e}, \citenamefont {Ciampini}, \citenamefont {Spagnolo}, \citenamefont
	  {Humphreys}, \citenamefont {Datta}, \citenamefont {Walmsley}, \citenamefont
	  {Barbieri}, \citenamefont {Sciarrino},\ and\ \citenamefont
	  {Smerzi}}]{Pezzeetal2017optimal}%
	  \BibitemOpen
	  \bibfield  {author} {\bibinfo {author} {\bibfnamefont {L.}~\bibnamefont
	  {Pezz\`e}}, \bibinfo {author} {\bibfnamefont {M.~A.}\ \bibnamefont
	  {Ciampini}}, \bibinfo {author} {\bibfnamefont {N.}~\bibnamefont {Spagnolo}},
	  \bibinfo {author} {\bibfnamefont {P.~C.}\ \bibnamefont {Humphreys}}, \bibinfo
	  {author} {\bibfnamefont {A.}~\bibnamefont {Datta}}, \bibinfo {author}
	  {\bibfnamefont {I.~A.}\ \bibnamefont {Walmsley}}, \bibinfo {author}
	  {\bibfnamefont {M.}~\bibnamefont {Barbieri}}, \bibinfo {author}
	  {\bibfnamefont {F.}~\bibnamefont {Sciarrino}}, \ and\ \bibinfo {author}
	  {\bibfnamefont {A.}~\bibnamefont {Smerzi}},\ }\href {\doibase 10.1103/PhysRevLett.119.130504} {\bibfield  {journal} {\bibinfo  {journal}
	  {Phys. Rev. Lett.}\ }\textbf {\bibinfo {volume} {119}},\ \bibinfo {pages}
	  {130504} (\bibinfo {year} {2017})}\BibitemShut {NoStop}%
	\bibitem [{\citenamefont {Yang}\ \emph
	  {et~al.}(2019{\natexlab{a}})\citenamefont {Yang}, \citenamefont {Pang},
	  \citenamefont {Zhou},\ and\ \citenamefont
	  {Jordan}}]{Yang2019partialcommutativity}%
	  \BibitemOpen
	  \bibfield  {author} {\bibinfo {author} {\bibfnamefont {J.}~\bibnamefont
	  {Yang}}, \bibinfo {author} {\bibfnamefont {S.}~\bibnamefont {Pang}}, \bibinfo
	  {author} {\bibfnamefont {Y.}~\bibnamefont {Zhou}}, \ and\ \bibinfo {author}
	  {\bibfnamefont {A.~N.}\ \bibnamefont {Jordan}},\ }\href {\doibase 10.1103/PhysRevA.100.032104} {\bibfield  {journal} {\bibinfo  {journal}
	  {Phys. Rev. A}\ }\textbf {\bibinfo {volume} {100}},\ \bibinfo {pages}
	  {032104} (\bibinfo {year} {2019}{\natexlab{a}})}\BibitemShut {NoStop}%
	\bibitem [{\citenamefont {Napoli}\ \emph {et~al.}(2019)\citenamefont {Napoli},
	  \citenamefont {Piano}, \citenamefont {Leach}, \citenamefont {Adesso},\ and\
	  \citenamefont {Tufarelli}}]{Napolietal2019}%
	  \BibitemOpen
	  \bibfield  {author} {\bibinfo {author} {\bibfnamefont {C.}~\bibnamefont
	  {Napoli}}, \bibinfo {author} {\bibfnamefont {S.}~\bibnamefont {Piano}},
	  \bibinfo {author} {\bibfnamefont {R.}~\bibnamefont {Leach}}, \bibinfo
	  {author} {\bibfnamefont {G.}~\bibnamefont {Adesso}}, \ and\ \bibinfo {author}
	  {\bibfnamefont {T.}~\bibnamefont {Tufarelli}},\ }\href {\doibase 10.1103/PhysRevLett.122.140505} {\bibfield  {journal} {\bibinfo  {journal}
	  {Phys. Rev. Lett.}\ }\textbf {\bibinfo {volume} {122}},\ \bibinfo {pages}
	  {140505} (\bibinfo {year} {2019})}\BibitemShut {NoStop}%
	\bibitem [{\citenamefont {Kukita}(2020)}]{Kukita2020}%
	  \BibitemOpen
	  \bibfield  {author} {\bibinfo {author} {\bibfnamefont {S.}~\bibnamefont
	  {Kukita}},\ }\href {\doibase 10.1088/1751-8121/ab6d3d} {\bibfield  {journal}
	  {\bibinfo  {journal} {J. Phys. A: Math. Theor.}\ }\textbf {\bibinfo {volume}
	  {53}},\ \bibinfo {pages} {095303} (\bibinfo {year} {2020})}\BibitemShut
	  {NoStop}%
	\bibitem [{\citenamefont {Belliardo}\ and\ \citenamefont
	  {Giovannetti}(2021)}]{BelliardoGiovannetti2021}%
	  \BibitemOpen
	  \bibfield  {author} {\bibinfo {author} {\bibfnamefont {F.}~\bibnamefont
	  {Belliardo}}\ and\ \bibinfo {author} {\bibfnamefont {V.}~\bibnamefont
	  {Giovannetti}},\ }\href {\doibase 10.1088/1367-2630/ac04ca} {\bibfield
	  {journal} {\bibinfo  {journal} {New J. Phys.}\ }\textbf {\bibinfo {volume}
	  {23}},\ \bibinfo {pages} {063055} (\bibinfo {year} {2021})}\BibitemShut
	  {NoStop}%
	\bibitem [{\citenamefont {Holevo}(1976)}]{Holevo1976}%
	  \BibitemOpen
	  \bibfield  {author} {\bibinfo {author} {\bibfnamefont {A.~S.}\ \bibnamefont
	  {Holevo}},\ }in\ \href {\doibase 10.1007/BFb0077491} {\emph {\bibinfo
	  {booktitle} {Proceedings of the Third Japan --- USSR Symposium on Probability
	  Theory}}},\ \bibinfo {editor} {edited by\ \bibinfo {editor} {\bibfnamefont
	  {G.}~\bibnamefont {Maruyama}}\ and\ \bibinfo {editor} {\bibfnamefont {J.~V.}\
	  \bibnamefont {Prokhorov}}}\ (\bibinfo  {publisher} {Springer Berlin
	  Heidelberg},\ \bibinfo {address} {Berlin, Heidelberg},\ \bibinfo {year}
	  {1976})\ pp.\ \bibinfo {pages} {194--222}\BibitemShut {NoStop}%
	\bibitem [{\citenamefont {Gu\c{t}\u{a}}\ and\ \citenamefont
	  {Kahn}(2006)}]{GutaKahn2006}%
	  \BibitemOpen
	  \bibfield  {author} {\bibinfo {author} {\bibfnamefont {M.}~\bibnamefont
	  {Gu\c{t}\u{a}}}\ and\ \bibinfo {author} {\bibfnamefont {J.}~\bibnamefont
	  {Kahn}},\ }\href {\doibase 10.1103/PhysRevA.73.052108} {\bibfield  {journal}
	  {\bibinfo  {journal} {Phys. Rev. A}\ }\textbf {\bibinfo {volume} {73}},\
	  \bibinfo {pages} {052108} (\bibinfo {year} {2006})}\BibitemShut {NoStop}%
	\bibitem [{\citenamefont {Hayashi}\ and\ \citenamefont
	  {Matsumoto}(2008)}]{HayashiMatsumoto2008}%
	  \BibitemOpen
	  \bibfield  {author} {\bibinfo {author} {\bibfnamefont {M.}~\bibnamefont
	  {Hayashi}}\ and\ \bibinfo {author} {\bibfnamefont {K.}~\bibnamefont
	  {Matsumoto}},\ }\href {\doibase 10.1063/1.2988130} {\bibfield  {journal}
	  {\bibinfo  {journal} {J. Math. Phys.}\ }\textbf {\bibinfo {volume} {49}},\
	  \bibinfo {pages} {102101} (\bibinfo {year} {2008})}\BibitemShut {NoStop}%
	\bibitem [{\citenamefont {Kahn}\ and\ \citenamefont
	  {Gu\c{t}\u{a}}(2009)}]{KahnGuta2009}%
	  \BibitemOpen
	  \bibfield  {author} {\bibinfo {author} {\bibfnamefont {J.}~\bibnamefont
	  {Kahn}}\ and\ \bibinfo {author} {\bibfnamefont {M.}~\bibnamefont
	  {Gu\c{t}\u{a}}},\ }\href {\doibase 10.1007/s00220-009-0787-3} {\bibfield
	  {journal} {\bibinfo  {journal} {Comm. Math. Phys}\ }\textbf {\bibinfo
	  {volume} {289}},\ \bibinfo {pages} {597} (\bibinfo {year}
	  {2009})}\BibitemShut {NoStop}%
	\bibitem [{\citenamefont {Yamagata}\ \emph {et~al.}(2013)\citenamefont
	  {Yamagata}, \citenamefont {Fujiwara},\ and\ \citenamefont
	  {Gill}}]{Yamagata2013}%
	  \BibitemOpen
	  \bibfield  {author} {\bibinfo {author} {\bibfnamefont {K.}~\bibnamefont
	  {Yamagata}}, \bibinfo {author} {\bibfnamefont {A.}~\bibnamefont {Fujiwara}},
	  \ and\ \bibinfo {author} {\bibfnamefont {R.~D.}\ \bibnamefont {Gill}},\
	  }\href {\doibase 10.1214/13-AOS1147} {\bibfield  {journal} {\bibinfo
	  {journal} {Ann. Stat.}\ }\textbf {\bibinfo {volume} {41}},\ \bibinfo {pages}
	  {2197} (\bibinfo {year} {2013})}\BibitemShut {NoStop}%
	\bibitem [{\citenamefont {Yang}\ \emph
	  {et~al.}(2019{\natexlab{b}})\citenamefont {Yang}, \citenamefont
	  {Chiribella},\ and\ \citenamefont {Hayashi}}]{Yangetal2019attaining}%
	  \BibitemOpen
	  \bibfield  {author} {\bibinfo {author} {\bibfnamefont {Y.}~\bibnamefont
	  {Yang}}, \bibinfo {author} {\bibfnamefont {G.}~\bibnamefont {Chiribella}}, \
	  and\ \bibinfo {author} {\bibfnamefont {M.}~\bibnamefont {Hayashi}},\ }\href
	  {\doibase 10.1007/s00220-019-03433-4} {\bibfield  {journal} {\bibinfo
	  {journal} {Comm. Math. Phys.}\ }\textbf {\bibinfo {volume} {368}},\ \bibinfo
	  {pages} {223} (\bibinfo {year} {2019}{\natexlab{b}})}\BibitemShut {NoStop}%
	\bibitem [{\citenamefont {Bradshaw}\ \emph {et~al.}(2017)\citenamefont
	  {Bradshaw}, \citenamefont {Assad},\ and\ \citenamefont
	  {Lam}}]{Bradshawetal2017}%
	  \BibitemOpen
	  \bibfield  {author} {\bibinfo {author} {\bibfnamefont {M.}~\bibnamefont
	  {Bradshaw}}, \bibinfo {author} {\bibfnamefont {S.~M.}\ \bibnamefont {Assad}},
	  \ and\ \bibinfo {author} {\bibfnamefont {P.~K.}\ \bibnamefont {Lam}},\ }\href
	  {\doibase 10.1016/j.physleta.2017.06.024} {\bibfield  {journal} {\bibinfo
	  {journal} {Phys. Lett. A}\ }\textbf {\bibinfo {volume} {381}},\ \bibinfo
	  {pages} {2598 } (\bibinfo {year} {2017})}\BibitemShut {NoStop}%
	\bibitem [{\citenamefont {Bradshaw}\ \emph {et~al.}(2018)\citenamefont
	  {Bradshaw}, \citenamefont {Lam},\ and\ \citenamefont
	  {Assad}}]{Bradshawetal2018}%
	  \BibitemOpen
	  \bibfield  {author} {\bibinfo {author} {\bibfnamefont {M.}~\bibnamefont
	  {Bradshaw}}, \bibinfo {author} {\bibfnamefont {P.~K.}\ \bibnamefont {Lam}}, \
	  and\ \bibinfo {author} {\bibfnamefont {S.~M.}\ \bibnamefont {Assad}},\ }\href
	  {\doibase 10.1103/PhysRevA.97.012106} {\bibfield  {journal} {\bibinfo
	  {journal} {Phys. Rev. A}\ }\textbf {\bibinfo {volume} {97}},\ \bibinfo
	  {pages} {012106} (\bibinfo {year} {2018})}\BibitemShut {NoStop}%
	\bibitem [{\citenamefont {G{\'{o}}recki}\ \emph {et~al.}(2020)\citenamefont
	  {G{\'{o}}recki}, \citenamefont {Zhou}, \citenamefont {Jiang},\ and\
	  \citenamefont {Demkowicz-Dobrza{\'{n}}ski}}]{Gorecki2020optimalprobeserror}%
	  \BibitemOpen
	  \bibfield  {author} {\bibinfo {author} {\bibfnamefont {W.}~\bibnamefont
	  {G{\'{o}}recki}}, \bibinfo {author} {\bibfnamefont {S.}~\bibnamefont {Zhou}},
	  \bibinfo {author} {\bibfnamefont {L.}~\bibnamefont {Jiang}}, \ and\ \bibinfo
	  {author} {\bibfnamefont {R.}~\bibnamefont {Demkowicz-Dobrza{\'{n}}ski}},\
	  }\href {\doibase 10.22331/q-2020-07-02-288} {\bibfield  {journal} {\bibinfo
	  {journal} {{Quantum}}\ }\textbf {\bibinfo {volume} {4}},\ \bibinfo {pages}
	  {288} (\bibinfo {year} {2020})}\BibitemShut {NoStop}%
	\bibitem [{\citenamefont {Carollo}\ \emph {et~al.}(2019)\citenamefont
	  {Carollo}, \citenamefont {Spagnolo}, \citenamefont {Dubkov},\ and\
	  \citenamefont {Valenti}}]{Carollo2019}%
	  \BibitemOpen
	  \bibfield  {author} {\bibinfo {author} {\bibfnamefont {A.}~\bibnamefont
	  {Carollo}}, \bibinfo {author} {\bibfnamefont {B.}~\bibnamefont {Spagnolo}},
	  \bibinfo {author} {\bibfnamefont {A.~A.}\ \bibnamefont {Dubkov}}, \ and\
	  \bibinfo {author} {\bibfnamefont {D.}~\bibnamefont {Valenti}},\ }\href
	  {\doibase 10.1088/1742-5468/ab3ccb} {\bibfield  {journal} {\bibinfo
	  {journal} {J. Stat. Mech.}\ }\textbf {\bibinfo {volume} {2019}},\ \bibinfo
	  {pages} {094010} (\bibinfo {year} {2019})}\BibitemShut {NoStop}%
	\bibitem [{\citenamefont {Yamagata}(2021)}]{Yamagata2021}%
	  \BibitemOpen
	  \bibfield  {author} {\bibinfo {author} {\bibfnamefont {K.}~\bibnamefont
	  {Yamagata}},\ }\href {\doibase 10.1063/5.0047496} {\bibfield  {journal}
	  {\bibinfo  {journal} {J. Math. Phys.}\ }\textbf {\bibinfo {volume} {62}},\
	  \bibinfo {pages} {062203} (\bibinfo {year} {2021})}\BibitemShut {NoStop}%
	\bibitem [{\citenamefont {Sidhu}\ \emph {et~al.}(2021)\citenamefont {Sidhu},
	  \citenamefont {Ouyang}, \citenamefont {Campbell},\ and\ \citenamefont
	  {Kok}}]{Sidhuetal2021}%
	  \BibitemOpen
	  \bibfield  {author} {\bibinfo {author} {\bibfnamefont {J.~S.}\ \bibnamefont
	  {Sidhu}}, \bibinfo {author} {\bibfnamefont {Y.}~\bibnamefont {Ouyang}},
	  \bibinfo {author} {\bibfnamefont {E.~T.}\ \bibnamefont {Campbell}}, \ and\
	  \bibinfo {author} {\bibfnamefont {P.}~\bibnamefont {Kok}},\ }\href {\doibase 10.1103/PhysRevX.11.011028} {\bibfield  {journal} {\bibinfo  {journal} {Phys.
	  Rev. X}\ }\textbf {\bibinfo {volume} {11}},\ \bibinfo {pages} {011028}
	  (\bibinfo {year} {2021})}\BibitemShut {NoStop}%
	\bibitem [{\citenamefont {Goldberg}\ \emph {et~al.}(2021)\citenamefont
	  {Goldberg}, \citenamefont {S\'anchez-Soto},\ and\ \citenamefont
	  {Ferretti}}]{Goldbergetal2021}%
	  \BibitemOpen
	  \bibfield  {author} {\bibinfo {author} {\bibfnamefont {A.~Z.}\ \bibnamefont
	  {Goldberg}}, \bibinfo {author} {\bibfnamefont {L.~L.}\ \bibnamefont
	  {S\'anchez-Soto}}, \ and\ \bibinfo {author} {\bibfnamefont {H.}~\bibnamefont
	  {Ferretti}},\ }\href {\doibase 10.1103/PhysRevLett.127.110501} {\bibfield
	  {journal} {\bibinfo  {journal} {Phys. Rev. Lett.}\ }\textbf {\bibinfo
	  {volume} {127}},\ \bibinfo {pages} {110501} (\bibinfo {year}
	  {2021})}\BibitemShut {NoStop}%
	\bibitem [{\citenamefont {Tsang}\ \emph {et~al.}(2016)\citenamefont {Tsang},
	  \citenamefont {Nair},\ and\ \citenamefont {Lu}}]{Tsangetal2016}%
	  \BibitemOpen
	  \bibfield  {author} {\bibinfo {author} {\bibfnamefont {M.}~\bibnamefont
	  {Tsang}}, \bibinfo {author} {\bibfnamefont {R.}~\bibnamefont {Nair}}, \ and\
	  \bibinfo {author} {\bibfnamefont {X.-M.}\ \bibnamefont {Lu}},\ }\href
	  {\doibase 10.1103/PhysRevX.6.031033} {\bibfield  {journal} {\bibinfo
	  {journal} {Phys. Rev. X}\ }\textbf {\bibinfo {volume} {6}},\ \bibinfo {pages}
	  {031033} (\bibinfo {year} {2016})}\BibitemShut {NoStop}%
	\bibitem [{\citenamefont {\v{R}eh\'{a}\v{c}ek}\ \emph
	  {et~al.}(2018)\citenamefont {\v{R}eh\'{a}\v{c}ek}, \citenamefont {Hradil},
	  \citenamefont {Koutn\'y}, \citenamefont {Grover}, \citenamefont {Krzic},\
	  and\ \citenamefont {S\'anchez-Soto}}]{Rehaceketal2018optmeas}%
	  \BibitemOpen
	  \bibfield  {author} {\bibinfo {author} {\bibfnamefont {J.}~\bibnamefont
	  {\v{R}eh\'{a}\v{c}ek}}, \bibinfo {author} {\bibfnamefont {Z.}~\bibnamefont
	  {Hradil}}, \bibinfo {author} {\bibfnamefont {D.}~\bibnamefont {Koutn\'y}},
	  \bibinfo {author} {\bibfnamefont {J.}~\bibnamefont {Grover}}, \bibinfo
	  {author} {\bibfnamefont {A.}~\bibnamefont {Krzic}}, \ and\ \bibinfo {author}
	  {\bibfnamefont {L.~L.}\ \bibnamefont {S\'anchez-Soto}},\ }\href {\doibase 10.1103/PhysRevA.98.012103} {\bibfield  {journal} {\bibinfo  {journal} {Phys.
	  Rev. A}\ }\textbf {\bibinfo {volume} {98}},\ \bibinfo {pages} {012103}
	  (\bibinfo {year} {2018})}\BibitemShut {NoStop}%
	\bibitem [{\citenamefont {Len}(2021)}]{Len2021}%
	  \BibitemOpen
	  \bibfield  {author} {\bibinfo {author} {\bibfnamefont {Y.~L.}\ \bibnamefont
	  {Len}},\ }\href {https://doi.org/10.48550/arXiv.2109.07430} {\bibfield
	  {journal} {\bibinfo  {journal} {arXiv preprint arXiv:2109.07430}\ } (\bibinfo
	  {year} {2021})}\BibitemShut {NoStop}%
	\bibitem [{\citenamefont {de~Almeida}\ \emph {et~al.}(2021)\citenamefont
	  {de~Almeida}, \citenamefont {Lewenstein},\ and\ \citenamefont
	  {Skotiniotis}}]{Almeidaetal2021collective}%
	  \BibitemOpen
	  \bibfield  {author} {\bibinfo {author} {\bibfnamefont {J.}~\bibnamefont
	  {de~Almeida}}, \bibinfo {author} {\bibfnamefont {M.}~\bibnamefont
	  {Lewenstein}}, \ and\ \bibinfo {author} {\bibfnamefont {M.}~\bibnamefont
	  {Skotiniotis}},\ }\href {https://doi.org/10.48550/arXiv.2110.00986}
	  {\bibfield  {journal} {\bibinfo  {journal} {arXiv preprint arXiv:2110.00986}\
	  } (\bibinfo {year} {2021})}\BibitemShut {NoStop}%
	\bibitem [{\citenamefont {Lee}\ \emph {et~al.}(2002)\citenamefont {Lee},
	  \citenamefont {Kok},\ and\ \citenamefont
	  {Dowling}}]{LeeKokDowling2002GHZequivalentNOON}%
	  \BibitemOpen
	  \bibfield  {author} {\bibinfo {author} {\bibfnamefont {H.}~\bibnamefont
	  {Lee}}, \bibinfo {author} {\bibfnamefont {P.}~\bibnamefont {Kok}}, \ and\
	  \bibinfo {author} {\bibfnamefont {J.~P.}\ \bibnamefont {Dowling}},\ }\href
	  {\doibase 10.1080/0950034021000011536} {\bibfield  {journal} {\bibinfo
	  {journal} {J. Mod. Opt.}\ }\textbf {\bibinfo {volume} {49}},\ \bibinfo
	  {pages} {2325} (\bibinfo {year} {2002})}\BibitemShut {NoStop}%
	\bibitem [{\citenamefont {Giovannetti}\ \emph {et~al.}(2004)\citenamefont
	  {Giovannetti}, \citenamefont {Lloyd},\ and\ \citenamefont
	  {Maccone}}]{Giovanettietal2004science}%
	  \BibitemOpen
	  \bibfield  {author} {\bibinfo {author} {\bibfnamefont {V.}~\bibnamefont
	  {Giovannetti}}, \bibinfo {author} {\bibfnamefont {S.}~\bibnamefont {Lloyd}},
	  \ and\ \bibinfo {author} {\bibfnamefont {L.}~\bibnamefont {Maccone}},\ }\href
	  {\doibase 10.1126/science.1104149} {\bibfield  {journal} {\bibinfo  {journal}
	  {Science}\ }\textbf {\bibinfo {volume} {306}},\ \bibinfo {pages} {1330}
	  (\bibinfo {year} {2004})}\BibitemShut {NoStop}%
	\bibitem [{\citenamefont {Gisin}\ and\ \citenamefont
	  {Popescu}(1999)}]{GisinPopescu1999}%
	  \BibitemOpen
	  \bibfield  {author} {\bibinfo {author} {\bibfnamefont {N.}~\bibnamefont
	  {Gisin}}\ and\ \bibinfo {author} {\bibfnamefont {S.}~\bibnamefont
	  {Popescu}},\ }\href {\doibase 10.1103/PhysRevLett.83.432} {\bibfield
	  {journal} {\bibinfo  {journal} {Phys. Rev. Lett.}\ }\textbf {\bibinfo
	  {volume} {83}},\ \bibinfo {pages} {432} (\bibinfo {year} {1999})}\BibitemShut
	  {NoStop}%
	\bibitem [{\citenamefont {Chang}\ \emph {et~al.}(2014)\citenamefont {Chang},
	  \citenamefont {Li}, \citenamefont {Luo},\ and\ \citenamefont
	  {Song}}]{Changetal2014}%
	  \BibitemOpen
	  \bibfield  {author} {\bibinfo {author} {\bibfnamefont {L.}~\bibnamefont
	  {Chang}}, \bibinfo {author} {\bibfnamefont {N.}~\bibnamefont {Li}}, \bibinfo
	  {author} {\bibfnamefont {S.}~\bibnamefont {Luo}}, \ and\ \bibinfo {author}
	  {\bibfnamefont {H.}~\bibnamefont {Song}},\ }\href {\doibase 10.1103/PhysRevA.89.042110} {\bibfield  {journal} {\bibinfo  {journal} {Phys.
	  Rev. A}\ }\textbf {\bibinfo {volume} {89}},\ \bibinfo {pages} {042110}
	  (\bibinfo {year} {2014})}\BibitemShut {NoStop}%
	\bibitem [{\citenamefont {Carollo}\ \emph
	  {et~al.}(2018{\natexlab{a}})\citenamefont {Carollo}, \citenamefont
	  {Spagnolo},\ and\ \citenamefont {Valenti}}]{Carollo2018}%
	  \BibitemOpen
	  \bibfield  {author} {\bibinfo {author} {\bibfnamefont {A.}~\bibnamefont
	  {Carollo}}, \bibinfo {author} {\bibfnamefont {B.}~\bibnamefont {Spagnolo}}, \
	  and\ \bibinfo {author} {\bibfnamefont {D.}~\bibnamefont {Valenti}},\ }\href
	  {\doibase 10.1038/s41598-018-27362-9} {\bibfield  {journal} {\bibinfo
	  {journal} {Sci. Rep.}\ }\textbf {\bibinfo {volume} {8}},\ \bibinfo {pages}
	  {1} (\bibinfo {year} {2018}{\natexlab{a}})}\BibitemShut {NoStop}%
	\bibitem [{\citenamefont {Carollo}\ \emph
	  {et~al.}(2018{\natexlab{b}})\citenamefont {Carollo}, \citenamefont
	  {Spagnolo},\ and\ \citenamefont {Valenti}}]{Carolloetal2018fermion}%
	  \BibitemOpen
	  \bibfield  {author} {\bibinfo {author} {\bibfnamefont {A.}~\bibnamefont
	  {Carollo}}, \bibinfo {author} {\bibfnamefont {B.}~\bibnamefont {Spagnolo}}, \
	  and\ \bibinfo {author} {\bibfnamefont {D.}~\bibnamefont {Valenti}},\ }\href
	  {\doibase 10.3390/e20070485} {\bibfield  {journal} {\bibinfo  {journal}
	  {Entropy}\ }\textbf {\bibinfo {volume} {20}},\ \bibinfo {pages} {485}
	  (\bibinfo {year} {2018}{\natexlab{b}})}\BibitemShut {NoStop}%
	\bibitem [{\citenamefont {Leonforte}\ \emph {et~al.}(2019)\citenamefont
	  {Leonforte}, \citenamefont {Valenti}, \citenamefont {Spagnolo}, \citenamefont
	  {Dubkov},\ and\ \citenamefont {Carollo}}]{Leonforte2019muc}%
	  \BibitemOpen
	  \bibfield  {author} {\bibinfo {author} {\bibfnamefont {L.}~\bibnamefont
	  {Leonforte}}, \bibinfo {author} {\bibfnamefont {D.}~\bibnamefont {Valenti}},
	  \bibinfo {author} {\bibfnamefont {B.}~\bibnamefont {Spagnolo}}, \bibinfo
	  {author} {\bibfnamefont {A.~A.}\ \bibnamefont {Dubkov}}, \ and\ \bibinfo
	  {author} {\bibfnamefont {A.}~\bibnamefont {Carollo}},\ }\href {\doibase 10.1088/1742-5468/ab33f8} {\bibfield  {journal} {\bibinfo  {journal} {J.
	  Stat. Mech.}\ }\textbf {\bibinfo {volume} {2019}},\ \bibinfo {pages} {094001}
	  (\bibinfo {year} {2019})}\BibitemShut {NoStop}%
	\bibitem [{\citenamefont {Bascone}\ \emph
	  {et~al.}(2019{\natexlab{a}})\citenamefont {Bascone}, \citenamefont
	  {Leonforte}, \citenamefont {Valenti}, \citenamefont {Spagnolo},\ and\
	  \citenamefont {Carollo}}]{Basconeetal2019muc}%
	  \BibitemOpen
	  \bibfield  {author} {\bibinfo {author} {\bibfnamefont {F.}~\bibnamefont
	  {Bascone}}, \bibinfo {author} {\bibfnamefont {L.}~\bibnamefont {Leonforte}},
	  \bibinfo {author} {\bibfnamefont {D.}~\bibnamefont {Valenti}}, \bibinfo
	  {author} {\bibfnamefont {B.}~\bibnamefont {Spagnolo}}, \ and\ \bibinfo
	  {author} {\bibfnamefont {A.}~\bibnamefont {Carollo}},\ }\href {\doibase 10.1088/1742-5468/ab35e9} {\bibfield  {journal} {\bibinfo  {journal} {J.
	  Stat. Mech.}\ }\textbf {\bibinfo {volume} {2019}},\ \bibinfo {pages} {094002}
	  (\bibinfo {year} {2019}{\natexlab{a}})}\BibitemShut {NoStop}%
	\bibitem [{\citenamefont {Bascone}\ \emph
	  {et~al.}(2019{\natexlab{b}})\citenamefont {Bascone}, \citenamefont
	  {Leonforte}, \citenamefont {Valenti}, \citenamefont {Spagnolo},\ and\
	  \citenamefont {Carollo}}]{Basconeetal2019mucPRB}%
	  \BibitemOpen
	  \bibfield  {author} {\bibinfo {author} {\bibfnamefont {F.}~\bibnamefont
	  {Bascone}}, \bibinfo {author} {\bibfnamefont {L.}~\bibnamefont {Leonforte}},
	  \bibinfo {author} {\bibfnamefont {D.}~\bibnamefont {Valenti}}, \bibinfo
	  {author} {\bibfnamefont {B.}~\bibnamefont {Spagnolo}}, \ and\ \bibinfo
	  {author} {\bibfnamefont {A.}~\bibnamefont {Carollo}},\ }\href {\doibase 10.1103/PhysRevB.99.205155} {\bibfield  {journal} {\bibinfo  {journal} {Phys.
	  Rev. B}\ }\textbf {\bibinfo {volume} {99}},\ \bibinfo {pages} {205155}
	  (\bibinfo {year} {2019}{\natexlab{b}})}\BibitemShut {NoStop}%
	\bibitem [{\citenamefont {Hickey}\ and\ \citenamefont
	  {Gour}(2018)}]{HickeyGour2018imaginarity}%
	  \BibitemOpen
	  \bibfield  {author} {\bibinfo {author} {\bibfnamefont {A.}~\bibnamefont
	  {Hickey}}\ and\ \bibinfo {author} {\bibfnamefont {G.}~\bibnamefont {Gour}},\
	  }\href {\doibase 10.1088/1751-8121/aabe9c} {\bibfield  {journal} {\bibinfo
	  {journal} {J. Phys. A: Math. Theor.}\ }\textbf {\bibinfo {volume} {51}},\
	  \bibinfo {pages} {414009} (\bibinfo {year} {2018})}\BibitemShut {NoStop}%
	\bibitem [{\citenamefont {Wu}\ \emph {et~al.}(2021{\natexlab{a}})\citenamefont
	  {Wu}, \citenamefont {Kondra}, \citenamefont {Rana}, \citenamefont {Scandolo},
	  \citenamefont {Xiang}, \citenamefont {Li}, \citenamefont {Guo},\ and\
	  \citenamefont {Streltsov}}]{Wu2021imaginarityPRL}%
	  \BibitemOpen
	  \bibfield  {author} {\bibinfo {author} {\bibfnamefont {K.-D.}\ \bibnamefont
	  {Wu}}, \bibinfo {author} {\bibfnamefont {T.~V.}\ \bibnamefont {Kondra}},
	  \bibinfo {author} {\bibfnamefont {S.}~\bibnamefont {Rana}}, \bibinfo {author}
	  {\bibfnamefont {C.~M.}\ \bibnamefont {Scandolo}}, \bibinfo {author}
	  {\bibfnamefont {G.-Y.}\ \bibnamefont {Xiang}}, \bibinfo {author}
	  {\bibfnamefont {C.-F.}\ \bibnamefont {Li}}, \bibinfo {author} {\bibfnamefont
	  {G.-C.}\ \bibnamefont {Guo}}, \ and\ \bibinfo {author} {\bibfnamefont
	  {A.}~\bibnamefont {Streltsov}},\ }\href {\doibase 10.1103/PhysRevLett.126.090401} {\bibfield  {journal} {\bibinfo  {journal}
	  {Phys. Rev. Lett.}\ }\textbf {\bibinfo {volume} {126}},\ \bibinfo {pages}
	  {090401} (\bibinfo {year} {2021}{\natexlab{a}})}\BibitemShut {NoStop}%
	\bibitem [{\citenamefont {Wu}\ \emph {et~al.}(2021{\natexlab{b}})\citenamefont
	  {Wu}, \citenamefont {Kondra}, \citenamefont {Rana}, \citenamefont {Scandolo},
	  \citenamefont {Xiang}, \citenamefont {Li}, \citenamefont {Guo},\ and\
	  \citenamefont {Streltsov}}]{Wu2021imaginarityPRA}%
	  \BibitemOpen
	  \bibfield  {author} {\bibinfo {author} {\bibfnamefont {K.-D.}\ \bibnamefont
	  {Wu}}, \bibinfo {author} {\bibfnamefont {T.~V.}\ \bibnamefont {Kondra}},
	  \bibinfo {author} {\bibfnamefont {S.}~\bibnamefont {Rana}}, \bibinfo {author}
	  {\bibfnamefont {C.~M.}\ \bibnamefont {Scandolo}}, \bibinfo {author}
	  {\bibfnamefont {G.-Y.}\ \bibnamefont {Xiang}}, \bibinfo {author}
	  {\bibfnamefont {C.-F.}\ \bibnamefont {Li}}, \bibinfo {author} {\bibfnamefont
	  {G.-C.}\ \bibnamefont {Guo}}, \ and\ \bibinfo {author} {\bibfnamefont
	  {A.}~\bibnamefont {Streltsov}},\ }\href {\doibase 10.1103/PhysRevA.103.032401} {\bibfield  {journal} {\bibinfo  {journal}
	  {Phys. Rev. A}\ }\textbf {\bibinfo {volume} {103}},\ \bibinfo {pages}
	  {032401} (\bibinfo {year} {2021}{\natexlab{b}})}\BibitemShut {NoStop}%
	\bibitem [{\citenamefont {Amari}\ and\ \citenamefont
	  {Nagaoka}(2000)}]{AmariNagaoka2000}%
	  \BibitemOpen
	  \bibfield  {author} {\bibinfo {author} {\bibfnamefont {S.}~\bibnamefont
	  {Amari}}\ and\ \bibinfo {author} {\bibfnamefont {H.}~\bibnamefont
	  {Nagaoka}},\ }\href {\doibase 10.1090/mmono/191} {\emph {\bibinfo {title}
	  {Methods of Information Geometry}}},\ \bibinfo {series} {Translations of
	  Mathematical Monographs}, Vol.\ \bibinfo {volume} {191}\ (\bibinfo
	  {publisher} {American Mathematical Society},\ \bibinfo {year}
	  {2000})\BibitemShut {NoStop}%
	\bibitem [{\citenamefont {Yang}(2021)}]{Yang2021}%
	  \BibitemOpen
	  \bibfield  {author} {\bibinfo {author} {\bibfnamefont {J.}~\bibnamefont
	  {Yang}},\ }\href@noop {} {}\bibinfo {howpublished} {private communication}
	  (\bibinfo {year} {2021})\BibitemShut {NoStop}%
	\bibitem [{\citenamefont {Uhlmann}(2016)}]{Uhlmann2016antilinear}%
	  \BibitemOpen
	  \bibfield  {author} {\bibinfo {author} {\bibfnamefont {A.}~\bibnamefont
	  {Uhlmann}},\ }\href {\doibase 10.1007/s11433-015-5777-1} {\bibfield
	  {journal} {\bibinfo  {journal} {Sci. China Phys. Mech. Astron.}\ }\textbf
	  {\bibinfo {volume} {59}},\ \bibinfo {pages} {630301} (\bibinfo {year}
	  {2016})}\BibitemShut {NoStop}%
	\bibitem [{\citenamefont {Fujiwara}(2001)}]{Fujiwara2001}%
	  \BibitemOpen
	  \bibfield  {author} {\bibinfo {author} {\bibfnamefont {A.}~\bibnamefont
	  {Fujiwara}},\ }\href {\doibase 10.1103/PhysRevA.65.012316} {\bibfield
	  {journal} {\bibinfo  {journal} {Phys. Rev. A}\ }\textbf {\bibinfo {volume}
	  {65}},\ \bibinfo {pages} {012316} (\bibinfo {year} {2001})}\BibitemShut
	  {NoStop}%
	\bibitem [{\citenamefont {Casanova}\ \emph {et~al.}(2011)\citenamefont
	  {Casanova}, \citenamefont {Sab\'{\i}n}, \citenamefont {Le\'on}, \citenamefont
	  {Egusquiza}, \citenamefont {Gerritsma}, \citenamefont {Roos}, \citenamefont
	  {Garc\'{\i}a-Ripoll},\ and\ \citenamefont {Solano}}]{Casanovaetal2011eqs}%
	  \BibitemOpen
	  \bibfield  {author} {\bibinfo {author} {\bibfnamefont {J.}~\bibnamefont
	  {Casanova}}, \bibinfo {author} {\bibfnamefont {C.}~\bibnamefont
	  {Sab\'{\i}n}}, \bibinfo {author} {\bibfnamefont {J.}~\bibnamefont {Le\'on}},
	  \bibinfo {author} {\bibfnamefont {I.~L.}\ \bibnamefont {Egusquiza}}, \bibinfo
	  {author} {\bibfnamefont {R.}~\bibnamefont {Gerritsma}}, \bibinfo {author}
	  {\bibfnamefont {C.~F.}\ \bibnamefont {Roos}}, \bibinfo {author}
	  {\bibfnamefont {J.~J.}\ \bibnamefont {Garc\'{\i}a-Ripoll}}, \ and\ \bibinfo
	  {author} {\bibfnamefont {E.}~\bibnamefont {Solano}},\ }\href {\doibase 10.1103/PhysRevX.1.021018} {\bibfield  {journal} {\bibinfo  {journal} {Phys.
	  Rev. X}\ }\textbf {\bibinfo {volume} {1}},\ \bibinfo {pages} {021018}
	  (\bibinfo {year} {2011})}\BibitemShut {NoStop}%
	\bibitem [{\citenamefont {Di~Candia}\ \emph {et~al.}(2013)\citenamefont
	  {Di~Candia}, \citenamefont {Mejia}, \citenamefont {Castillo}, \citenamefont
	  {Pedernales}, \citenamefont {Casanova},\ and\ \citenamefont
	  {Solano}}]{Candiaetal2013}%
	  \BibitemOpen
	  \bibfield  {author} {\bibinfo {author} {\bibfnamefont {R.}~\bibnamefont
	  {Di~Candia}}, \bibinfo {author} {\bibfnamefont {B.}~\bibnamefont {Mejia}},
	  \bibinfo {author} {\bibfnamefont {H.}~\bibnamefont {Castillo}}, \bibinfo
	  {author} {\bibfnamefont {J.~S.}\ \bibnamefont {Pedernales}}, \bibinfo
	  {author} {\bibfnamefont {J.}~\bibnamefont {Casanova}}, \ and\ \bibinfo
	  {author} {\bibfnamefont {E.}~\bibnamefont {Solano}},\ }\href {\doibase 10.1103/PhysRevLett.111.240502} {\bibfield  {journal} {\bibinfo  {journal}
	  {Phys. Rev. Lett.}\ }\textbf {\bibinfo {volume} {111}},\ \bibinfo {pages}
	  {240502} (\bibinfo {year} {2013})}\BibitemShut {NoStop}%
	\bibitem [{\citenamefont {Zhang}\ \emph {et~al.}(2015)\citenamefont {Zhang},
	  \citenamefont {Shen}, \citenamefont {Zhang}, \citenamefont {Casanova},
	  \citenamefont {Lamata}, \citenamefont {Solano}, \citenamefont {Yung},
	  \citenamefont {Zhang},\ and\ \citenamefont {Kim}}]{Zhangetal2015eqs}%
	  \BibitemOpen
	  \bibfield  {author} {\bibinfo {author} {\bibfnamefont {X.}~\bibnamefont
	  {Zhang}}, \bibinfo {author} {\bibfnamefont {Y.}~\bibnamefont {Shen}},
	  \bibinfo {author} {\bibfnamefont {J.}~\bibnamefont {Zhang}}, \bibinfo
	  {author} {\bibfnamefont {J.}~\bibnamefont {Casanova}}, \bibinfo {author}
	  {\bibfnamefont {L.}~\bibnamefont {Lamata}}, \bibinfo {author} {\bibfnamefont
	  {E.}~\bibnamefont {Solano}}, \bibinfo {author} {\bibfnamefont {M.-H.}\
	  \bibnamefont {Yung}}, \bibinfo {author} {\bibfnamefont {J.-N.}\ \bibnamefont
	  {Zhang}}, \ and\ \bibinfo {author} {\bibfnamefont {K.}~\bibnamefont {Kim}},\
	  }\href {\doibase 10.1038/ncomms8917} {\bibfield  {journal} {\bibinfo
	  {journal} {Nat. Commun.}\ }\textbf {\bibinfo {volume} {6}},\ \bibinfo {pages}
	  {1} (\bibinfo {year} {2015})}\BibitemShut {NoStop}%
	\bibitem [{\citenamefont {Chen}\ \emph {et~al.}(2016)\citenamefont {Chen},
	  \citenamefont {Wu}, \citenamefont {Su}, \citenamefont {Cai}, \citenamefont
	  {Wang}, \citenamefont {Yang}, \citenamefont {Li}, \citenamefont {Liu},
	  \citenamefont {Lu},\ and\ \citenamefont {Pan}}]{Chenetal2016eqs}%
	  \BibitemOpen
	  \bibfield  {author} {\bibinfo {author} {\bibfnamefont {M.-C.}\ \bibnamefont
	  {Chen}}, \bibinfo {author} {\bibfnamefont {D.}~\bibnamefont {Wu}}, \bibinfo
	  {author} {\bibfnamefont {Z.-E.}\ \bibnamefont {Su}}, \bibinfo {author}
	  {\bibfnamefont {X.-D.}\ \bibnamefont {Cai}}, \bibinfo {author} {\bibfnamefont
	  {X.-L.}\ \bibnamefont {Wang}}, \bibinfo {author} {\bibfnamefont
	  {T.}~\bibnamefont {Yang}}, \bibinfo {author} {\bibfnamefont {L.}~\bibnamefont
	  {Li}}, \bibinfo {author} {\bibfnamefont {N.-L.}\ \bibnamefont {Liu}},
	  \bibinfo {author} {\bibfnamefont {C.-Y.}\ \bibnamefont {Lu}}, \ and\ \bibinfo
	  {author} {\bibfnamefont {J.-W.}\ \bibnamefont {Pan}},\ }\href {\doibase 10.1103/PhysRevLett.116.070502} {\bibfield  {journal} {\bibinfo  {journal}
	  {Phys. Rev. Lett.}\ }\textbf {\bibinfo {volume} {116}},\ \bibinfo {pages}
	  {070502} (\bibinfo {year} {2016})}\BibitemShut {NoStop}%
	\bibitem [{\citenamefont {Loredo}\ \emph {et~al.}(2016)\citenamefont {Loredo},
	  \citenamefont {Almeida}, \citenamefont {Di~Candia}, \citenamefont
	  {Pedernales}, \citenamefont {Casanova}, \citenamefont {Solano},\ and\
	  \citenamefont {White}}]{Loredoetal2016eqs}%
	  \BibitemOpen
	  \bibfield  {author} {\bibinfo {author} {\bibfnamefont {J.~C.}\ \bibnamefont
	  {Loredo}}, \bibinfo {author} {\bibfnamefont {M.~P.}\ \bibnamefont {Almeida}},
	  \bibinfo {author} {\bibfnamefont {R.}~\bibnamefont {Di~Candia}}, \bibinfo
	  {author} {\bibfnamefont {J.~S.}\ \bibnamefont {Pedernales}}, \bibinfo
	  {author} {\bibfnamefont {J.}~\bibnamefont {Casanova}}, \bibinfo {author}
	  {\bibfnamefont {E.}~\bibnamefont {Solano}}, \ and\ \bibinfo {author}
	  {\bibfnamefont {A.~G.}\ \bibnamefont {White}},\ }\href {\doibase 10.1103/PhysRevLett.116.070503} {\bibfield  {journal} {\bibinfo  {journal}
	  {Phys. Rev. Lett.}\ }\textbf {\bibinfo {volume} {116}},\ \bibinfo {pages}
	  {070503} (\bibinfo {year} {2016})}\BibitemShut {NoStop}%
	\bibitem [{\citenamefont {Cheng}\ \emph {et~al.}(2017)\citenamefont {Cheng},
	  \citenamefont {Arrazola}, \citenamefont {Pedernales}, \citenamefont {Lamata},
	  \citenamefont {Chen},\ and\ \citenamefont {Solano}}]{Chengetal2017eqs}%
	  \BibitemOpen
	  \bibfield  {author} {\bibinfo {author} {\bibfnamefont {X.-H.}\ \bibnamefont
	  {Cheng}}, \bibinfo {author} {\bibfnamefont {I.~n.}\ \bibnamefont {Arrazola}},
	  \bibinfo {author} {\bibfnamefont {J.~S.}\ \bibnamefont {Pedernales}},
	  \bibinfo {author} {\bibfnamefont {L.}~\bibnamefont {Lamata}}, \bibinfo
	  {author} {\bibfnamefont {X.}~\bibnamefont {Chen}}, \ and\ \bibinfo {author}
	  {\bibfnamefont {E.}~\bibnamefont {Solano}},\ }\href {\doibase 10.1103/PhysRevA.95.022305} {\bibfield  {journal} {\bibinfo  {journal} {Phys.
	  Rev. A}\ }\textbf {\bibinfo {volume} {95}},\ \bibinfo {pages} {022305}
	  (\bibinfo {year} {2017})}\BibitemShut {NoStop}%
	\bibitem [{\citenamefont {Matsumoto}(1997)}]{Matsumoto1997}%
	  \BibitemOpen
	  \bibfield  {author} {\bibinfo {author} {\bibfnamefont {K.}~\bibnamefont
	  {Matsumoto}},\ }\emph {\bibinfo {title} {A Geometrical Approach to Quantum
	  Estimation Theory}},\ \href {\doibase 10.11501/3163169} {Ph.D. thesis},\
	  \bibinfo  {school} {The University of Tokyo} (\bibinfo {year} {1997}),\
	  \bibinfo {note} {preprint avairable at
	  \url{https://doi.org/10.48550/arXiv.2111.09667}}\BibitemShut {NoStop}%
	\bibitem [{\citenamefont {Matsumoto}(2005)}]{Matsumoto2005}%
	  \BibitemOpen
	  \bibfield  {author} {\bibinfo {author} {\bibfnamefont {K.}~\bibnamefont
	  {Matsumoto}},\ }\enquote {\bibinfo {title} {A geometrical approach to quantum
	  estimation theory},}\ in\ \href {\doibase 10.1142/9789812563071_0021} {\emph
	  {\bibinfo {booktitle} {Asymptotic Theory of Quantum Statistical
	  Inference}}},\ \bibinfo {editor} {edited by\ \bibinfo {editor} {\bibfnamefont
	  {M.}~\bibnamefont {Hayashi}}}\ (\bibinfo  {publisher} {WORLD SCIENTIFIC},\
	  \bibinfo {year} {2005})\ Chap.~\bibinfo {chapter} {20}, pp.\ \bibinfo {pages}
	  {305--350}\BibitemShut {NoStop}%
	\bibitem [{\citenamefont {Stueckelberg}(1960)}]{Stueckelberg1960real}%
	  \BibitemOpen
	  \bibfield  {author} {\bibinfo {author} {\bibfnamefont {E.~C.~G.}\
	  \bibnamefont {Stueckelberg}},\ }\href@noop {} {\bibfield  {journal} {\bibinfo
	   {journal} {Helv. Phys. Acta}\ }\textbf {\bibinfo {volume} {33}},\ \bibinfo
	  {pages} {458} (\bibinfo {year} {1960})}\BibitemShut {NoStop}%
	\bibitem [{\citenamefont {Lahti}\ and\ \citenamefont
	  {Maczynski}(1987)}]{LahtiMaczynski1987real}%
	  \BibitemOpen
	  \bibfield  {author} {\bibinfo {author} {\bibfnamefont {P.~J.}\ \bibnamefont
	  {Lahti}}\ and\ \bibinfo {author} {\bibfnamefont {M.~J.}\ \bibnamefont
	  {Maczynski}},\ }\href {\doibase 10.1063/1.527822} {\bibfield  {journal}
	  {\bibinfo  {journal} {J. Math. Phys.}\ }\textbf {\bibinfo {volume} {28}},\
	  \bibinfo {pages} {1764} (\bibinfo {year} {1987})}\BibitemShut {NoStop}%
	\bibitem [{\citenamefont {Lu}\ and\ \citenamefont
	  {Wang}(2021)}]{LuWang2021heisenberg}%
	  \BibitemOpen
	  \bibfield  {author} {\bibinfo {author} {\bibfnamefont {X.-M.}\ \bibnamefont
	  {Lu}}\ and\ \bibinfo {author} {\bibfnamefont {X.}~\bibnamefont {Wang}},\
	  }\href {\doibase 10.1103/PhysRevLett.126.120503} {\bibfield  {journal}
	  {\bibinfo  {journal} {Phys. Rev. Lett.}\ }\textbf {\bibinfo {volume} {126}},\
	  \bibinfo {pages} {120503} (\bibinfo {year} {2021})}\BibitemShut {NoStop}%
	\bibitem [{\citenamefont {Uhlmann}(1986)}]{Uhlmann1986holonomy}%
	  \BibitemOpen
	  \bibfield  {author} {\bibinfo {author} {\bibfnamefont {A.}~\bibnamefont
	  {Uhlmann}},\ }\href {\doibase 10.1016/0034-4877(86)90055-8} {\bibfield
	  {journal} {\bibinfo  {journal} {Rep. Math. Phys.}\ }\textbf {\bibinfo
	  {volume} {24}},\ \bibinfo {pages} {229} (\bibinfo {year} {1986})}\BibitemShut
	  {NoStop}%
	\bibitem [{\citenamefont {Massar}(2000)}]{Massar2000}%
	  \BibitemOpen
	  \bibfield  {author} {\bibinfo {author} {\bibfnamefont {S.}~\bibnamefont
	  {Massar}},\ }\href {\doibase 10.1103/PhysRevA.62.040101} {\bibfield
	  {journal} {\bibinfo  {journal} {Phys. Rev. A}\ }\textbf {\bibinfo {volume}
	  {62}},\ \bibinfo {pages} {040101(R)} (\bibinfo {year} {2000})}\BibitemShut
	  {NoStop}%
	\bibitem [{\citenamefont {Polino}\ \emph {et~al.}(2020)\citenamefont {Polino},
	  \citenamefont {Valeri}, \citenamefont {Spagnolo},\ and\ \citenamefont
	  {Sciarrino}}]{Polinoetal2020}%
	  \BibitemOpen
	  \bibfield  {author} {\bibinfo {author} {\bibfnamefont {E.}~\bibnamefont
	  {Polino}}, \bibinfo {author} {\bibfnamefont {M.}~\bibnamefont {Valeri}},
	  \bibinfo {author} {\bibfnamefont {N.}~\bibnamefont {Spagnolo}}, \ and\
	  \bibinfo {author} {\bibfnamefont {F.}~\bibnamefont {Sciarrino}},\ }\href
	  {\doibase 10.1116/5.0007577} {\bibfield  {journal} {\bibinfo  {journal} {AVS
	  Quantum Sci.}\ }\textbf {\bibinfo {volume} {2}},\ \bibinfo {pages} {024703}
	  (\bibinfo {year} {2020})}\BibitemShut {NoStop}%
	\bibitem [{\citenamefont {Bennett}\ and\ \citenamefont
	  {Wiesner}(1992)}]{Bennettetal1992}%
	  \BibitemOpen
	  \bibfield  {author} {\bibinfo {author} {\bibfnamefont {C.~H.}\ \bibnamefont
	  {Bennett}}\ and\ \bibinfo {author} {\bibfnamefont {S.~J.}\ \bibnamefont
	  {Wiesner}},\ }\href {\doibase 10.1103/PhysRevLett.69.2881} {\bibfield
	  {journal} {\bibinfo  {journal} {Phys. Rev. Lett.}\ }\textbf {\bibinfo
	  {volume} {69}},\ \bibinfo {pages} {2881} (\bibinfo {year}
	  {1992})}\BibitemShut {NoStop}%
	\bibitem [{\citenamefont {Gross}\ and\ \citenamefont
	  {Caves}(2020)}]{GrossCaves2020onefrommany}%
	  \BibitemOpen
	  \bibfield  {author} {\bibinfo {author} {\bibfnamefont {J.~A.}\ \bibnamefont
	  {Gross}}\ and\ \bibinfo {author} {\bibfnamefont {C.~M.}\ \bibnamefont
	  {Caves}},\ }\href {\doibase 10.1088/1751-8121/abb9ed} {\bibfield  {journal}
	  {\bibinfo  {journal} {J. Phys. A: Math. Theor.}\ }\textbf {\bibinfo {volume}
	  {54}},\ \bibinfo {pages} {014001} (\bibinfo {year} {2020})}\BibitemShut
	  {NoStop}%
	\bibitem [{\citenamefont {Braunstein}\ and\ \citenamefont
	  {Caves}(1994)}]{BraunsteinCaves1994}%
	  \BibitemOpen
	  \bibfield  {author} {\bibinfo {author} {\bibfnamefont {S.~L.}\ \bibnamefont
	  {Braunstein}}\ and\ \bibinfo {author} {\bibfnamefont {C.~M.}\ \bibnamefont
	  {Caves}},\ }\href {\doibase 10.1103/PhysRevLett.72.3439} {\bibfield
	  {journal} {\bibinfo  {journal} {Phys. Rev. Lett.}\ }\textbf {\bibinfo
	  {volume} {72}},\ \bibinfo {pages} {3439} (\bibinfo {year}
	  {1994})}\BibitemShut {NoStop}%
	\bibitem [{\citenamefont {Braunstein}\ \emph {et~al.}(1996)\citenamefont
	  {Braunstein}, \citenamefont {Caves},\ and\ \citenamefont
	  {Milburn}}]{Braunsteinetal1996}%
	  \BibitemOpen
	  \bibfield  {author} {\bibinfo {author} {\bibfnamefont {S.~L.}\ \bibnamefont
	  {Braunstein}}, \bibinfo {author} {\bibfnamefont {C.~M.}\ \bibnamefont
	  {Caves}}, \ and\ \bibinfo {author} {\bibfnamefont {G.}~\bibnamefont
	  {Milburn}},\ }\href {\doibase 10.1006/aphy.1996.0040} {\bibfield  {journal}
	  {\bibinfo  {journal} {Ann. Phys.}\ }\textbf {\bibinfo {volume} {247}},\
	  \bibinfo {pages} {135} (\bibinfo {year} {1996})}\BibitemShut {NoStop}%
	\bibitem [{\citenamefont {Cerf}\ and\ \citenamefont
	  {Iblisdir}(2001)}]{CerfIblisdir2001}%
	  \BibitemOpen
	  \bibfield  {author} {\bibinfo {author} {\bibfnamefont {N.~J.}\ \bibnamefont
	  {Cerf}}\ and\ \bibinfo {author} {\bibfnamefont {S.}~\bibnamefont
	  {Iblisdir}},\ }\href {\doibase 10.1103/PhysRevLett.87.247903} {\bibfield
	  {journal} {\bibinfo  {journal} {Phys. Rev. Lett.}\ }\textbf {\bibinfo
	  {volume} {87}},\ \bibinfo {pages} {247903} (\bibinfo {year}
	  {2001})}\BibitemShut {NoStop}%
	\bibitem [{\citenamefont {Braunstein}\ \emph {et~al.}(2007)\citenamefont
	  {Braunstein}, \citenamefont {Ghosh},\ and\ \citenamefont
	  {Severini}}]{BraunsteinGhoshSeverini2007}%
	  \BibitemOpen
	  \bibfield  {author} {\bibinfo {author} {\bibfnamefont {S.~L.}\ \bibnamefont
	  {Braunstein}}, \bibinfo {author} {\bibfnamefont {S.}~\bibnamefont {Ghosh}}, \
	  and\ \bibinfo {author} {\bibfnamefont {S.}~\bibnamefont {Severini}},\ }\href
	  {\doibase 10.1088/1751-8113/40/8/009} {\bibfield  {journal} {\bibinfo
	  {journal} {J. Phys. A: Math. Theor.}\ }\textbf {\bibinfo {volume} {40}},\
	  \bibinfo {pages} {1809} (\bibinfo {year} {2007})}\BibitemShut {NoStop}%
	\bibitem [{\citenamefont {Kato}(2009)}]{Kato2009}%
	  \BibitemOpen
	  \bibfield  {author} {\bibinfo {author} {\bibfnamefont {G.}~\bibnamefont
	  {Kato}},\ }\href {\doibase 10.1103/PhysRevA.79.032315} {\bibfield  {journal}
	  {\bibinfo  {journal} {Phys. Rev. A}\ }\textbf {\bibinfo {volume} {79}},\
	  \bibinfo {pages} {032315} (\bibinfo {year} {2009})}\BibitemShut {NoStop}%
	\bibitem [{\citenamefont {Miyazaki}(2020)}]{Miyazaki2020strongly}%
	  \BibitemOpen
	  \bibfield  {author} {\bibinfo {author} {\bibfnamefont {J.}~\bibnamefont
	  {Miyazaki}},\ }\href {https://doi.org/10.48550/arXiv.2005.06685} {\bibfield
	  {journal} {\bibinfo  {journal} {arXiv preprint arXiv:2005.06685}\ } (\bibinfo
	  {year} {2020})}\BibitemShut {NoStop}%
	\bibitem [{\citenamefont {Tang}\ \emph {et~al.}(2020)\citenamefont {Tang},
	  \citenamefont {Hou}, \citenamefont {Shang}, \citenamefont {Zhu},
	  \citenamefont {Xiang}, \citenamefont {Li},\ and\ \citenamefont
	  {Guo}}]{Tangetal2020}%
	  \BibitemOpen
	  \bibfield  {author} {\bibinfo {author} {\bibfnamefont {J.-F.}\ \bibnamefont
	  {Tang}}, \bibinfo {author} {\bibfnamefont {Z.}~\bibnamefont {Hou}}, \bibinfo
	  {author} {\bibfnamefont {J.}~\bibnamefont {Shang}}, \bibinfo {author}
	  {\bibfnamefont {H.}~\bibnamefont {Zhu}}, \bibinfo {author} {\bibfnamefont
	  {G.-Y.}\ \bibnamefont {Xiang}}, \bibinfo {author} {\bibfnamefont {C.-F.}\
	  \bibnamefont {Li}}, \ and\ \bibinfo {author} {\bibfnamefont {G.-C.}\
	  \bibnamefont {Guo}},\ }\href {\doibase 10.1103/PhysRevLett.124.060502}
	  {\bibfield  {journal} {\bibinfo  {journal} {Phys. Rev. Lett.}\ }\textbf
	  {\bibinfo {volume} {124}},\ \bibinfo {pages} {060502} (\bibinfo {year}
	  {2020})}\BibitemShut {NoStop}%
	\bibitem [{\citenamefont {Bagan}\ \emph {et~al.}(2000)\citenamefont {Bagan},
	  \citenamefont {Baig}, \citenamefont {Brey}, \citenamefont {Mu\~noz Tapia},\
	  and\ \citenamefont {Tarrach}}]{Baganetal2000}%
	  \BibitemOpen
	  \bibfield  {author} {\bibinfo {author} {\bibfnamefont {E.}~\bibnamefont
	  {Bagan}}, \bibinfo {author} {\bibfnamefont {M.}~\bibnamefont {Baig}},
	  \bibinfo {author} {\bibfnamefont {A.}~\bibnamefont {Brey}}, \bibinfo {author}
	  {\bibfnamefont {R.}~\bibnamefont {Mu\~noz Tapia}}, \ and\ \bibinfo {author}
	  {\bibfnamefont {R.}~\bibnamefont {Tarrach}},\ }\href {\doibase 10.1103/PhysRevLett.85.5230} {\bibfield  {journal} {\bibinfo  {journal}
	  {Phys. Rev. Lett.}\ }\textbf {\bibinfo {volume} {85}},\ \bibinfo {pages}
	  {5230} (\bibinfo {year} {2000})}\BibitemShut {NoStop}%
	\bibitem [{\citenamefont {Jozsa}\ and\ \citenamefont
	  {Schlienz}(2000)}]{JoszaSchlienz2000}%
	  \BibitemOpen
	  \bibfield  {author} {\bibinfo {author} {\bibfnamefont {R.}~\bibnamefont
	  {Jozsa}}\ and\ \bibinfo {author} {\bibfnamefont {J.}~\bibnamefont
	  {Schlienz}},\ }\href {\doibase 10.1103/PhysRevA.62.012301} {\bibfield
	  {journal} {\bibinfo  {journal} {Phys. Rev. A}\ }\textbf {\bibinfo {volume}
	  {62}},\ \bibinfo {pages} {012301} (\bibinfo {year} {2000})}\BibitemShut
	  {NoStop}%
	\bibitem [{\citenamefont {Baumgratz}\ \emph {et~al.}(2014)\citenamefont
	  {Baumgratz}, \citenamefont {Cramer},\ and\ \citenamefont
	  {Plenio}}]{BaumgratzPlenio2014coherence}%
	  \BibitemOpen
	  \bibfield  {author} {\bibinfo {author} {\bibfnamefont {T.}~\bibnamefont
	  {Baumgratz}}, \bibinfo {author} {\bibfnamefont {M.}~\bibnamefont {Cramer}}, \
	  and\ \bibinfo {author} {\bibfnamefont {M.~B.}\ \bibnamefont {Plenio}},\
	  }\href {\doibase 10.1103/PhysRevLett.113.140401} {\bibfield  {journal}
	  {\bibinfo  {journal} {Phys. Rev. Lett.}\ }\textbf {\bibinfo {volume} {113}},\
	  \bibinfo {pages} {140401} (\bibinfo {year} {2014})}\BibitemShut {NoStop}%
	\bibitem [{\citenamefont {Chitambar}\ and\ \citenamefont
	  {Gour}(2016)}]{ChitambarGour2016coherence}%
	  \BibitemOpen
	  \bibfield  {author} {\bibinfo {author} {\bibfnamefont {E.}~\bibnamefont
	  {Chitambar}}\ and\ \bibinfo {author} {\bibfnamefont {G.}~\bibnamefont
	  {Gour}},\ }\href {\doibase 10.1103/PhysRevLett.117.030401} {\bibfield
	  {journal} {\bibinfo  {journal} {Phys. Rev. Lett.}\ }\textbf {\bibinfo
	  {volume} {117}},\ \bibinfo {pages} {030401} (\bibinfo {year}
	  {2016})}\BibitemShut {NoStop}%
	\bibitem [{\citenamefont {Winter}\ and\ \citenamefont
	  {Yang}(2016)}]{WinterDong2016coherence}%
	  \BibitemOpen
	  \bibfield  {author} {\bibinfo {author} {\bibfnamefont {A.}~\bibnamefont
	  {Winter}}\ and\ \bibinfo {author} {\bibfnamefont {D.}~\bibnamefont {Yang}},\
	  }\href {\doibase 10.1103/PhysRevLett.116.120404} {\bibfield  {journal}
	  {\bibinfo  {journal} {Phys. Rev. Lett.}\ }\textbf {\bibinfo {volume} {116}},\
	  \bibinfo {pages} {120404} (\bibinfo {year} {2016})}\BibitemShut {NoStop}%
	\bibitem [{\citenamefont {Designolle}\ \emph {et~al.}(2021)\citenamefont
	  {Designolle}, \citenamefont {Uola}, \citenamefont {Luoma},\ and\
	  \citenamefont {Brunner}}]{DesignolleBrunner2021setcoherence}%
	  \BibitemOpen
	  \bibfield  {author} {\bibinfo {author} {\bibfnamefont {S.}~\bibnamefont
	  {Designolle}}, \bibinfo {author} {\bibfnamefont {R.}~\bibnamefont {Uola}},
	  \bibinfo {author} {\bibfnamefont {K.}~\bibnamefont {Luoma}}, \ and\ \bibinfo
	  {author} {\bibfnamefont {N.}~\bibnamefont {Brunner}},\ }\href {\doibase 10.1103/PhysRevLett.126.220404} {\bibfield  {journal} {\bibinfo  {journal}
	  {Phys. Rev. Lett.}\ }\textbf {\bibinfo {volume} {126}},\ \bibinfo {pages}
	  {220404} (\bibinfo {year} {2021})}\BibitemShut {NoStop}%
	\bibitem [{\citenamefont {Chrostowski}\ \emph {et~al.}(2017)\citenamefont
	  {Chrostowski}, \citenamefont {Demkowicz-Dobrza\'{n}ski}, \citenamefont
	  {Jarzyna},\ and\ \citenamefont {Banaszek}}]{Chrostowskietal2017}%
	  \BibitemOpen
	  \bibfield  {author} {\bibinfo {author} {\bibfnamefont {A.}~\bibnamefont
	  {Chrostowski}}, \bibinfo {author} {\bibfnamefont {R.}~\bibnamefont
	  {Demkowicz-Dobrza\'{n}ski}}, \bibinfo {author} {\bibfnamefont
	  {M.}~\bibnamefont {Jarzyna}}, \ and\ \bibinfo {author} {\bibfnamefont
	  {K.}~\bibnamefont {Banaszek}},\ }\href {\doibase 10.1142/S0219749917400056}
	  {\bibfield  {journal} {\bibinfo  {journal} {Int. J. Quantum Inf.}\ }\textbf
	  {\bibinfo {volume} {15}},\ \bibinfo {pages} {1740005} (\bibinfo {year}
	  {2017})}\BibitemShut {NoStop}%
	\bibitem [{\citenamefont {\v{R}eh\'{a}\v{c}ek}\ \emph
	  {et~al.}(2017)\citenamefont {\v{R}eh\'{a}\v{c}ek}, \citenamefont {Hradil},
	  \citenamefont {Stoklasa}, \citenamefont {Pa\'ur}, \citenamefont {Grover},
	  \citenamefont {Krzic},\ and\ \citenamefont
	  {S\'anchez-Soto}}]{Rehaceketal2017}%
	  \BibitemOpen
	  \bibfield  {author} {\bibinfo {author} {\bibfnamefont {J.}~\bibnamefont
	  {\v{R}eh\'{a}\v{c}ek}}, \bibinfo {author} {\bibfnamefont {Z.}~\bibnamefont
	  {Hradil}}, \bibinfo {author} {\bibfnamefont {B.}~\bibnamefont {Stoklasa}},
	  \bibinfo {author} {\bibfnamefont {M.}~\bibnamefont {Pa\'ur}}, \bibinfo
	  {author} {\bibfnamefont {J.}~\bibnamefont {Grover}}, \bibinfo {author}
	  {\bibfnamefont {A.}~\bibnamefont {Krzic}}, \ and\ \bibinfo {author}
	  {\bibfnamefont {L.~L.}\ \bibnamefont {S\'anchez-Soto}},\ }\href {\doibase 10.1103/PhysRevA.96.062107} {\bibfield  {journal} {\bibinfo  {journal} {Phys.
	  Rev. A}\ }\textbf {\bibinfo {volume} {96}},\ \bibinfo {pages} {062107}
	  (\bibinfo {year} {2017})}\BibitemShut {NoStop}%
	\bibitem [{\citenamefont {Yu}\ and\ \citenamefont
	  {Prasad}(2018)}]{YuPrasad2018}%
	  \BibitemOpen
	  \bibfield  {author} {\bibinfo {author} {\bibfnamefont {Z.}~\bibnamefont
	  {Yu}}\ and\ \bibinfo {author} {\bibfnamefont {S.}~\bibnamefont {Prasad}},\
	  }\href {\doibase 10.1103/PhysRevLett.121.180504} {\bibfield  {journal}
	  {\bibinfo  {journal} {Phys. Rev. Lett.}\ }\textbf {\bibinfo {volume} {121}},\
	  \bibinfo {pages} {180504} (\bibinfo {year} {2018})}\BibitemShut {NoStop}%
	\bibitem [{\citenamefont {Suzuki}(2019)}]{Suzuki2019}%
	  \BibitemOpen
	  \bibfield  {author} {\bibinfo {author} {\bibfnamefont {J.}~\bibnamefont
	  {Suzuki}},\ }\href {\doibase 10.3390/e21070703} {\bibfield  {journal}
	  {\bibinfo  {journal} {Entropy}\ }\textbf {\bibinfo {volume} {21}},\ \bibinfo
	  {pages} {703} (\bibinfo {year} {2019})}\BibitemShut {NoStop}%
	\end{thebibliography}
\end{document}